\documentclass[showpacs,prd,aps,nofootinbib]{revtex4}

\pdfoutput=1

\usepackage[colorlinks,urlcolor=blue,citecolor=blue]{hyperref}

\usepackage{amsmath}
\usepackage{amssymb}
\usepackage{latexsym}
\usepackage{amsfonts}
\usepackage{epsfig}
\usepackage{psfrag}
\usepackage{graphicx}

\def\hhref#1{\href{http://arxiv.org/abs/#1}{arXiv:#1}} 

\usepackage{color}

\newcommand{\bea}{\begin{eqnarray}}
\newcommand{\ea}{\end{eqnarray}}
\newcommand{\eea}{\end{eqnarray}}

\begin{document}

\title{Super-Adiabatic Particle Number in Schwinger and de Sitter Particle Production}

\author{Robert Dabrowski$^{1,2}$ and Gerald~V.~Dunne$^{1,2,3}$}

\affiliation{
$^1$Theoretisch-Physikalisches Institut, Friedrich-Schiller-Universit\"at Jena,
Max-Wien-Platz 1,  D-07743 Jena, Germany\\
$^2$Department of Physics, College of Liberal Arts and Sciences, University of Connecticut,
Storrs CT 06269-3046, USA\\
$^3$Physics Department, Technion, Israel Institute of Technology, Haifa 32000, Israel}

\begin{abstract}
We consider the time evolution of the adiabatic particle number in both time-dependent electric fields and in de Sitter spaces, and  define a super-adiabatic particle number in which the (divergent) adiabatic expansion is truncated at optimal order. In this {\it super-adiabatic basis}, the particle number evolves smoothly in time, according to Berry's universal adiabatic smoothing of the Stokes phenomenon. This super-adiabatic basis also illustrates clearly the quantum interference effects associated with  particle production, in particular for sequences of time-dependent electric field pulses, and in eternal de Sitter space where there is constructive interference in even dimensions, and destructive interference in odd dimensions.

\end{abstract}



\maketitle

\section{Introduction}

Under the influence of certain gauge or gravitational curvatures, particles can be produced from vacuum. This unusual feature of quantum field theory  has far-reaching consequences for cosmology and astrophysics \cite{linde,liddle,mukhanov,blhu,Traschen:1990sw,Peebles:1998qn,Vachaspati:2006ki,Antoniadis:2006wq}, as well as for fundamental properties of particle physics, which may be probed with intense electromagnetic \cite{greiner,dittrich-gies,mourou,dunne-eli} or chromo-electromagnetic fields \cite{McLerran:1994vd,dima,raju,gelis}, and also in analogue systems \cite{nation}. Well-known examples related to the  particle production phenomenon include: 
\\
(i) the Schwinger effect, the non-perturbative production of electron-positron pairs when an external electric field is applied to  the quantum electrodynamical (QED) vacuum \cite{he,schwinger,Ringwald:2001ib,dunne,Ruffini:2009hg};
\\
(ii) cosmological particle production due to expanding cosmologies \cite{parker,Parker:1972kp,zeldovich} and de Sitter spacetime \cite{emil,Bousso:2001mw,Polyakov:2007mm,padman,greenwood,Anderson:2013ila,Anderson:2013zia};
\\
(iii) Hawking radiation due to black holes and gravitational horizon effects 
  \cite{hawking-ellis,Gibbons:1977mu,Birrell:1982ix,Fulling:1989nb,Ford:1986sy,Parikh:1999mf,Schutzhold:2010ig};
\\
(iv) Unruh radiation, particle number as seen by an accelerating observer \cite{Unruh:1976db};
\\
The quantitative description of  particle production processes requires a well-defined concept of particle number, and these processes can be analyzed in various equivalent formalisms, for example:  Bogoliubov transformations \cite{parker,Fulling:1989nb,emil,lapedes,kluger,Anderson:2013ila,Anderson:2013zia}, Green's function methods \cite{Gavrilov:1996pz},  semiclassical tunneling \cite{feynman-positron,brezin,popov,Parikh:1999mf,KeskiVakkuri:1996gn,Kim:2000un}, worldline instantons \cite{Affleck:1981bma,Dunne:2005sx}, quantum kinetic equations \cite{kluger,Schmidt:1998vi,Alkofer:2001ik}, the Dirac-Heisenberg-Wigner formalism \cite{BialynickiBirula:1991tx,Aarts:1999zn,Hebenstreit:2010vz}, classical-statistical formulations \cite{Gelis:2013oca,Hebenstreit:2013qxa,Fukushima:2014sia}, amongst others.  In addition to technical computational issues, these processes raise challenging conceptual questions about quantum information, entanglement entropy, quantum back reaction, radiation reaction and the possibility of a maximum attainable electric field.

One of the main interesting aspects of vacuum pair production is the lack of a unique definition of the vacuum in a curved spacetime or in a time-dependent electromagnetic field \cite{DeWitt:1975ys}. An essential difficulty is that one is trying to relate a final  vacuum to an initial vacuum, with the vacua connected by an intervening time-dependent perturbation of the system. While a Fock vacuum may be defined at past- and future-infinity when the perturbation vanishes, there is no unique separation into positive and negative energy states at intermediate times. This is just like the fact that there is no unique separation into left- and right-going waves in an inhomogeneous dielectric medium \cite{budden,Berry:1972na}. A standard approach is to define particle number with respect to an adiabatic basis \cite{parker,Parker:1972kp,Birrell:1982ix,Fulling:1989nb,blhu,Parker:1974qw,kluger,Kim:2011jw,Chung:1998bt,Habib:1999cs,Winitzki:2005rw,Anderson:2013ila,Anderson:2013zia}.
In the adiabatic approach, particle number is defined with respect to a basis of approximate states which reduce to free states at past infinity. Such a  construction is usually based on a semiclassical WKB-type approximation. However,  such WKB approximations are only defined within  certain Stokes regions, which typically do not include both past- and future-infinity. Thus, the evaluation of the  final particle number requires evolving across  Stokes lines, which means that particle production is a direct manifestation of the Stokes phenomenon \cite{dd1,Akkermans:2011yn}, the jumps that arise in certain quantities when Stokes lines are crossed. This Stokes phenomenon approach has been used to provide a simple quantitative understanding of the quantum interference effects that arise in the Schwinger effect when the probe laser pulse has non-trivial temporal structure \cite{Hebenstreit:2009km}, and also to analyze particle production in de Sitter space \cite{Kim:2010xm}. Such interference phenomena due to sub-cycle pulse structure are well known, for example, in strong-field atomic-molecular-optical (AMO) physics \cite{popov-amo}, as well as in condensed matter physics \cite{oka,shevchenko}. 

In this paper  we extend the Stokes phenomenon interpretation of particle production to the question of the {\it time evolution} of the particle number, not just the particle number at asymptotically early and late times, but also at intermediate times. This  immediately confronts the aforementioned problem of the non-uniqueness of the adiabatic basis. However, there is a remarkable general result in semiclassical asymptotic analysis, due to Dingle \cite{dingle} and Berry \cite{berry,berry-uniform}, that describes a universal smoothing of the Stokes jumps in a special {\it super-adiabatic basis} in which the (divergent and asymptotic) adiabatic expansion is truncated at optimal order in the vicinity of the turning points. This mathematical result has previously  been applied to the quantum evolution of non-relativistic two-level systems \cite{lim-berry,wimberger}. Here we apply this concept of a super-adiabatic basis to the problem of particle production in quantum field theory. We illustrate the idea with applications to the Schwinger effect in sequences of time-dependent electric field pulses, and also to particle production in de Sitter spacetime. The main results are:
\begin{enumerate}
\item
the time evolution of the super-adiabatic particle number is significantly smoother than in other adiabatic bases, with much smaller fluctuations;
\item
 the super-adiabatic basis reveals more clearly the quantum interference effects which previously have been identified only in the asymptotically late time particle number.
\end{enumerate}

In Section II we recall the Bogoliubov transformation approach to particle number, and its relation to the adiabatic basis. In Section III we review the Dingle-Berry results concerning the large-order behavior of the adiabatic expansion, and their implications for a universal smoothing of the Stokes phenomenon. In Sections IV and V we apply this super-adiabatic basis to the Schwinger effect, and to particle production in de Sitter space. Section VI is devoted to a brief discussion of these results.

\section{Adiabatic  Particle Number}

\subsection{Bogoliubov Transformation and Adiabatic Particle Number}

The production of particles from vacuum may be characterized by a Bogoliubov transformation that connects asymptotic vacua before and after the effect of an external field \cite{Birrell:1982ix,Fulling:1989nb,parker,Parker:1972kp,Anderson:2013ila,Anderson:2013zia}. For the Schwinger effect in a time-dependent background electric field, and for a cosmological background Friedmann-Robertson-Walker (FRW) gravitational field (the Parker-Zeldovich effect), the computational problem reduces to the study of an oscillator problem with a time-dependent  frequency:
\begin{eqnarray}
\left(\frac{d^2}{dt^2}+\omega_k^2(t)\right)f_k(t)=0
\label{osc}
\end{eqnarray}
The  modes $k$ are decoupled due to the spatial homogeneity of the background.

For the Schwinger effect, the oscillator equation (\ref{osc})  comes from a mode reduction of the Klein-Gordon equation [for notational simplicity we consider scalar QED, but spinor QED is similar \cite{dd1}], $\left(-D_\mu^2+m^2\right)\Phi=0$,  for modes $\Phi\sim e^{i\vec{k}\cdot\vec{x}}f_k(t)$, in the gauge with time-dependent vector potential $A_\parallel(t)$. Then the effective time-dependent frequency is \cite{popov,brezin}
\begin{eqnarray}
\omega_k^2(t)=m^2+k_\perp^2+\left(k_\parallel- A_\parallel(t)\right)^2
\label{schwinger-osc}
\end{eqnarray}
For the Parker-Zeldovich effect, the oscillator equation  (\ref{osc}) comes from a a mode reduction of the Klein-Gordon equation, $\left(-\frac{1}{\sqrt{-g}}\,\frac{\partial}{\partial x^a}\left(\sqrt{-g}g^{ab}\frac{\partial}{\partial x^b}\right)+M^2\right)\Phi=0$, for modes $\Phi=Y_{k}(\Sigma)\, f_k(t)/a(t)^{(d-1)/2}$, for the FRW metric $ds^2=-dt^2+a^2(t)\,d\Sigma^2$, with expansion parameter $a(t)$. Here the effective mass-squared is $M^2=m^2+\xi R$, with coupling parameter $\xi$ and curvature $R = d(d-1) H^2$. de Sitter space is described in global coordinates by the scale parameter $a(t)=\frac{1}{H}\cosh(H\, t)$. Then the effective time-dependent frequency is \cite{parker,zeldovich}
\begin{align}
\omega_k^2(t) &= H^2\left( \gamma^2 + \left( \frac{2k + d - 3}{2} \right) \left( \frac{2k+d-1}{2} \right) \text{sech}^2(H t) \right)
\label{parker-osc1} \\
\gamma^2 &= \frac{m^2}{H^2} + d(d-1) \left( \xi - \frac{d-2}{4(d-1)} \right) - \frac{1}{4}
\label{parker-osc2}
\end{align}
A conformally coupled scalar has $\xi=\frac{1}{6}$ in four dimensional spacetime, and $\xi=\frac{1}{8}$ in three dimensional spacetime.

The Bogoliubov transformation defines a set of time-dependent creation and annihilation operators, $\tilde{a}_{k}(t)$ and $\tilde{b}_{k}(t)$, 
related to the original time-independent operators, $a_{k}$ and $b_{k}$, by:
\begin{eqnarray}
\begin{pmatrix}
\tilde{a}_{k}(t)\cr
\tilde{b}_{ - k}^\dagger(t)
\end{pmatrix}
=\begin{pmatrix}
\alpha_{ k}(t) & \beta_{k}^*(t) \cr
\beta_{ k}(t)  & \alpha_{k}^*(t) 
\end{pmatrix}
\begin{pmatrix}
a_{k}\cr
b_{ -k}^\dag
\end{pmatrix}
\label{bog2}
\end{eqnarray}
For scalar fields $|\alpha_{k}(t)|^2-|\beta_{k}(t)|^2=1$, for all $t$. [For an uncharged scalar we replace $(b_{k}, b_{ -k}^\dagger)$ with $(a_{k}, a_{ -k}^\dagger)$.]

The time-dependent adiabatic particle number, $\tilde{\mathcal N}_k(t)$, is defined for each mode 
$k$ by the expectation value in the original vacuum state $|0\rangle$ of the time-dependent number operator 
$ \tilde{a}_k^\dagger(t)\,\tilde{a}_k(t)$ 
\begin{eqnarray}
\tilde{\mathcal N}_k(t)\equiv \langle 0| \tilde{a}_k^\dagger(t)\,\tilde{a}_k(t)\,| 0\rangle =|\beta_k(t)|^2
\label{an}
\end{eqnarray}
In the last step we have used the fact that the vacuum is annihilated by the original creation and annihilation operators, $a_k|0\rangle=0=b_{-k}|0\rangle$, assuming no particles are present initially.
The total number of particles produced in the mode $k$ is given by the final value of  $\tilde{\mathcal N}_k(t)$:
\begin{eqnarray}
\tilde{N}_k\equiv \tilde{\mathcal N}_k(t=+\infty)  =|\beta_k(t=+\infty)|^2
\label{anfinal}
\end{eqnarray}
In this paper we are interested in the full time evolution of the adiabatic particle number $\tilde{\mathcal N}_k(t)$, as it evolves from an initial value of zero to some final asymptotic value $\tilde{N}_k\equiv \tilde{\mathcal N}_k(t=+\infty)$.

\subsection{Basis Dependence of  the Adiabatic Particle Number}

In a time-dependent background field there is no unique separation into positive and negative energy states \cite{DeWitt:1975ys,parker}. However, for a {\it slowly varying} time-dependent background we can define an adiabatic particle number with respect to a reference basis that has a semiclassical limit corresponding to ordinary positive and negative energy plane waves when the background is constant. 

The adiabatic particle number thus depends on this choice of basis, and is specified as follows. We choose a set of reference mode functions 
\begin{eqnarray}
\tilde{f}_k(t)\equiv \frac{1}{\sqrt{2\, W_k(t)}}\, e^{-i\int^t W_k}
\label{modes}
\end{eqnarray}
where $W_k(t)$ is specified below. We then express the exact solution to the oscillator equation (\ref{osc}) as
\begin{eqnarray}
f_k(t)=\alpha_k(t)\, \tilde{f}_k(t)+\beta_k(t)\, \tilde{f}_{-k}^*(t)
\label{decomp}
\end{eqnarray}
This can be interpreted equivalently as a change of basis of creation and annihilation operators, writing the  mode-decomposition of the field operator as
\begin{eqnarray}
\phi_k(t)&=&a_k f_k(t)+b_{-k}^\dagger f_{-k}^*(t) \\
&=&\tilde{a}_k(t)\tilde{f}_k(t)+\tilde{b}_{-k}^\dagger(t) \tilde{f}^*_{-k}(t)
\label{osc2}
\end{eqnarray}
Thus, the Bogoliubov transformation of creation and annihilation operators in (\ref{bog2}) is consistent with the linear transformation (\ref{decomp}) between the exact solution $f_k(t)$ and the mode functions $\tilde{f}_k(t)$.

We must also specify the decomposition of the first derivative of $f_k(t)$ in terms of the mode functions, and the general form consistent with unitarity (preserving the Wronskian relation: $f^*\, \dot{f}-\dot{f}^*\, f=i$) is
\begin{eqnarray}
\dot{f}_k(t)=\left(-i\, W_k(t)+\frac{1}{2} V_k(t)\right)\alpha_k(t)\, \tilde{f}_k(t)+\left(i\, W_k(t)+\frac{1}{2} V_k(t)\right)\beta_k(t)\, \tilde{f}_{-k}^*(t)
\label{decomp2}
\end{eqnarray}
where $V_k(t)$ is to be chosen. The freedom  in the choice of $W_k(t)$ and $V_k(t)$ encodes the arbitrariness in defining positive and negative energy states at intermediate times.

Conventional choices are based on a WKB approximation, taking $W_k(t)=\omega_k(t)$. Two common choices are
\begin{enumerate}
\item $W_k(t)=\omega_k(t)$ and $V_k(t)=0$. This choice is made for example in \cite{brezin,popov,dd1}. Then substituting (\ref{decomp}, \ref{decomp2}) into the oscillator equation (\ref{osc}) we obtain the time evolution of the Bogoliubov coefficients $\alpha_k(t)$ and $\beta_k(t)$ as
\begin{eqnarray}
\begin{pmatrix}
\dot{\alpha}_{k}(t)\cr
\dot{\beta}_{k}(t)
\end{pmatrix}
=
\frac{\dot{\omega}_{{k}}(t)}{2\omega_{{k}}(t)}
\begin{pmatrix}
0 &&  e^{2i\int^t \omega_{{k}}}\\
e^{-2i\int^t \omega_{{k}}} && 0
\end{pmatrix}
\begin{pmatrix}
{\alpha}_{{k}}(t)\cr
{\beta}_{{k}}(t)
\end{pmatrix} 
\label{abdot1}
\end{eqnarray}

\item $W_k(t)=\omega_k(t)$ and $V_k(t)= - \frac{\dot{W}_k(t)}{W_k(t)}= - \frac{\dot{\omega}_k(t)}{\omega_k(t)}$. This choice is made for example in \cite{berry,kluger}. 
Then substituting (\ref{decomp}, \ref{decomp2}) into the oscillator equation (\ref{osc}) we obtain the time evolution of the Bogoliubov coefficients $\alpha_k(t)$ and $\beta_k(t)$ as
\begin{eqnarray}
\begin{pmatrix}
\dot{\alpha}_{k}(t)\cr
\dot{\beta}_{k}(t)
\end{pmatrix}
=
\frac{1}{4 i} \left( \frac{3}{2} \frac{\dot{\omega}_{k}^2(t)}{\omega_{k}^3(t)} - \frac{\ddot{\omega}_{{k}}(t)}{\omega_{{k}}^2(t)} \right)
\begin{pmatrix}
1 &&  e^{2i\int^t \omega_{{k}}}\\
-e^{-2i\int^t \omega_{{k}}} && -1
\end{pmatrix}
\begin{pmatrix}
{\alpha}_{{k}}(t)\cr
{\beta}_{{k}}(t)
\end{pmatrix} 
\label{abdot2}
\end{eqnarray}

\end{enumerate}

\noindent Similar to the derivation of (\ref{abdot1}) and (\ref{abdot2}), the time evolution of the Bogoliubov coefficients $\alpha_k(t)$ and $\beta_k(t)$ for arbitrary $W_k(t)$ and $V_k(t)$ can be found and takes the form
\begin{align}
\begin{pmatrix}
\dot{\alpha}_{k}(t) \\
\dot{\beta}_{k}(t)
\end{pmatrix}
=
\begin{pmatrix}
\delta_{k}(t)					& \big[ \Delta_{k}(t) + \delta_{k}(t) \big] e^{2 i \int^t W_{k}}		\\
\big[ \Delta_{k}(t) - \delta_{k}(t) \big] e^{-2 i \int^t W_{k}}	& -\delta_{k}(t)					\\
\end{pmatrix}
\begin{pmatrix}
\alpha_{k}(t) \\
\beta_{k}(t)
\end{pmatrix}
\label{evogen}
\end{align}
such that
\begin{align}
\delta_{k}(t) &\equiv \frac{1}{2 i W_{k}(t)} \left[ \omega_{k}^2(t) - W_{k}^2(t) + \left(\frac{\dot{V}_{k}(t)}{2} + \frac{V_{k}^2(t)}{4} \right) \right] \label{evogen1} \\
\Delta_{k}(t) &\equiv \frac{\dot{W}_{k}(t)}{2W_{k}(t)} + \frac{V_{k}(t)}{2}
\label{evogen2}
\end{align}
The basis choice $W_k(t) = \omega_k(t)$ and $V_k(t) = 0$ then implies that $\delta_k(t) = 0$ and $\Delta_k(t)=\frac{\dot{\omega}_k(t)}{2 \omega_k(t)}$, while the second choice $W_k(t) = \omega_k(t)$ and $V_k(t)=-\frac{\dot{W}_k(t)}{W_k(t)}=-\frac{\dot{\omega}_k(t)}{\omega_k(t)}$ implies that $\Delta_k(t) = 0$ and $\delta_k(t) = \frac{1}{4 i} \left( \frac{3}{2} \frac{\dot{\omega}_{k}^2(t)}{\omega_{k}^3(t)}-\frac{\ddot{\omega}_{{k}}(t)}{\omega_{{k}}^2(t)} \right)$.

From (\ref{an}) we see that the time evolution of the adiabatic particle number $\tilde{\mathcal N}_k(t)$ is given by the time evolution of (the magnitude squared of) the Bogoliubov coefficient $\beta_k(t)$ that enters the Bogoliubov transformation (\ref{bog2}). We thus solve the evolution equations (\ref{abdot1}) or (\ref{abdot2}) for $\beta_k(t)$, with initial conditions $\alpha_k(t=-\infty)=1$, and $\beta_k(t=-\infty)=0$. Since the evolution equations depend on the choice made for $W_k(t)$ and $V_k(t)$ in the mode functions, the time evolution of $\alpha_k(t)$ and  $\beta_k(t)$, and hence of the adiabatic particle number $\tilde{\mathcal N}_k(t)$, will depend on the basis choice.

\begin{figure}[h!]
\begin{tabular}{cc}
\centering
\includegraphics[width=8.5cm]{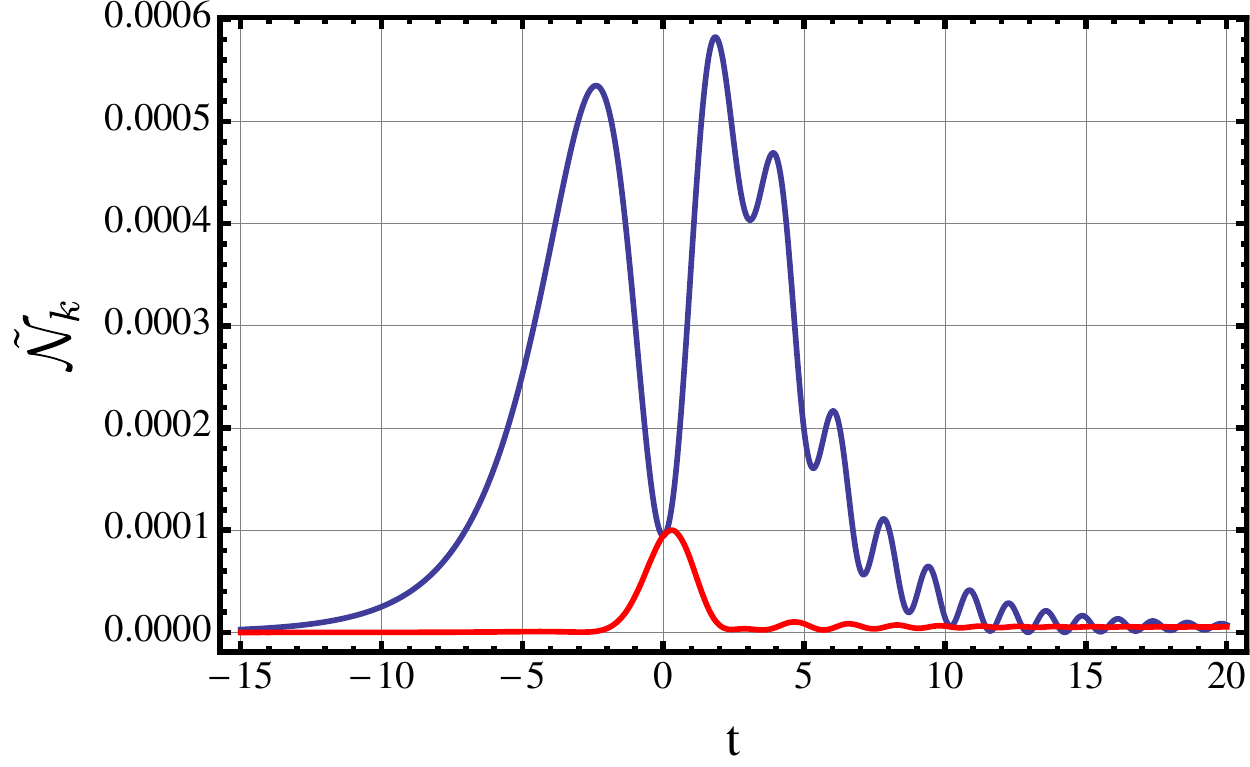} 
& 
\includegraphics[width=8.5cm]{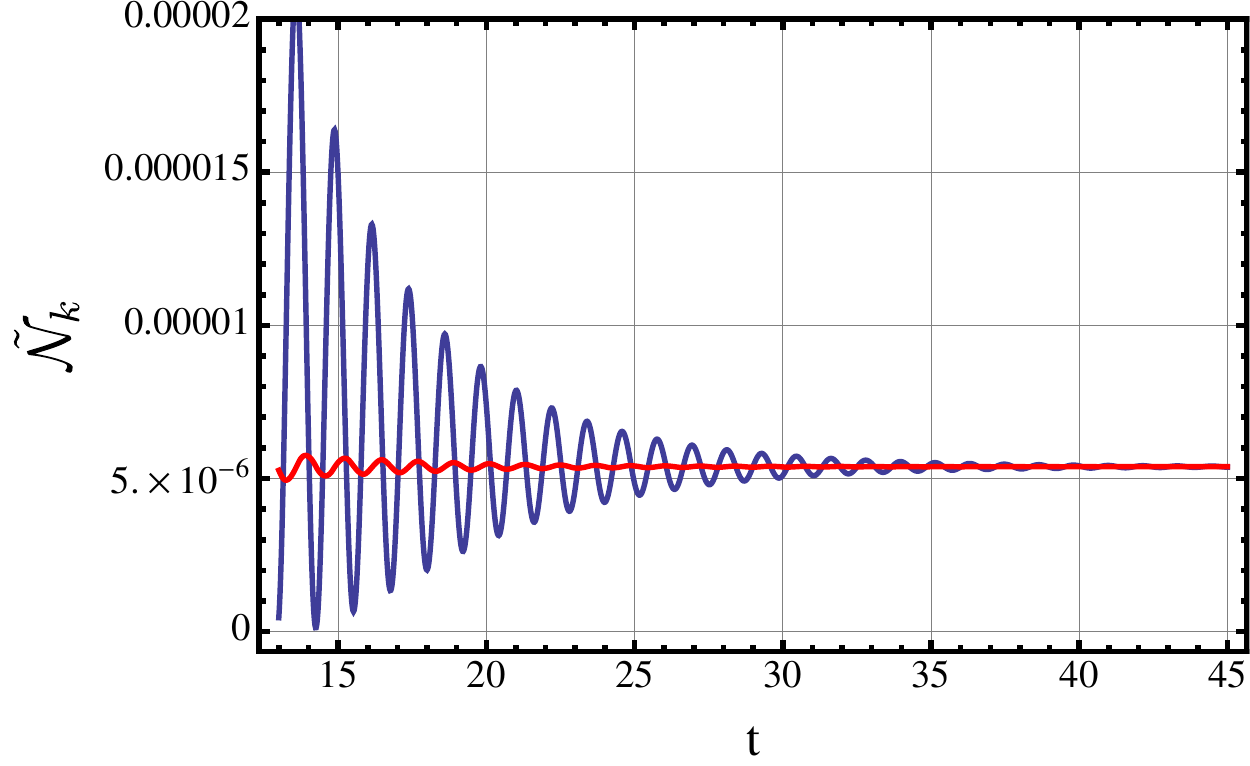}
\end{tabular}
\caption{The adiabatic particle number $\tilde{\mathcal N}_k(t)$ in (\ref{an}), defined with respect to two different adiabatic bases, for the Schwinger effect in a time-dependent electric field with longitudinal vector potential $A_\parallel = - \frac{E}{a} \tanh(a t)$, corresponding to a single electric field pulse, $E(t) = E\, {\rm sech}^2(at)$. The pulse parameters for this plot are: $E = 0.25, a = 0.1, k_\perp = 0, k_\parallel = 0$, all in units with $m=1$. The blue curves correspond to the basis $W_k(t) = \omega_k(t)$ and $V_k(t) = 0$, with associated evolution equations (\ref{abdot1}), while the red curves correspond to the basis $W_k(t) = \omega_k(t)$ and $V_k(t) = - \frac{\dot{\omega}_k(t)}{\omega_k(t)}$, with associated evolution equations (\ref{abdot2}). 
The right-hand figure shows the late-time evolution [notice the different vertical scale]. Notice that at  intermediate times there is a large difference in the time evolution profile of $\tilde{\mathcal N}_k(t)$, both in the scale and form of the oscillations, while the  final asymptotic value of the particle number is the same in the two bases.
}
\label{1pulsebasisdependence}
\end{figure}
\begin{figure}[h!]
\begin{tabular}{cc}
\centering
\includegraphics[width=8.5cm]{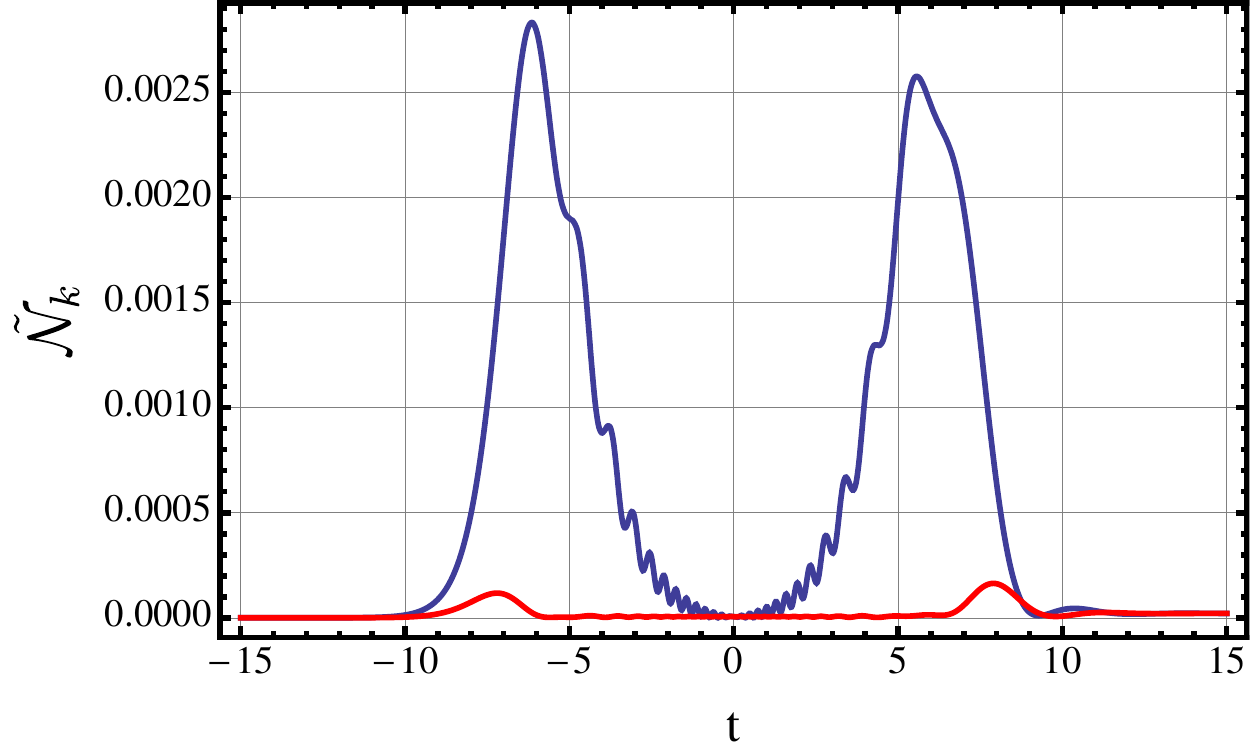} 
& 
\includegraphics[width=8.5cm]{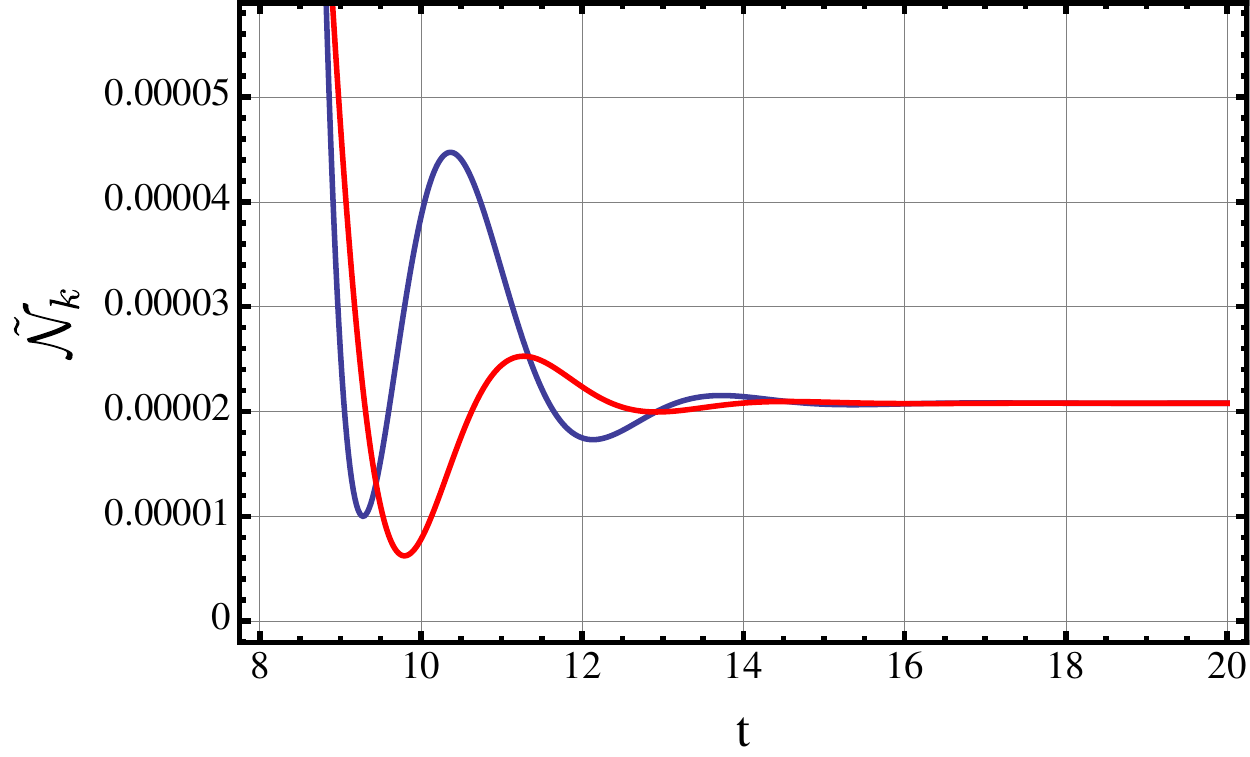}
\end{tabular}
\caption{The adiabatic particle number $\tilde{\mathcal N}_k(t)$ in (\ref{an}), defined with respect to two different adiabatic bases, for particle creation in 4 dimensional de Sitter space with conformal coupling. The blue curves correspond to the basis $W_k(t) = \omega_k(t)$ and $V_k(t) = 0$, with associated evolution equations (\ref{abdot1}), while the red curves correspond to the basis $W_k(t) = \omega_k(t)$ and $V_k(t) = - \frac{\dot{\omega}_k(t)}{\omega_k(t)}$, with associated evolution equations (\ref{abdot2}). The de Sitter space has physical parameters: $H = 0.5, k = 25$, in units with mass scale $m = 1$. 
The right-hand figure shows the late-time evolution [notice the very different vertical scale]. Notice that at  intermediate times there is a very large difference in the time evolution profile of $\tilde{\mathcal N}_k(t)$, both in the scale and form of the oscillations, while the  final asymptotic value of the particle number is the same in the two bases.}
\label{dS4basisdependence}
\end{figure}
Nevertheless, a crucial observation is that the final asymptotic value, the total particle number $N_k=\tilde{\mathcal N}_k(t=+\infty)$ is independent of the choice of basis. This is illustrated in Figure \ref{1pulsebasisdependence} for the case of Schwinger particle production in a single-pulse electric field, and in  Figure \ref{dS4basisdependence} for particle production in dS$_4$, using the two different choices of mode functions listed above in  (\ref{abdot1}) and  (\ref{abdot2}). The final particle number is the same for the two bases, but the adiabatic particle number $\tilde{\mathcal N}_k(t)$ is different at intermediate times. These differences include both differences in the scale, and in the actual form of the oscillatory behavior at intermediate times.

\section{Super-Adiabatic Particle Number}

In this section we review how to generate higher orders of the adiabatic expansion, and we introduce the concept of the super-adiabatic basis.

\subsection{The Adiabatic Expansion}

The adiabatic expansion is obtained by substituting into the basic time-dependent oscillator equation (\ref{osc}) an ansatz:
\begin{eqnarray}
f_k(t) = \frac{1}{\sqrt{2 W_k(t)}}\,e^{-i \int^t W_k}
\label{aexp1}
\end{eqnarray}
This gives a solution to (\ref{osc}) provided that $W_k(t)$ is related to the time-dependent frequency $\omega_k(t)$ as:
\begin{eqnarray}
W_k^2 =
\omega_k^2-\left[ \frac{\ddot{W}_k}{2 W_k} -\frac{3}{4} \!  \left( \frac{\dot{W}_k}{W_k} \right)^{\!\! 2}\right]
\label{aexp2}
\end{eqnarray}
Successive orders of the adiabatic expansion are obtained by truncating the expansion for $W_k(t)$ at a certain order of derivatives of $\omega_k(t)$:
\begin{enumerate}
\item
Leading order: 
\begin{eqnarray}
W_k^{(0)}(t)=\omega_k(t)
\label{ae1}
\end{eqnarray}
\item
Next-to-leading order: 
\begin{eqnarray}
W_k^{(1)}(t)=\omega_k(t)-\frac{1}{4}\left(\frac{\ddot{\omega}_k}{\omega_k^2}-\frac{3}{2}\frac{\dot{\omega}_k^2}{\omega_k^3}\right)
\label{ae2}
\end{eqnarray}
\item
Next-to-next-to-leading order: 
\begin{eqnarray}
W_k^{(2)}(t) =
\omega_k(t) - \frac{1}{4}\left(\frac{\ddot{\omega}_k}{\omega_k^2}-\frac{3}{2}\frac{\dot{\omega}_k^2}{\omega_k^3}\right) 
 - 	\frac{1}{8} \left( 
		\frac{13}{4} \frac{\ddot{\omega}_k^2}{\omega_k^5}
		- \frac{99}{4} \frac{\dot{\omega}_k^2 \ddot{\omega}_k}{\omega_k^6} 
		+ 5 \frac{\dot{\omega}_k \dddot{\omega \hspace{0pt}}_{\! k}}{\omega_k^5} 
		+ \frac{1}{2} \frac{ \stackrel{\tiny \mbox{(4)}}{\omega}_{\! k}}{\omega_k} 
		- \frac{297}{16} \frac{\dot{\omega}_k^4}{\omega_k^7} \right)
\label{ae3}
\end{eqnarray}

\item
At $(j+1)$-th order, we obtain the expansion of 
\begin{eqnarray}
W_k^{(j+1)} = \sqrt{\omega_k^2 - \left[ \frac{\ddot{W}_k^{(j)}}{2 W_k^{(j)}} - \frac{3}{4} \! \left( \frac{\dot{W}_k^{(j)}}{W_k^{(j)}} \right)^{\!\! 2} \right]}
\end{eqnarray}
truncated at terms involving at most $2j$ derivatives with respect to $t$.

\end{enumerate}

In this paper we will adopt the choice $V_k(t) = - \frac{\dot{W}_k(t)}{W_k(t)}$,  for the function relating the function $V_k(t)$ in (\ref{decomp2}) to the function $W_k(t)$. 
The final super-adiabatic universal time behavior derived below  is actually independent of this choice \cite{berry}. Thus, the time evolution of the Bogoliubov coefficients $\alpha_k(t)$ and $\beta_k(t)$, from which we deduce the adiabatic particle number using (\ref{an}), is given by the coupled equations (\ref{evogen}, \ref{evogen1}, \ref{evogen2}) evaluated with $V_k(t) = - \frac{\dot{W}_k(t)}{W_k(t)}$, with $W_k(t)$ replaced by $W_k^{(j)}(t)$ for the entire expression. 
At the $j$-th adiabatic order it reads (recall that the subscript $k$ labels the longitudinal momentum, while the superscript $(j)$ labels the order of the adiabatic expansion):
\begin{align}
\begin{pmatrix}
\dot{\alpha}_{k}^{(j)}(t)     \cr
\dot{\beta}_{k}^{(j)}(t)
\end{pmatrix}
&= \Lambda_k^{(j)}(t) 
\begin{pmatrix}
1 &&  e^{2i\int^t W_{k}^{(j)}}\\
-e^{-2i\int^t W_{k}^{(j)}} && -1
\end{pmatrix}
\begin{pmatrix}
{\alpha}_{{k}}^{(j)}(t)\cr
{\beta}_{{k}}^{(j)}(t)
\end{pmatrix} \label{abWdot2} \\
 \Lambda_k^{(j)}(t) &= \frac{1}{2 i W_k^{(j)}(t)} \left[ \omega_k^2(t) - \left( W_k^{(j)}(t) \right)^2 + \left( \frac{3}{4} \left( \frac{\dot{W}_k^{(j)}(t)}{W_k^{(j)}(t)} \right)^{\!\! 2} - \frac{\ddot{W}_k^{(j)}(t)}{2 W_k^{(j)}(t)}
\right)
\right]
\label{abWdot2factor}
\end{align}
In this paper we investigate how the time evolution of the adiabatic particle number $\tilde{\mathcal N}_k(t)$  changes as we change the order of truncation of the adiabatic expansion. Note that the final asymptotic value for the particle number, $N_k\equiv \tilde{\mathcal N}_k(t=+\infty)$, is independent of the choice of truncation order.

\subsection{The  Super-Adiabatic Basis}

\begin{figure}[htb]
\includegraphics[scale=0.4]{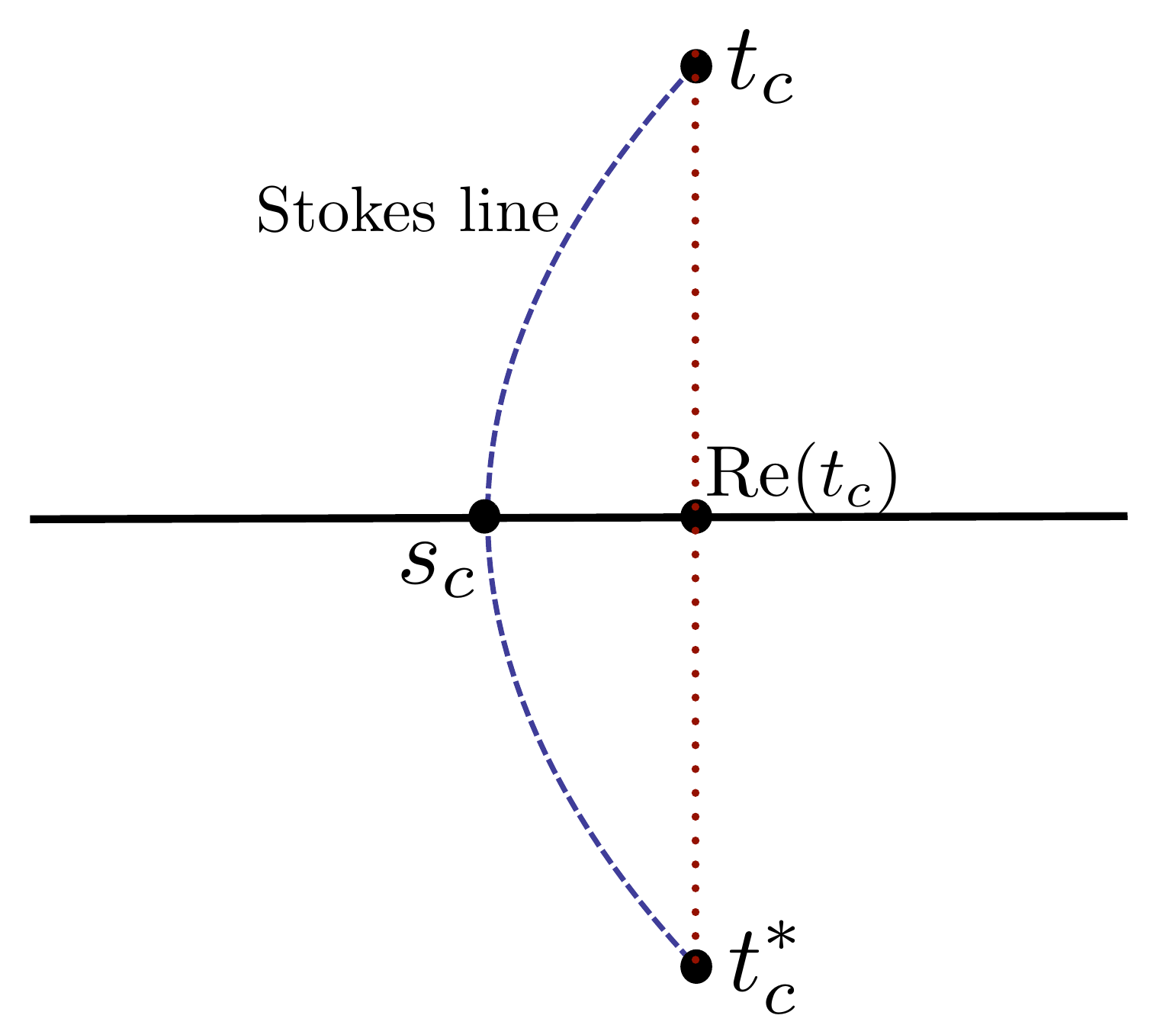}
\caption{For a single complex conjugate pair of turning points, $(t_c, t_c^*)$ joined by a Stokes line [dashed blue line], we define $s_c$ as the time at which the Stokes line crosses the real axis axis.  Particle production  associated with this pair $(t_c, t_c^*)$ corresponds to the jump across the Stokes line at $s_c$.}
\label{stokes}
\end{figure}

The adiabatic expansion is a divergent asymptotic expansion, and at higher orders $j$ the expressions for $W_k^{(j)}(t)$ in terms of the original frequency $\omega_k(t)$ rapidly become more and more complicated: see equations (\ref{ae1})-(\ref{ae3}). This makes  the situation look uninteresting and suggests that a  study of high orders is hopeless. But Dingle discovered a remarkable universal large-order behavior for the adiabatic expansion \cite{dingle}. Define the ``singulant'' variable
\begin{eqnarray}
F_k(t) = 2i \int^t_{t_c} \omega_k(t^\prime) \, dt^\prime 
\label{sing1}
\end{eqnarray}
where $t_c$ is a turning point, a solution of $\omega_k(t_c) = 0$ that is closest to the real axis and located in the upper half plane. 
Then if we characterize the higher-orders of the adiabatic expansion through the terms generated in the phase-integral approach to WKB \cite{froman}, 
\begin{eqnarray}
W_k(t)=\omega_k(t)\sum_{l=0}^\infty \varphi_k^{(2l)}(t)
\label{pi}
\end{eqnarray}
then there is a simple and universal large-order behavior in terms of the singulant:
\begin{eqnarray}
\varphi_k^{(2l+2)}\sim - \frac{(2l+1)!}{\pi \, F_k^{2l+2}}\qquad, \quad l\gg 1
\label{magic}
\end{eqnarray}
Berry used this large-order behavior to resum the adiabatic expansion, to give a {\it universal} time-dependent form of the transition across a turning point \cite{berry}. Each (complex) turning point can be identified with a particle creation event (equivalently, in the scattering language, the ``birth of a reflection'' \cite{berry}). Berry's result can be stated as follows. For real $\omega_k^2(t)$, the turning points occur in complex conjugate pairs. Consider the situation of a single dominant  complex conjugate pair of turning points, as sketched in Figure \ref{stokes}. Then Berry found that when  the adiabatic expansion is truncated at optimal order, the Bogoliubov coefficient $\beta_k(t)$ evolves across the associated Stokes line according to  the {\it universal} approximate expression
\begin{eqnarray}
\beta_k(t)\approx \frac{i}{2}\,  {\rm Erfc}\left(-\sigma_k(t)\right) \, e^{- F_k^{(0)}}\quad ,
\label{beta-answer}
\end{eqnarray}
where Erfc is the error function \cite{nist}, and the natural time evolution parameter $\sigma_k(t)$ is expressed in terms of the real and imaginary parts of the singulant function:
\begin{eqnarray}
\sigma_k(t)\equiv \frac{{\rm Im}\, F_k(t)}{\sqrt{2\, {\rm Re}\, F_k(t)}}
\label{sigma}
\end{eqnarray}
The amplitude $e^{- F_k^{(0)}}$ is determined by the singulant between the complex conjugate turning points:
\begin{eqnarray}
F_k^{(0)} = i \int_{t_c}^{t_c^*} \omega_k(t) \, dt \quad,
\label{f0}
\end{eqnarray}
where the integral is taken along the Stokes line connecting the two turning points. In fact, the integral can be taken along the straight line connecting $t_c$ and $t_c^*$, and with proper choices of branches the result is real and positive \cite{froman,berry}. Thus, the Bogoliubov coefficient makes a smooth jump of universal shape when crossing a Stokes line, which suggests the interpretation of the ``time of particle creation'' as the time $s_c$ at which the Stokes line connecting $t_c$ and $t_c^*$ crosses the real axis, as sketched in Figure \ref{stokes}; this is the time at which $\sigma_k(t)$ vanishes.

Recalling (\ref{an}), we immediately deduce that in the context of particle production, the time evolution of the adiabatic
particle number $\tilde{\mathcal N}_k(t)$ is given by the {\it universal} approximate expression
\begin{eqnarray}
\tilde{\mathcal N}_k(t)\approx \frac{1}{4} \left|{\rm Erfc}\left(-\sigma_k(t)\right) e^{- \, F_k^{(0)}} \right|^2
\label{answer}
\end{eqnarray}
Furthermore, in the vicinity of the real crossing  point $s_c$ of the Stokes line, 
\begin{align}
F_k(t) \approx F_k(s_c) + 2 i \, \omega_k(s_c) (t - s_c)
\label{sing2}
\end{align}
which leads to the simplified approximation:
\begin{align}
\sigma_k(t) \approx \frac{ 2 \, \omega_k(s_c) \, (t - s_c)}{\sqrt{2 \, F_k(s_c)}} 
\label{sigma2}
\end{align}
Finally, the order $j$ at which the adiabatic approximation should be truncated depends on the parameters associated with the turning points, and can be estimated in terms of $F_k^{(0)}$ defined in (\ref{f0}):
the optimal order $j$ is the integer closest to
\begin{eqnarray}
j\approx {\rm Int}\left[\frac{1}{2}\left( \left| F_k^{(0)} \right| - 1 \right)\right]
\label{joptimal}
\end{eqnarray}
In practice it is often a good approximation to estimate this optimal order by computing the absolute value of the singulant (\ref{sing1}), evaluated at the real part of $t_c$.

We take these remarkable results as our definition of the super-adiabatic basis: the basis in which the adiabatic expansion is truncated at optimal order. The corresponding adiabatic particle number is then defined to be the {\it super-adiabatic particle number}, and it has a universal time evolution in the vicinity of a turning point.

 It is the {\it universality} of this result  that makes this definition of a super-adiabatic particle number a useful and well-defined concept. Because of this universality, we do not need to make the explicit large-order truncation of the adiabatic expansion, which is very complicated at high orders, and moreover would be truncated at different orders for different parameters. But the results of Dingle and Berry imply that this is not necessary: the universal form of time dependence in (\ref{answer}) applies in general with optimal truncation.
 
\begin{figure}[h!]
\begin{tabular}{cc}
\centering
\includegraphics[width=8.5cm]{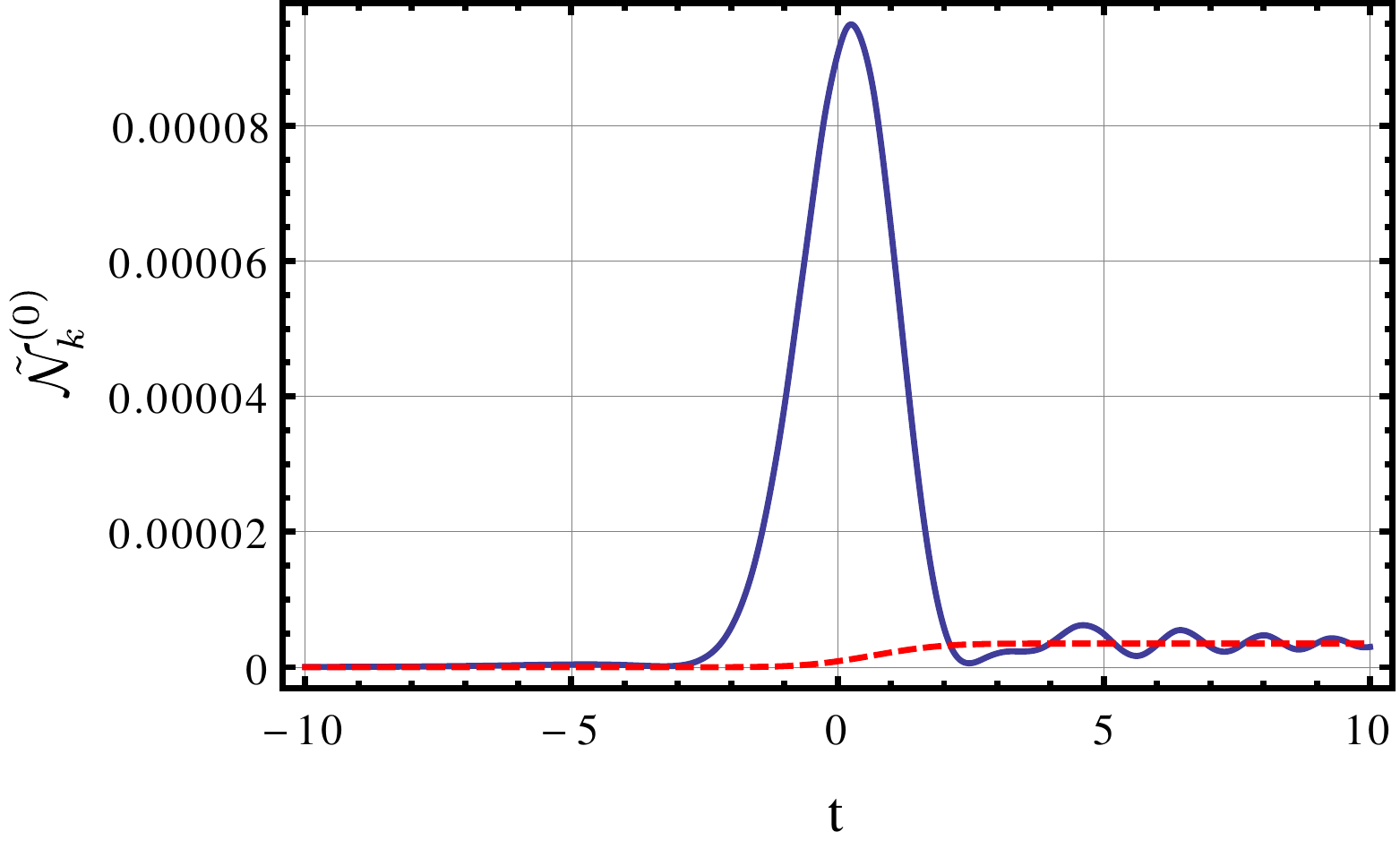} 
& 
\includegraphics[width=8.5cm]{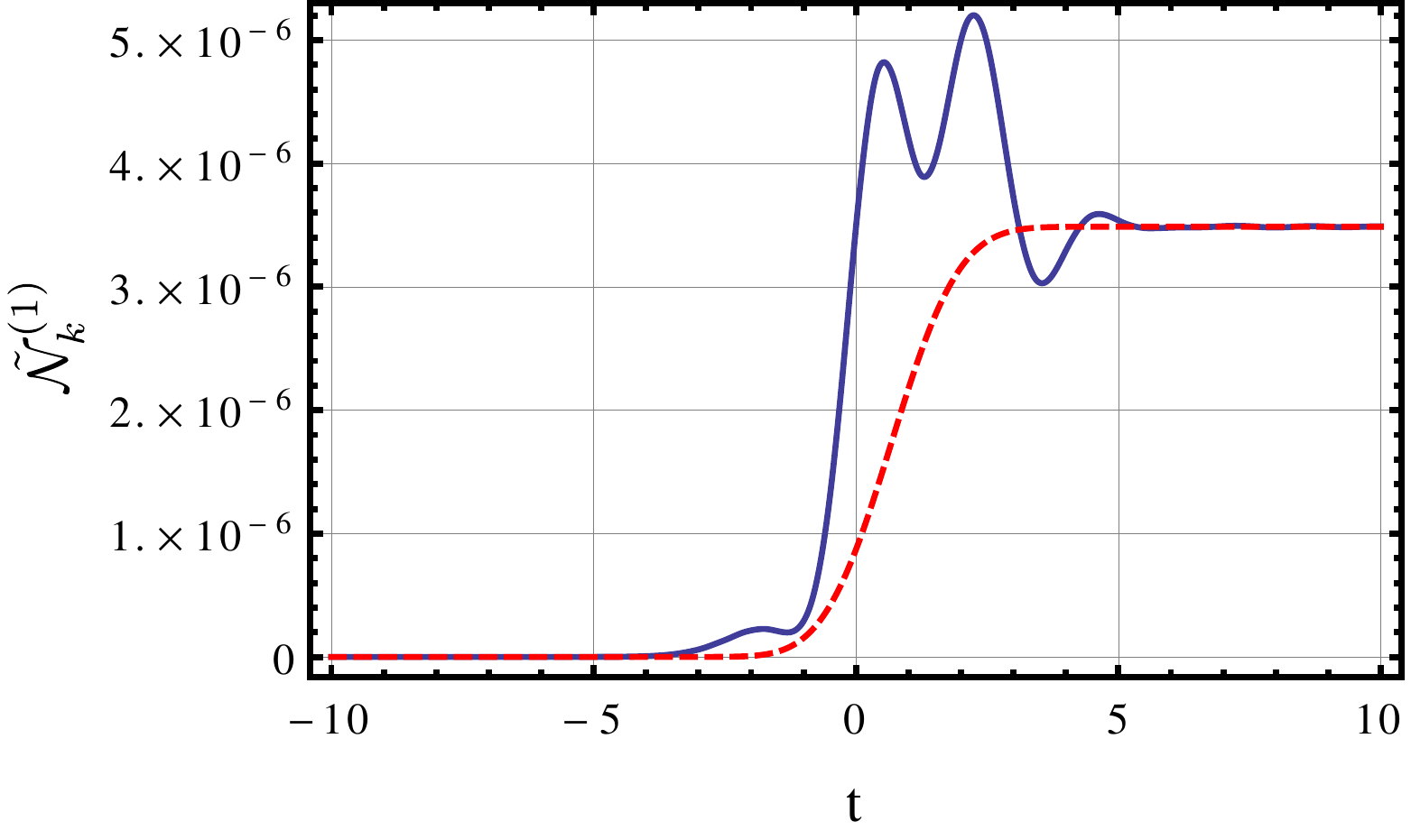} \\
\includegraphics[width=8.5cm]{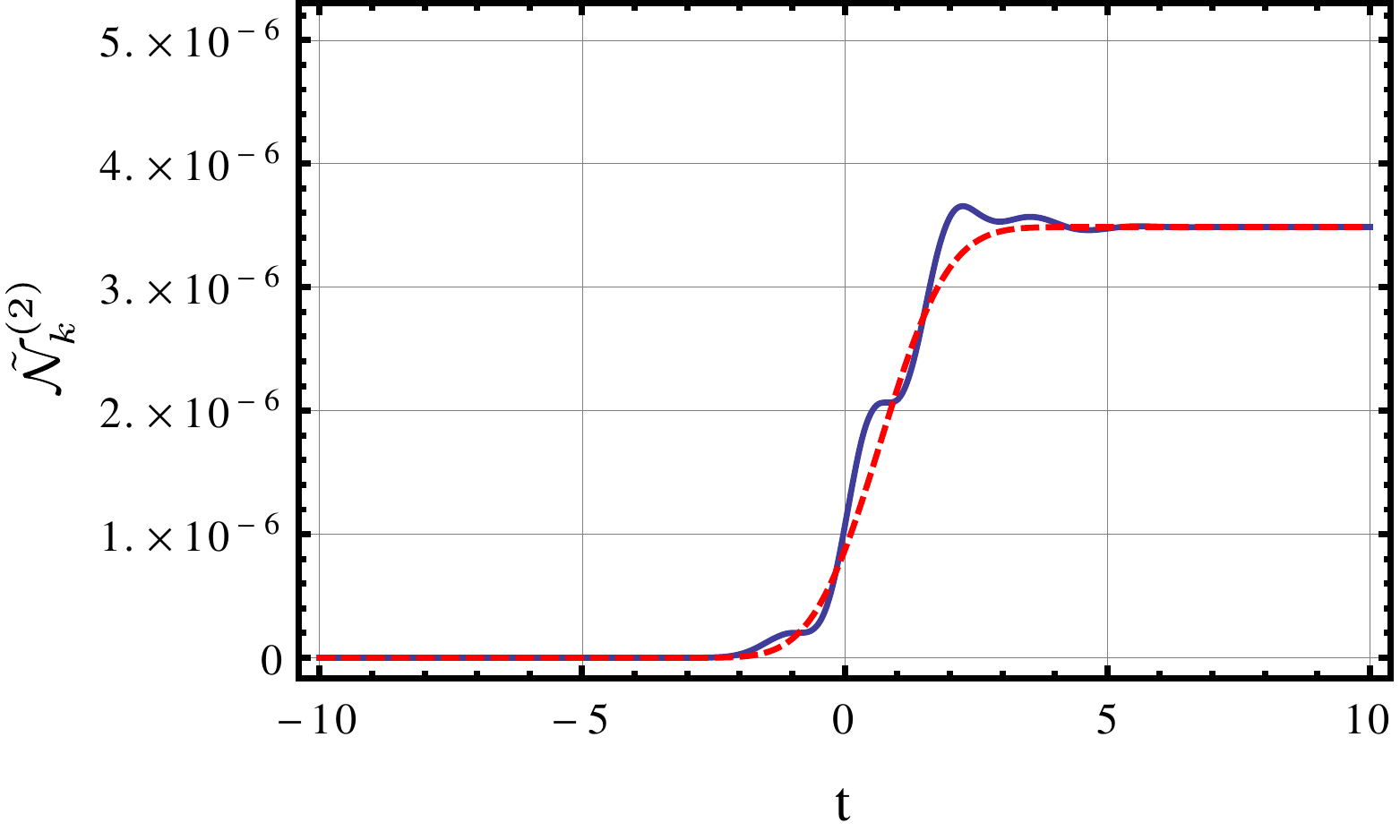} 
 & 
\includegraphics[width=8.5cm]{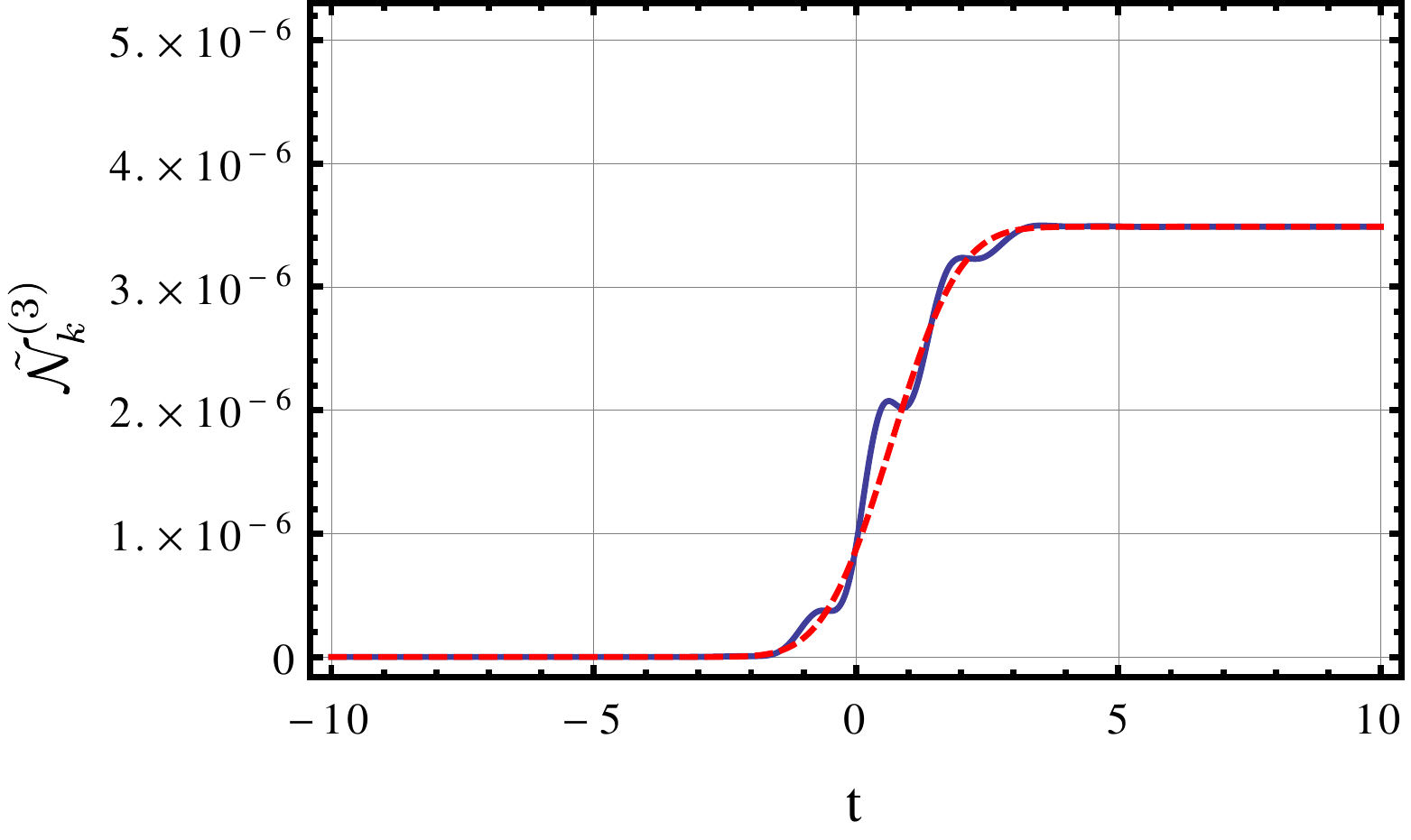} \\
\includegraphics[width=8.5cm]{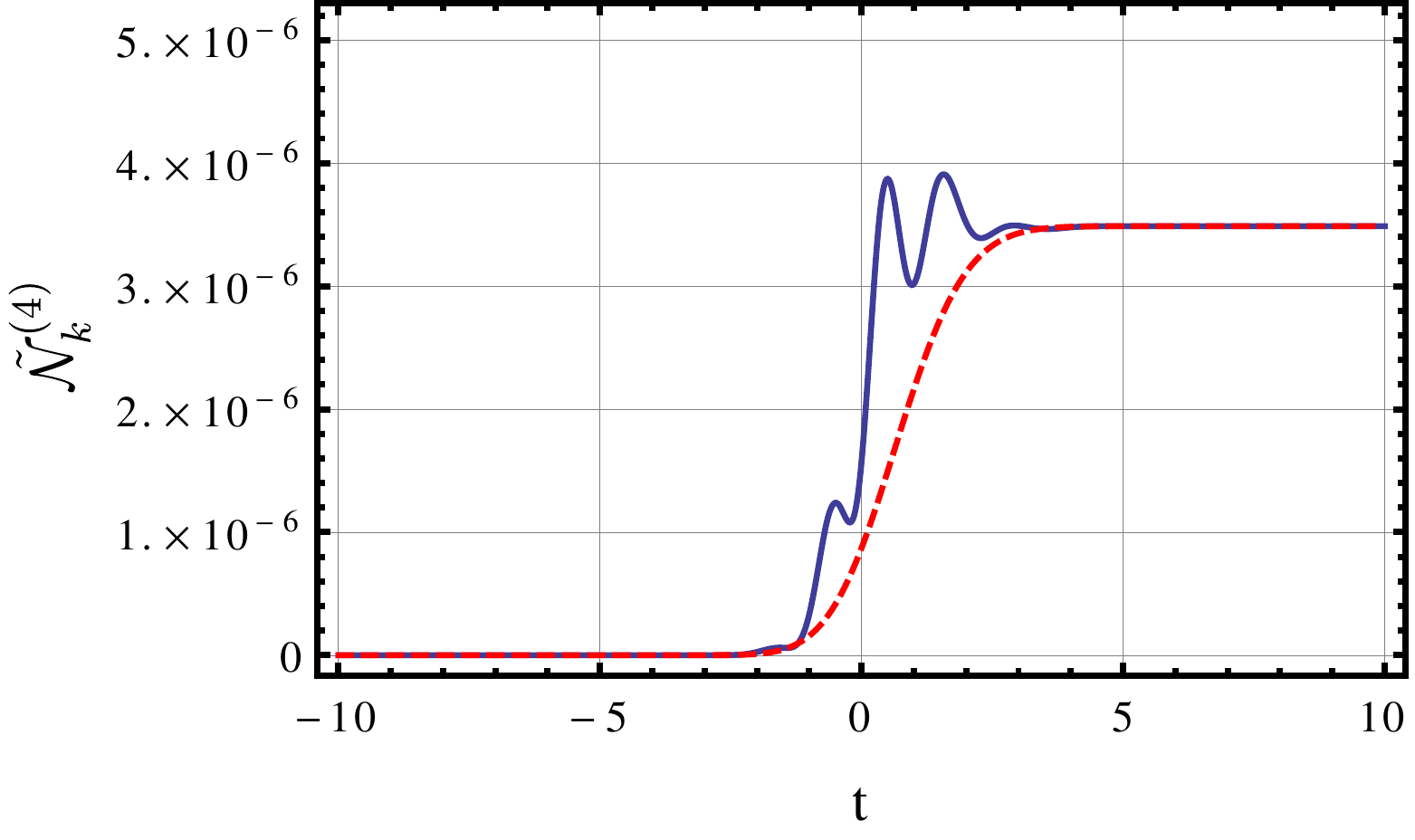} 
& 
\includegraphics[width=8.5cm]{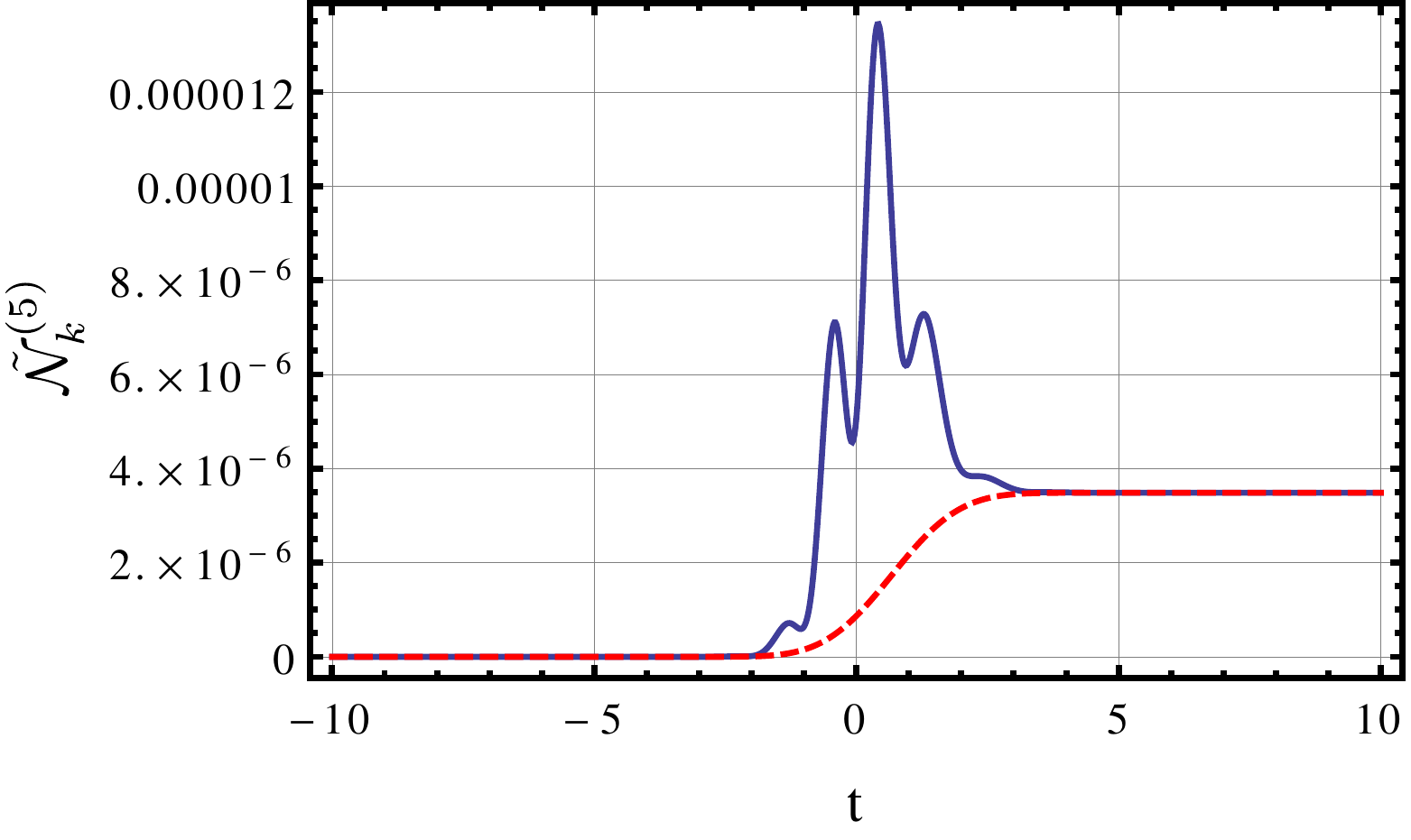} \\
\end{tabular}
\caption{Time evolution of the adiabatic particle number for the first 6 orders of the adiabatic expansion, for  Schwinger particle production in a constant electric field of magnitude $E=0.25$, with transverse momentum $k_{\perp}=0$ and longitudinal momentum $k_\parallel=0$, in units with $m=1$. The numerical results, from integrating (\ref{abWdot2}, \ref{abWdot2factor}) are plotted in solid-blue lines, and Berry's universal form (\ref{answer}) is plotted as a red-dashed curve in each sub-plot. Note the different vertical scales. The final asymptotic value of the particle number, at future infinity, is the same for all orders of truncation. At intermediate times there are large oscillations in the particle number, which become much smaller as the optimal order ($j=3$) is reached, and then grow again rapidly beyond this optimal order of truncation. Such behavior is characteristic of {\it asymptotic} expansions, where the order of truncation depends  on the size of the expansion parameter, and going beyond this optimal order typically yields increasingly worse results.
}
\label{constant-E-orders}
\end{figure}

\subsection{Illustrations of Super-Adiabatic Particle Number for Simple Fields}

We first illustrate Berry's result in the context of Schwinger particle production for simple electric fields that have just a single pair of dominant complex-conjugate turning points. 

\subsubsection{Constant Electric Field}

We first consider the simplest case, that of a constant electric field: $E(t)=E$. There is one pair of complex-conjugate turning points (see, for example, \cite{Anderson:2013ila,Anderson:2013zia} for a thorough discussion).
In Figure \ref{constant-E-orders} we plot the time evolution of the adiabatic particle number for the first 6 orders of the adiabatic expansion. The numerical results, from integrating (\ref{abWdot2}, \ref{abWdot2factor}) are plotted in solid-blue lines, and Berry's universal form (\ref{answer}) is plotted as a red-dashed curve in each plot. The final asymptotic value of the particle number, at future infinity, is the same for all orders of truncation [note the different vertical scales]. At zeroth order of the adiabatic expansion we see at intermediate times large oscillations in the particle number,  roughly 30 times the scale of the final value. The magnitude of the oscillations decreases as we approach the optimal order, $j=3$, and then they rapidly grow again if we continue beyond the optimal order. Recall that such behavior is characteristic of {\it asymptotic} expansions, where the order of truncation depends crucially on the size of the expansion parameter, and going beyond this optimal order typically yields increasingly worse results.
For the physical parameters used in Figure \ref{constant-E-orders}, we have $F_k^{(0)}\approx 6.283$, consistent with the estimate (\ref{joptimal}) for the optimal truncation order, and $\exp(-2\,F_k^{(0)})\approx 3.49\times 10^{-6}$, consistent with the universal formula (\ref{answer})  for the particle number, at late times.


\subsubsection{Single-pulse Electric Field}

\begin{figure}[h!]
\begin{tabular}{cc}
\centering
\includegraphics[width=8.5cm]{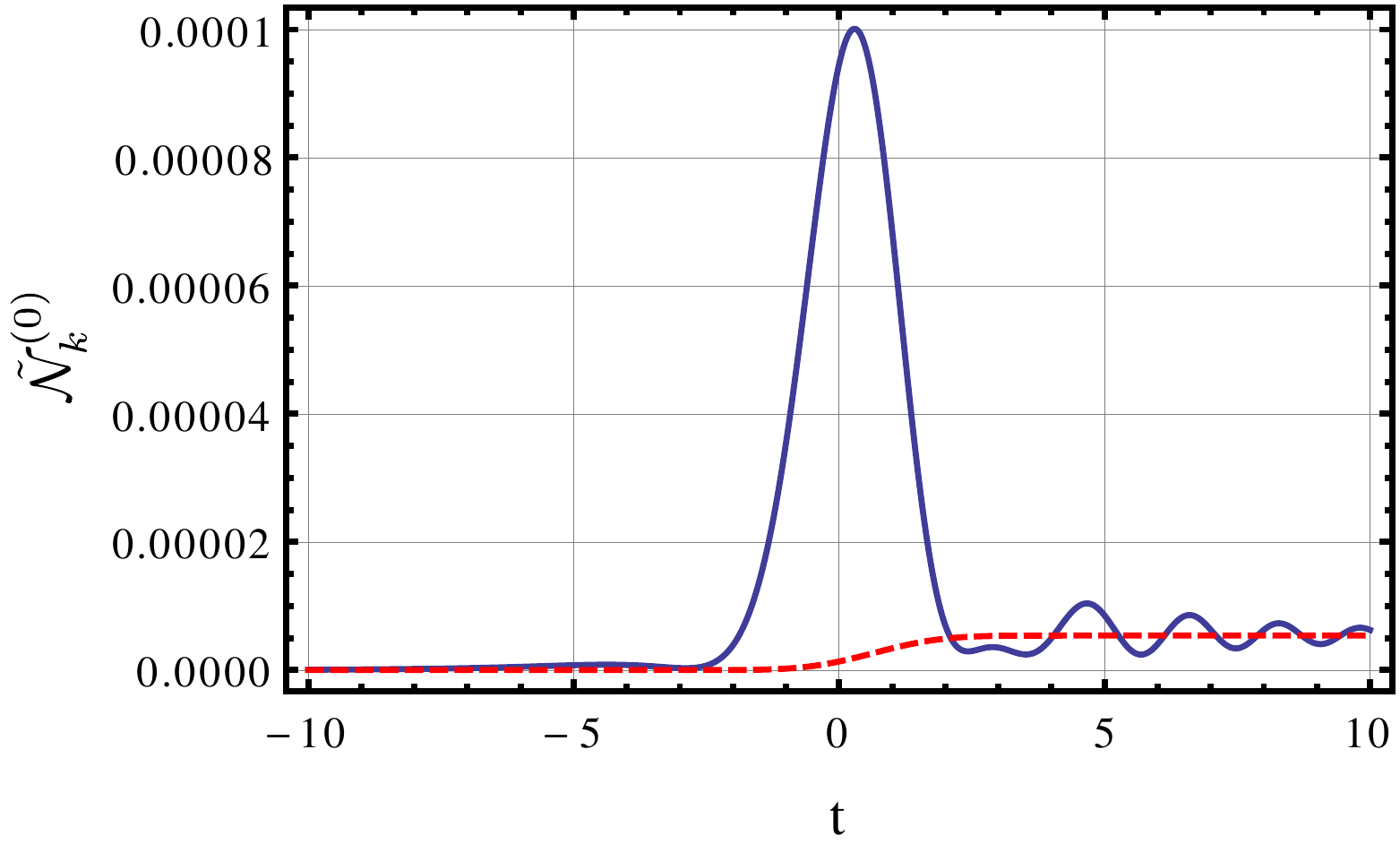} 
 & 
\includegraphics[width=8.5cm]{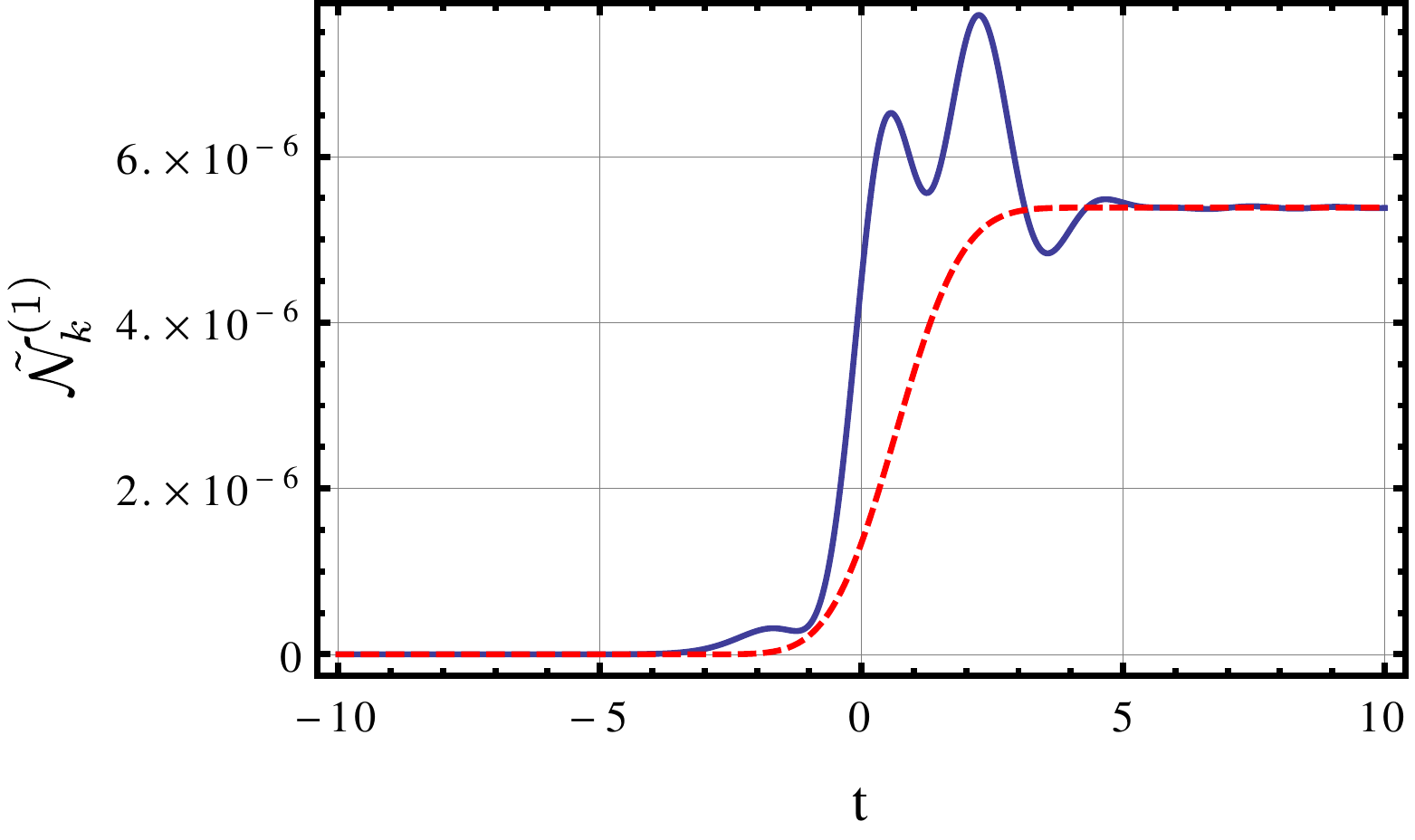} \\
\includegraphics[width=8.5cm]{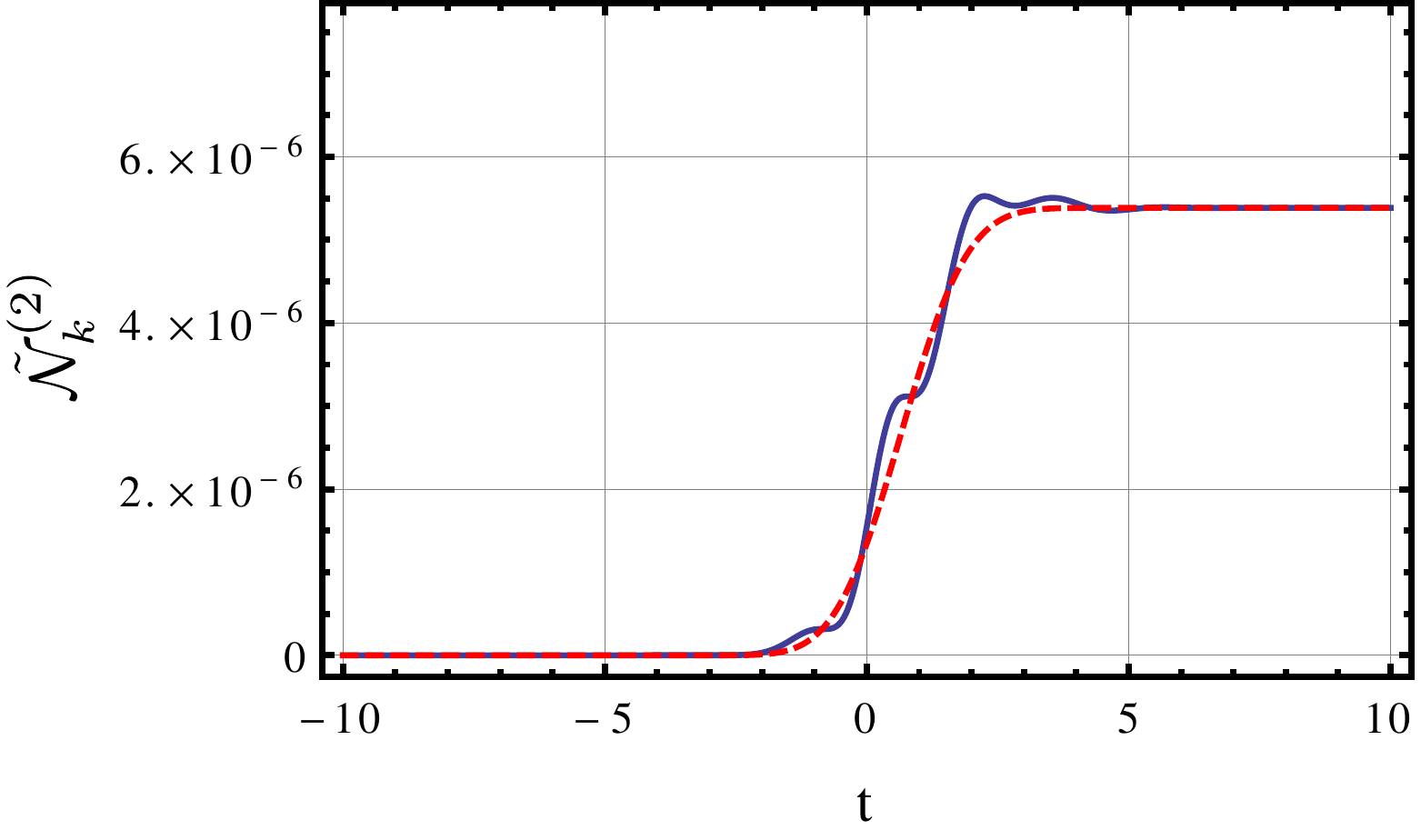} 
 & 
\includegraphics[width=8.5cm]{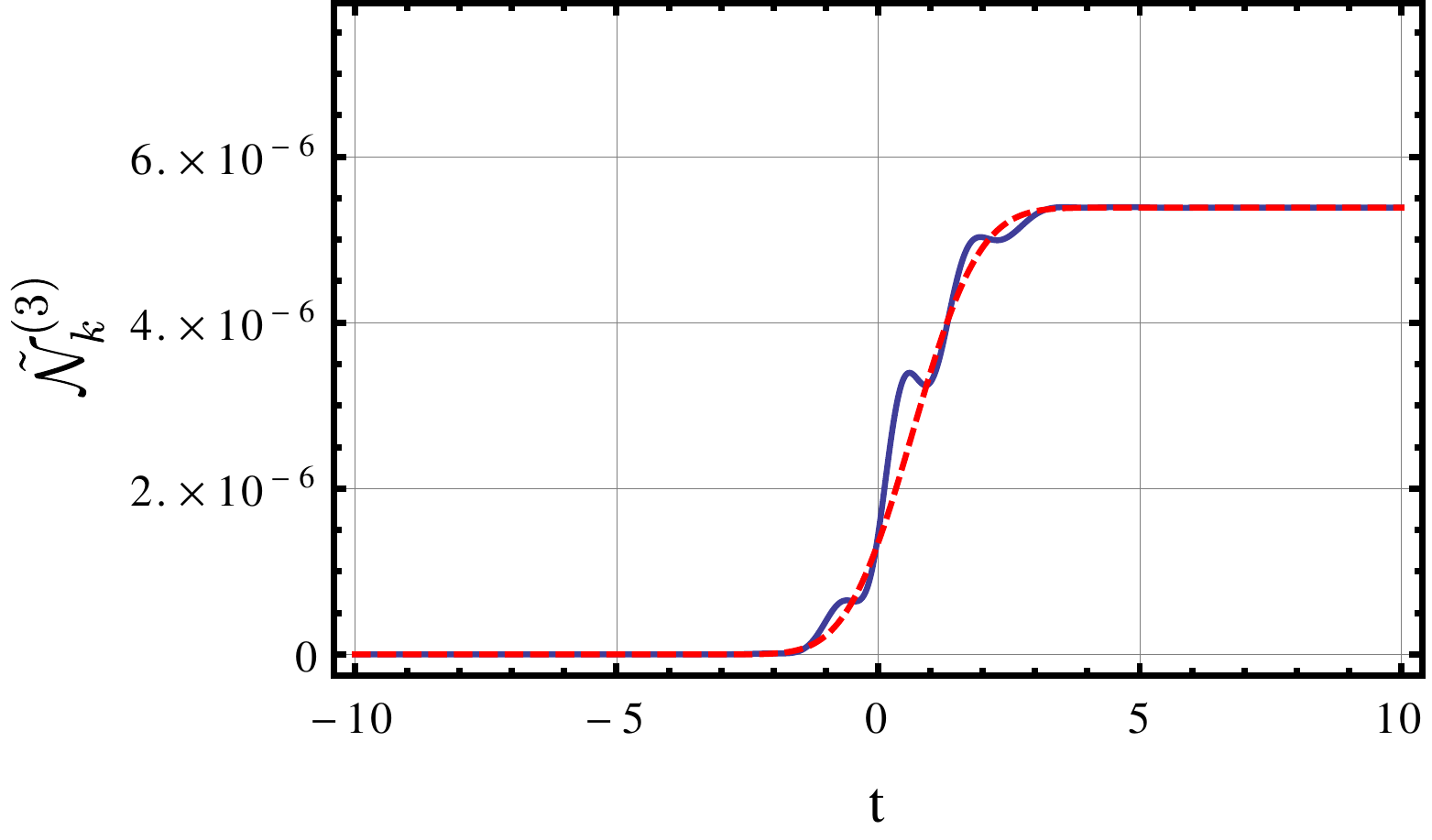} \\
\includegraphics[width=8.5cm]{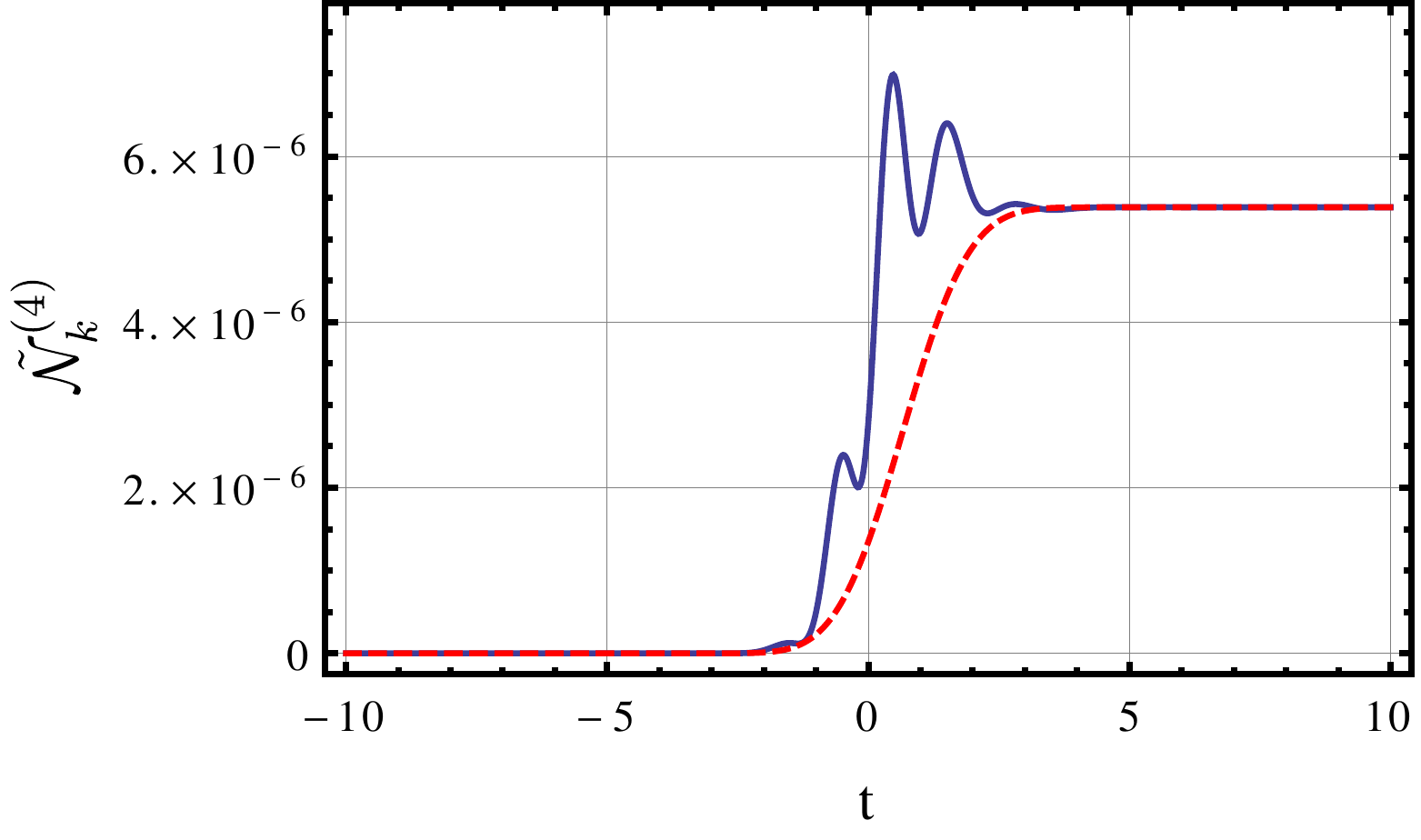} 
& 
\includegraphics[width=8.5cm]{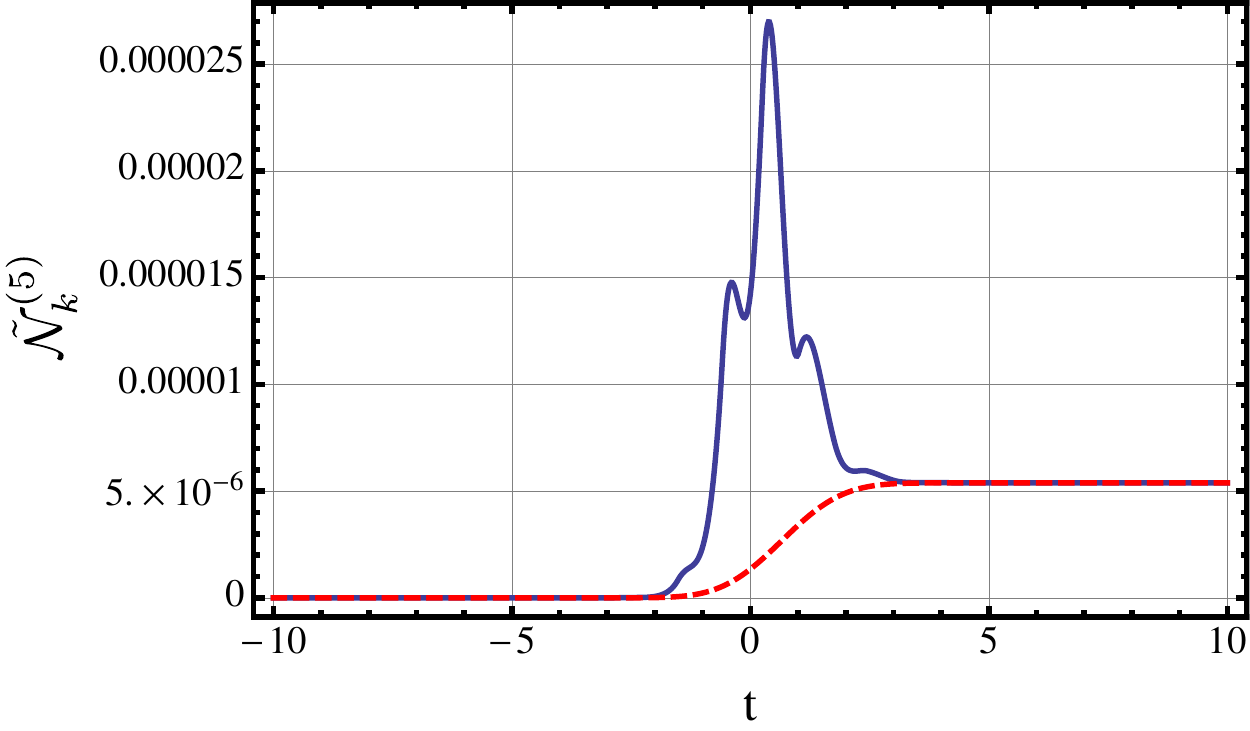}
\end{tabular}
\caption{Time evolution of the adiabatic particle number for the first 6 orders of the adiabatic expansion, for  Schwinger particle production in a time-dependent single-pulse  electric field $E(t)=E\, {\rm sech}^2(a t)$, with magnitude $E=0.25$, $a=0.1$, transverse momentum $k_{\perp}=0$, and longitudinal momentum $k_\parallel=0$, in units with $m=1$. The numerical results, from integrating (\ref{abWdot2}, \ref{abWdot2factor}) are plotted in solid-blue lines, and Berry's universal form (\ref{answer}) is plotted as a red-dashed curve in each sub-plot. Note the different vertical scales. The final asymptotic value of the particle number, at future infinity, is the same for all orders of truncation. At intermediate times there are large oscillations in the particle number, which become much smaller as the optimal order ($j=3$) is reached, and then grow again rapidly beyond this optimal order of truncation. Such behavior is characteristic of {\it asymptotic} expansions, where the order of truncation depends  on the size of the expansion parameter, and going beyond this optimal order typically yields increasingly worse results.
}
\label{tanh-E-orders}
\end{figure}

A slightly more physical example is that of a single-pulse electric field, $E(t)=E\, {\rm sech}^2(a t)$, for which we use the time-dependent vector potential:
\begin{eqnarray}
A_\parallel(t) = - \frac{E}{a} \tanh (at)
\label{one-pulse}
\end{eqnarray}
This field leads to an infinite tower of pairs of complex-conjugate turning points, but  in the semiclassical regime  where $E\ll m^2$ and $a\ll m$, the effect is dominated by the pair closest to the real axis  \cite{dd1}. Thus, the behavior is quite similar to that of the constant $E$ field. The results are shown in Figure \ref{tanh-E-orders}, and we observe the close similarity to the constant $E$ field results in Figure \ref{constant-E-orders}. Again there are large oscillations at intermediate times for low orders of the adiabatic expansion. These subside as the optimal order ($j=3$) is reached, and then grow again as one goes to higher orders in the adiabatic expansion. The red-dashed curves show Berry's universal error-function form (\ref{answer}), with excellent agreement with the optimal order of truncation. For the physical parameters used in Figure \ref{tanh-E-orders}, we have $F_k^{(0)}\approx 6.050$, consistent with the estimate (\ref{joptimal}) for the optimal truncation order, and $\exp(-2\,F_k^{(0)})\approx 5.558\times 10^{-6}$, consistent with the universal formula (\ref{answer}) for the particle number, at late times.

\begin{figure}[h!]
\centering
\includegraphics[width=17cm]{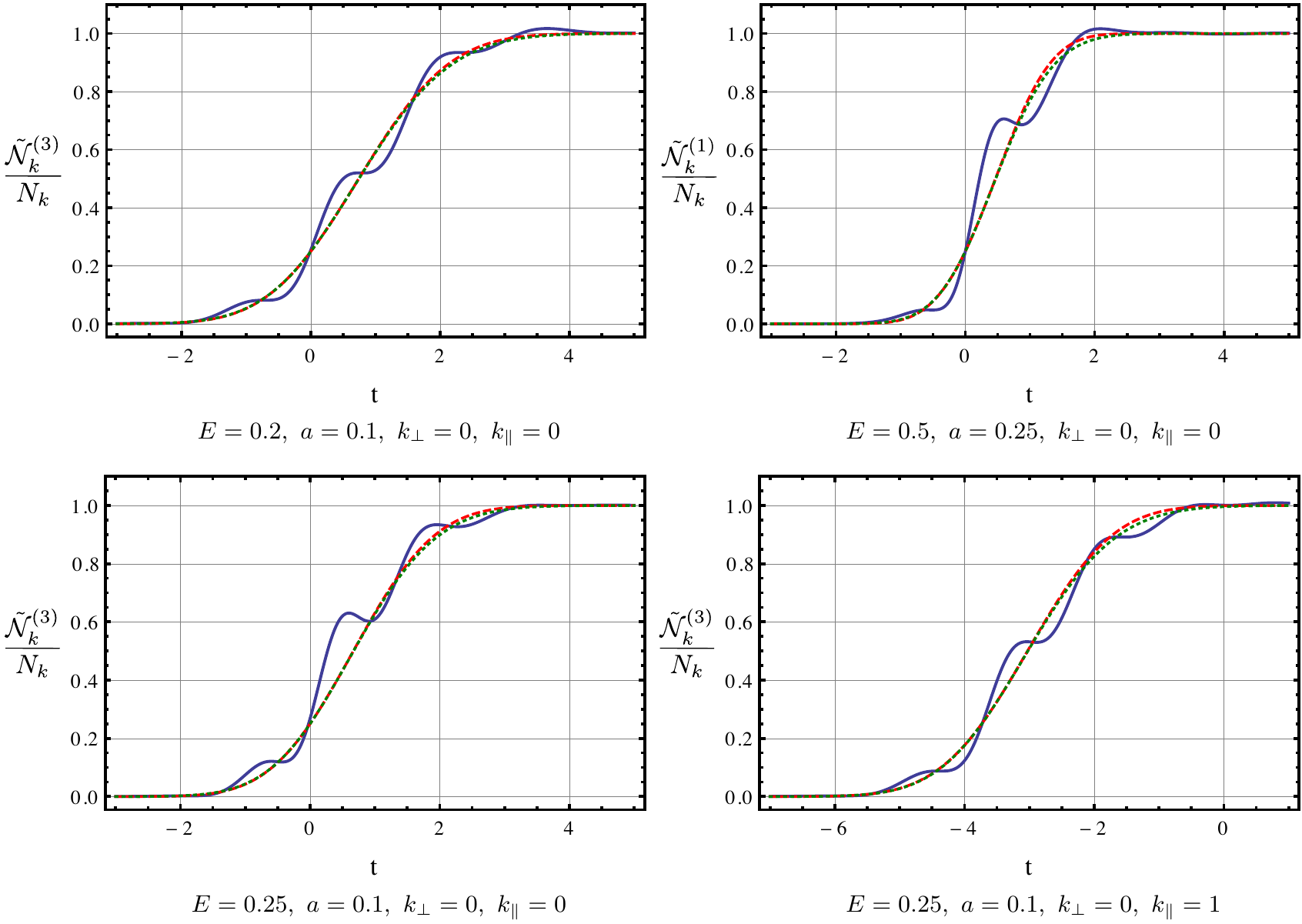} 
\caption{Illustration of the universal error-function form (\ref{answer}) of the time evolution of the adiabatic particle number in the super-adiabatic basis. These plots show the time evolution of the optimally truncated adiabatic particle number (solid blue curves) for the Schwinger effect, with the single-pulse time-dependent electric field with vector potential $A_\parallel = - \frac{E}{a} \tanh(a t)$. The four plots correspond to different choices of field parameters $E$ and $a$, and also different longitudinal momenta $k_\parallel$ of the produced particles, in units with $m=1$. In order to show all plots on the same scale, we have normalized the adiabatic particle number by its final value $N_k\equiv \tilde{\mathcal N}_k(t=+\infty)$. In each plot the order $j$ of the optimal truncation is indicated by the superscript $(j)$ on $\tilde{\mathcal N}_k^{(j)}(t)$. The red-dashed curves show the universal error function form in (\ref{answer}), while the green-dot-dashed curves show the universal form with the approximate expression (\ref{sigma2}) for the function $\sigma_k(t)$. For the parameter values used in these plots, $F_k^{(0)}$ takes values $7.416$, $2.967$, $6.050$, and $6.956$, respectively, consistent with the estimate (\ref{joptimal}) for the optimal truncation order.}
\label{universal}
\end{figure}

In Figure \ref{universal} we illustrate the universal nature of the optimally truncated form (\ref{answer}), by comparing the optimally truncated order, for different field and momentum parameters, for Schwinger pair production in a single-pulse electric field. In order to show all plots on the same scale, we have normalized the adiabatic particle number by its final value $N_k\equiv \tilde{\mathcal N}_k(t=+\infty)$. In each plot the order $j$ of the optimal truncation is indicated by the superscript $(j)$ on $\tilde{\mathcal N}_k^{(j)}(t)$. The red-dashed curves show the universal error function form in (\ref{answer}), while the green-dot-dashed curves show the universal form with the approximate expression (\ref{sigma2}) for the function $\sigma_k(t)$. These results illustrate that across a wide range of different field parameters, the universal super-adiabatic form (\ref{answer}) agrees well with the particle number at the optimal truncation of the adiabatic expansion. Furthermore, the approximation (\ref{sigma2}), valid in the vicinity of a simple turning point, works very well. Finally, the optimal order gives a clear physical picture of the particle creation event as a single smoothed jump across the Stokes line, in contrast to the leading order of the adiabatic expansion which has large unphysical oscillations in this time region. This feature is even more pronounced for the fields studied in the subsequent sections, as particle production in these fields involves quantum interference between different creation events.

\section{Super-Adiabatic Particle Production for the Schwinger Effect in Sequences of Time-dependent Electric Field Pulses}

In this Section we show how the super-adiabatic particle number evolves in time for various classes of time-dependent electric fields having nontrivial temporal structure, to illustrate the time-dependence of the associated quantum interference effects. We concentrate on sequences of alternating-sign pulses, as these have been shown to permit maximal constructive interference, via an analogy with the Ramsey effect of atomic physics, leading to a coherent enhancement of the Schwinger effect \cite{Akkermans:2011yn}. For certain longitudinal momenta of the produced particles, the final particle number is enhanced, while for others it is reduced by destructive interference, producing a momentum spectrum that is a precise time-domain analogue of the interference pattern of the multiple slit interferometer \cite{Akkermans:2011yn}. We note that such interferometric effects have been experimentally observed in analogue atomic ionization systems \cite{dalibard,paulus,remetter,mansten}, and more recently in a polariton system with two-color fields \cite{gabelli}.


\subsection{Two Alternating-sign Electric Field Pulses}
\begin{figure}[h!]
\begin{tabular}{cc}
\centering
\includegraphics[width=8cm]{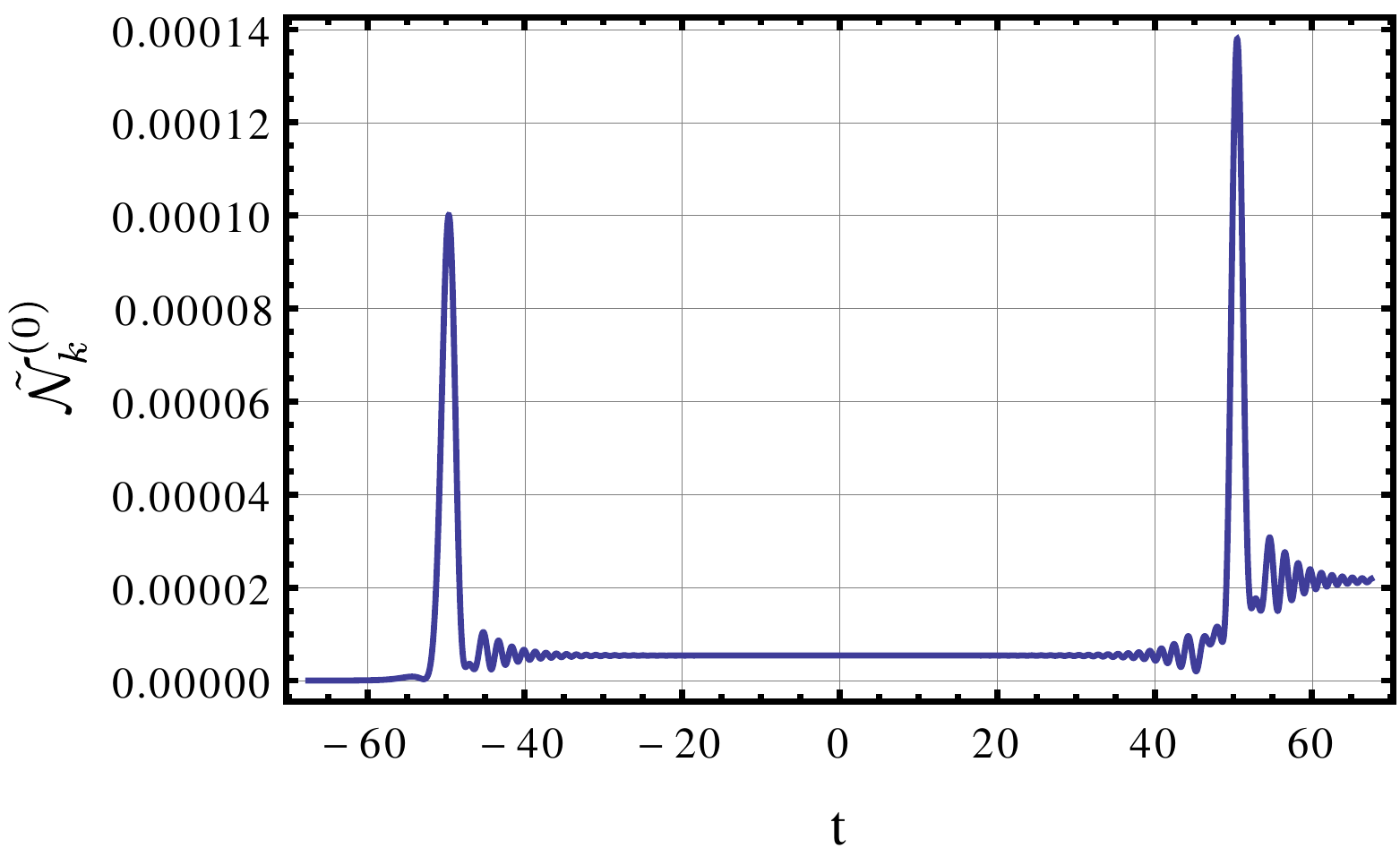} 
& 
\includegraphics[width=8cm]{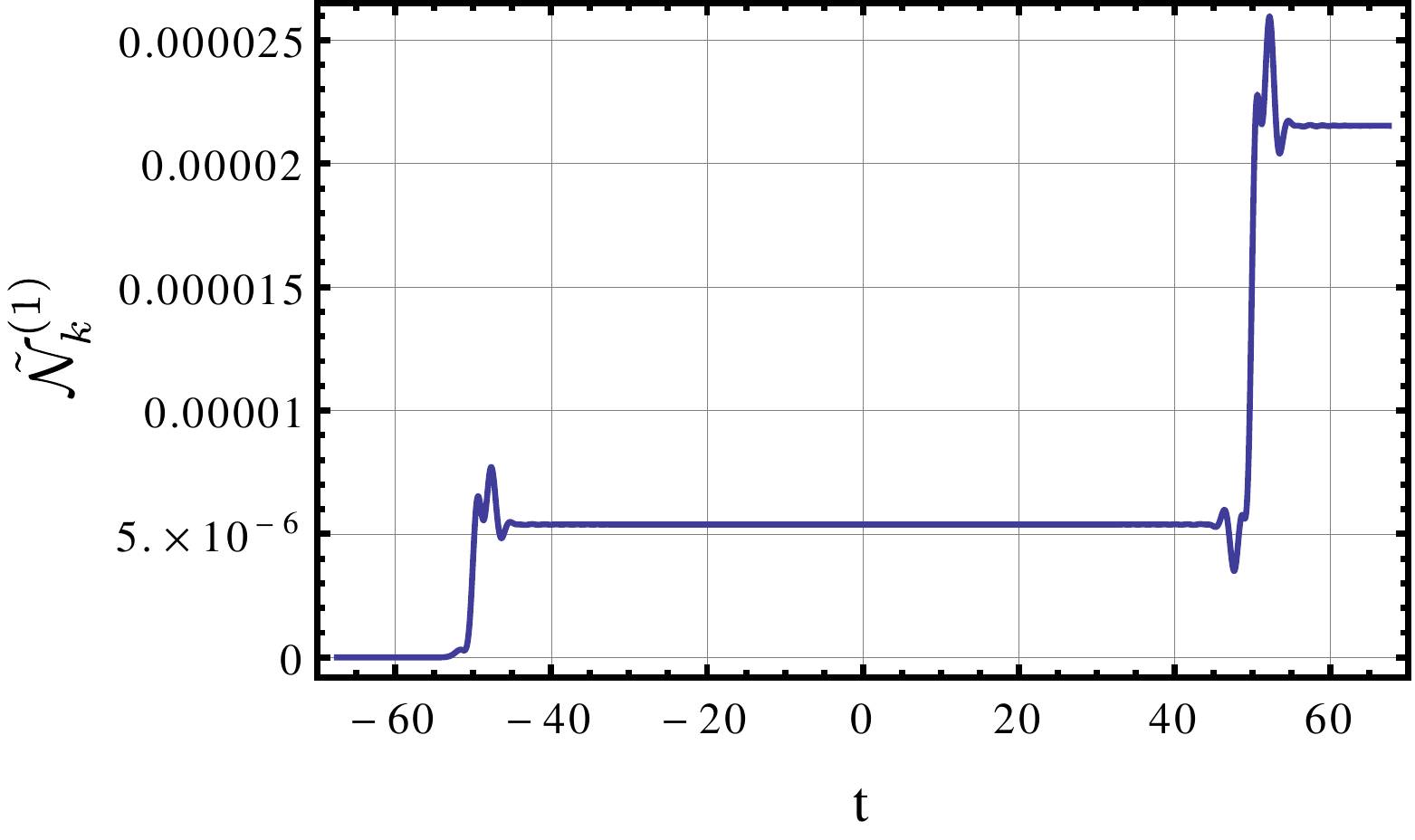} \\
\includegraphics[width=8cm]{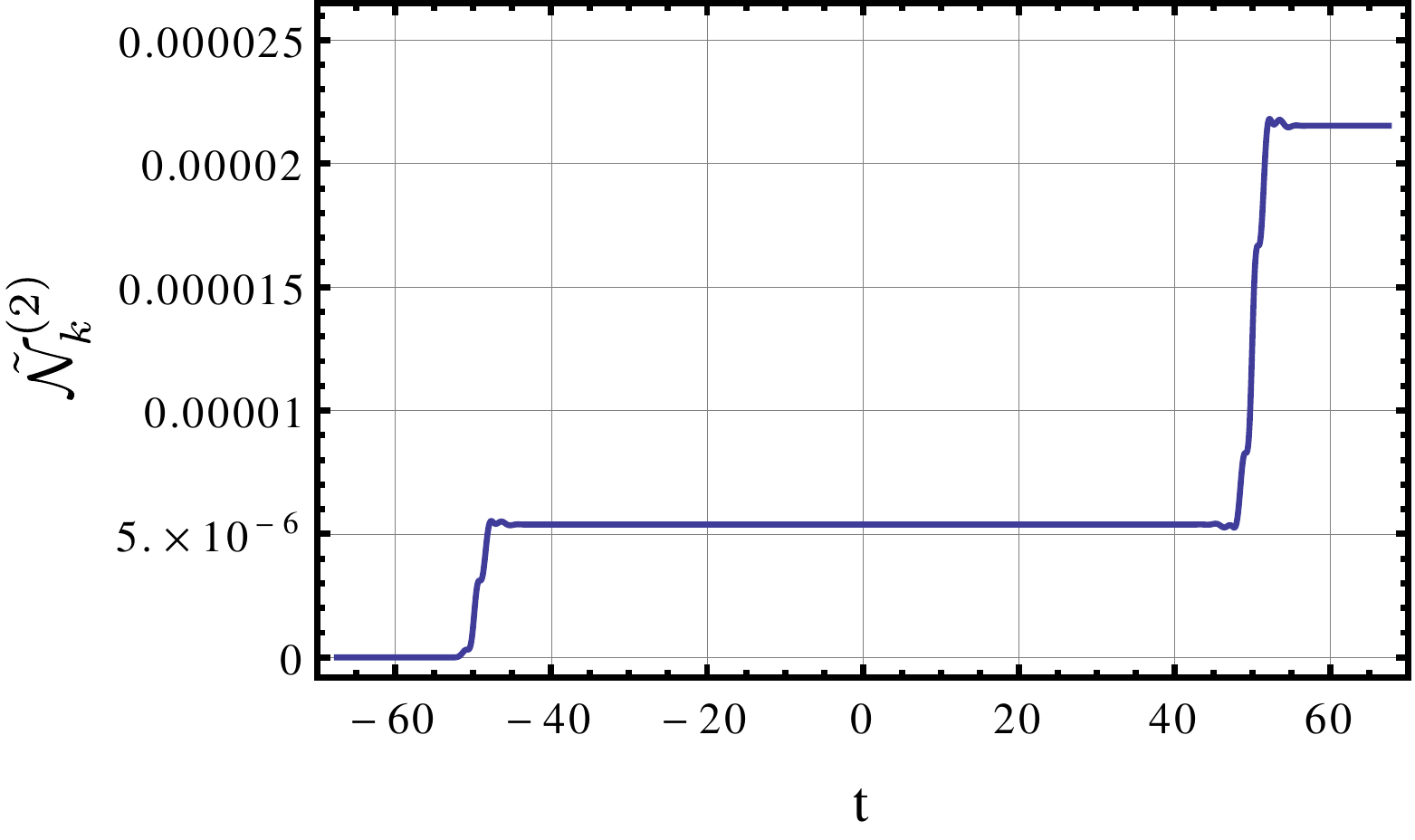} 
& 
\includegraphics[width=8cm]{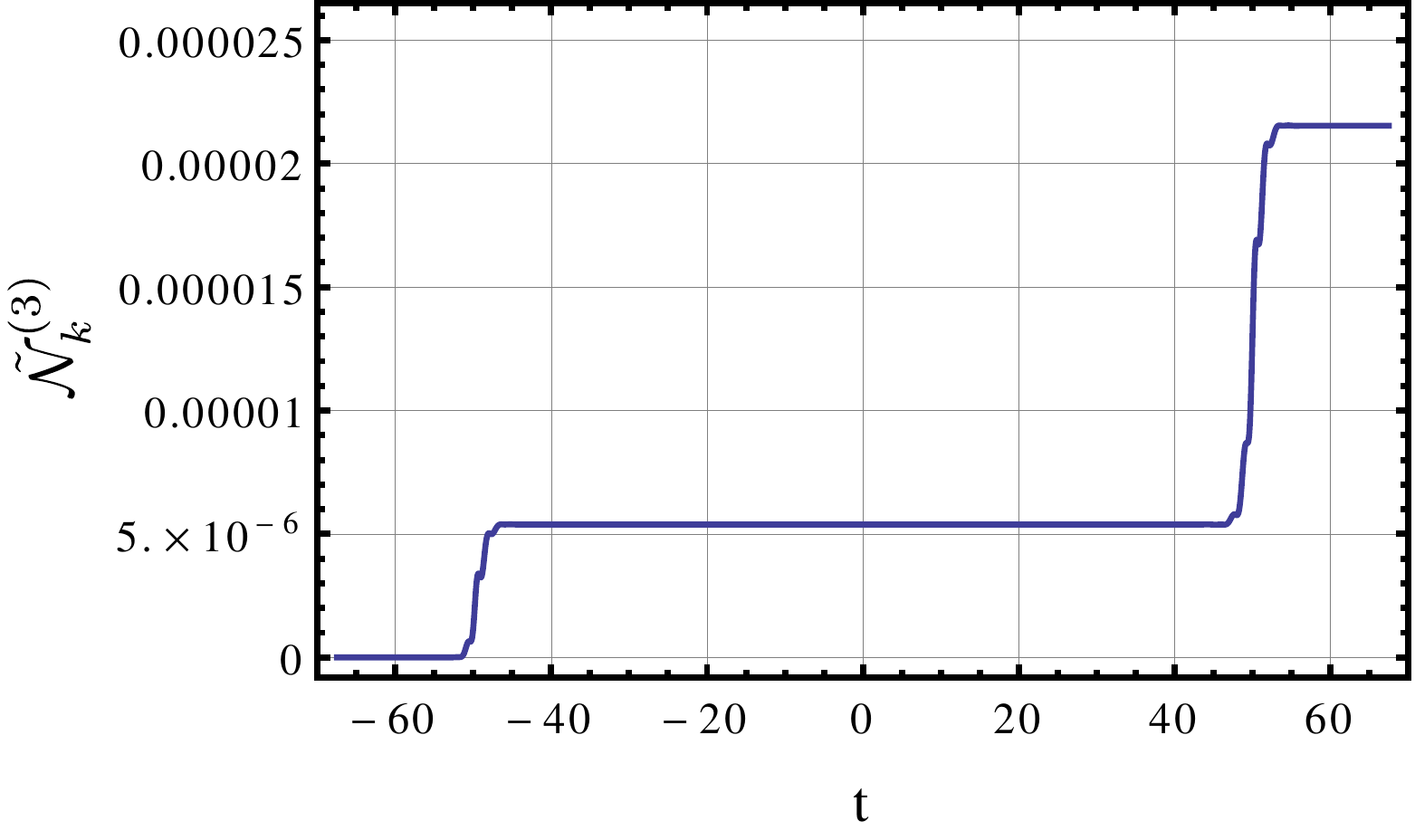} \\
\includegraphics[width=8cm]{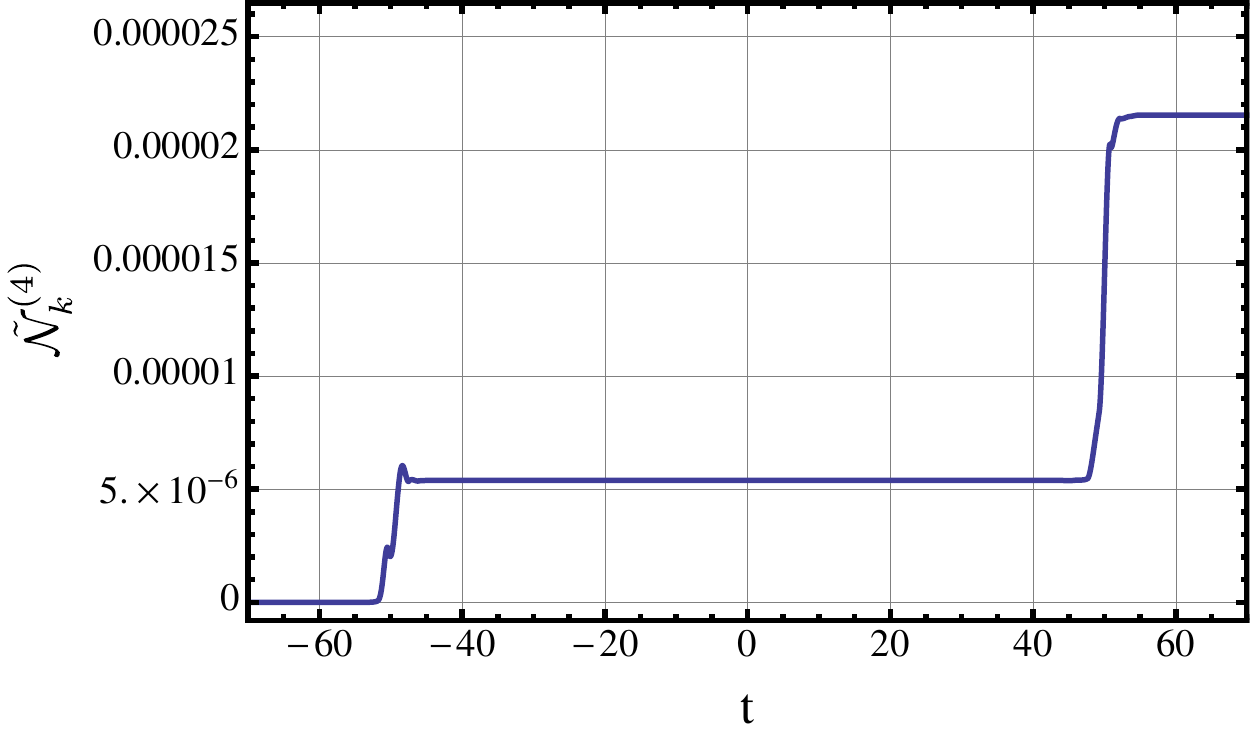} 
& 
\includegraphics[width=8cm]{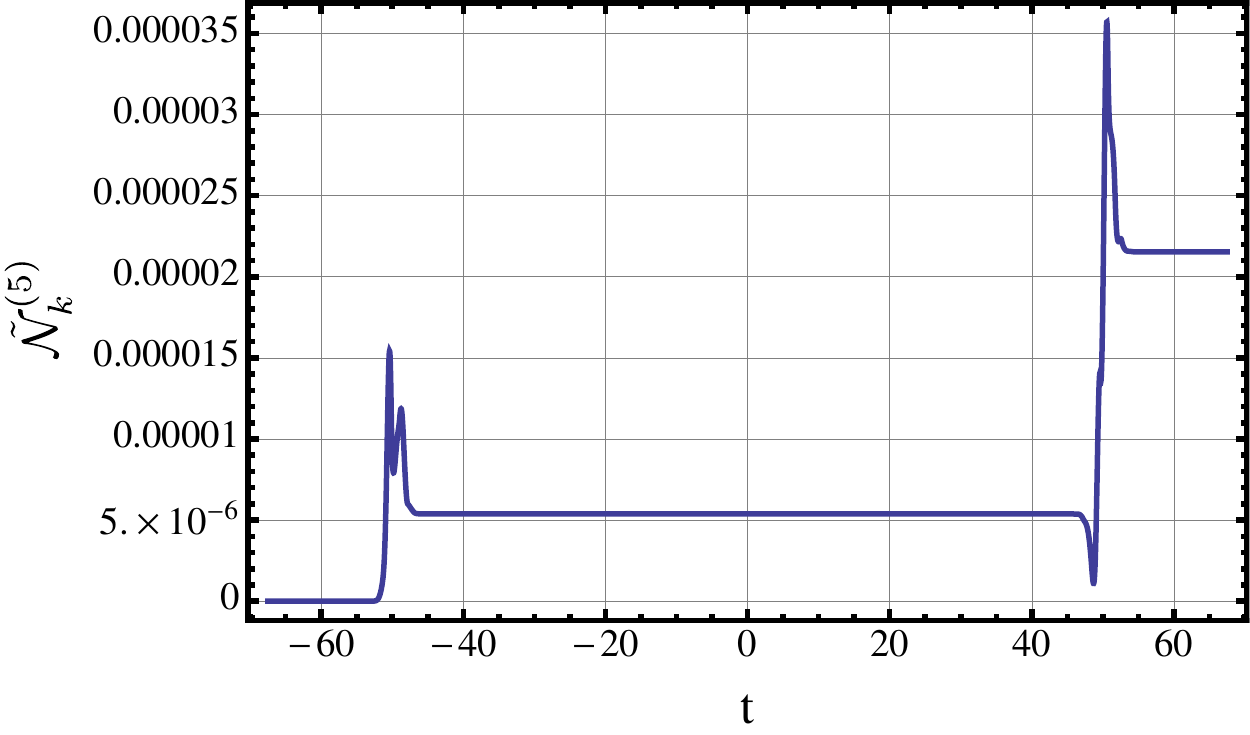}
\end{tabular}
\caption{
Time evolution of the adiabatic particle number for the first 6 orders of the adiabatic expansion, for  Schwinger particle production in a time-dependent double-pulse  electric field (\ref{2-pulse-field}), with parameters: $E=0.25$, $a=0.1$, $b=50$, transverse momentum $k_{\perp}=0$, and longitudinal momentum $k_\parallel=2.51555$, in units with $m=1$. The longitudinal momentum is chosen to correspond to constructive interference \cite{dd1,Akkermans:2011yn}. The final asymptotic value of the particle number, at future infinity, is the same for all orders of truncation. Note that the final asymptotic value of the particle number is 4 times that of the intermediate plateau, which is the $n^2$ enhancement factor for coherent constructive quantum interference \cite{Akkermans:2011yn}.
At intermediate times there are large oscillations in the particle number, which become much smaller as the optimal order ($j=3$) is reached, and then grow again rapidly beyond this optimal order of truncation. Such behavior is characteristic of {\it asymptotic} expansions, where the order of truncation depends  on the size of the expansion parameter, and going beyond this optimal order typically yields increasingly worse results.  
}
\label{2-pulse-constructive} 
\end{figure}
\begin{figure}[h!]
\begin{tabular}{cc}
\centering
\includegraphics[width=8cm]{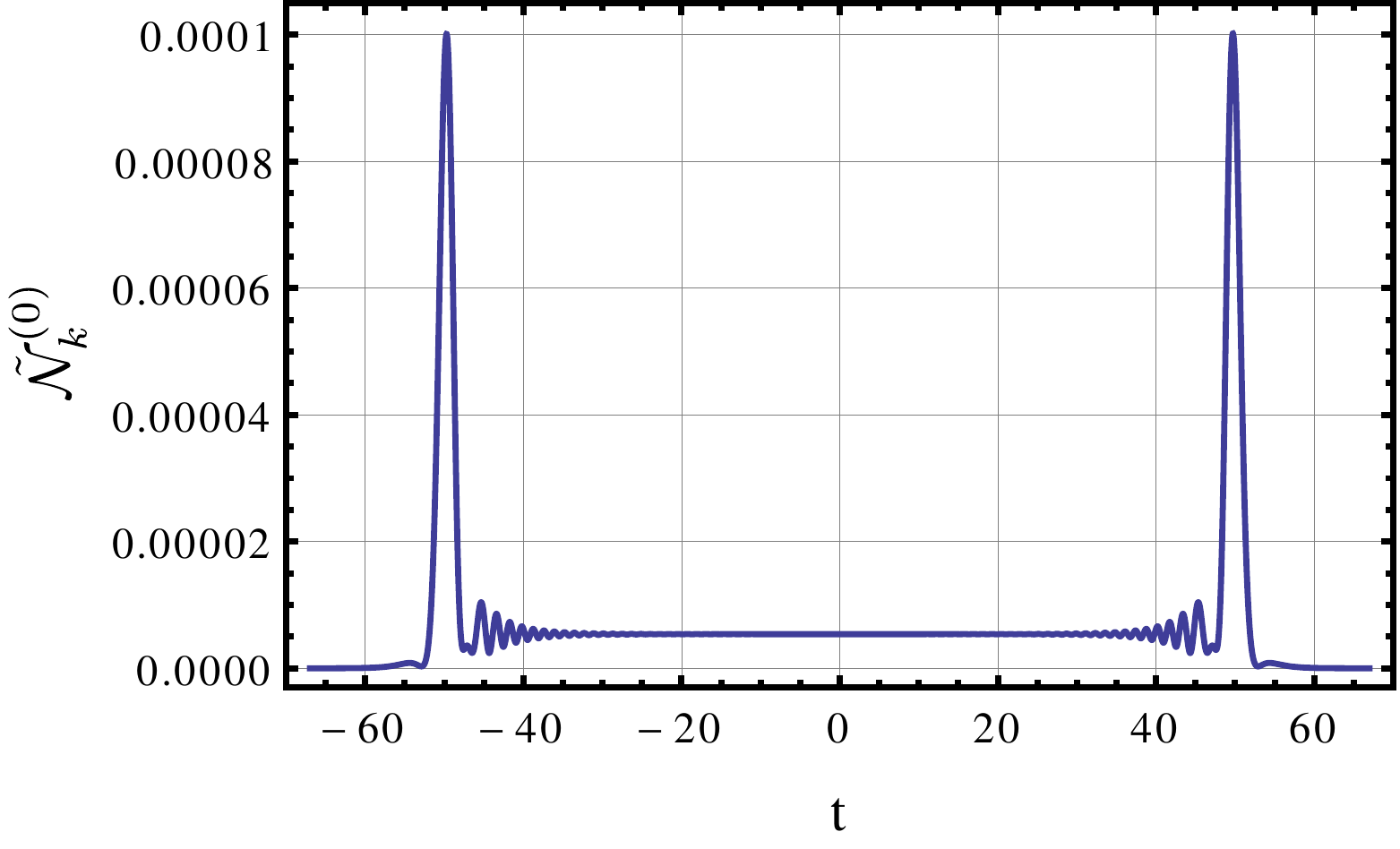} 
& 
\includegraphics[width=8cm]{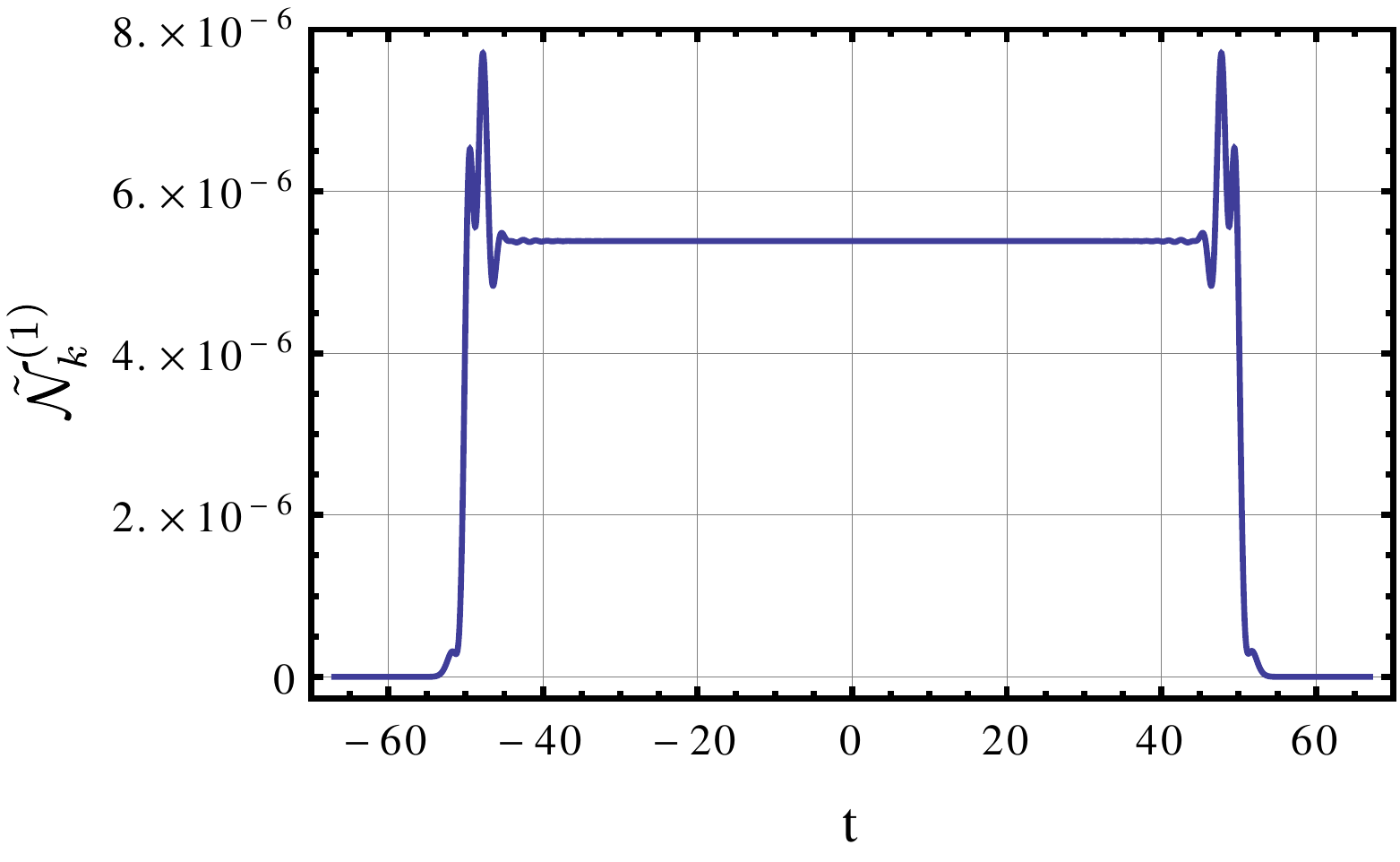} \\
\includegraphics[width=8cm]{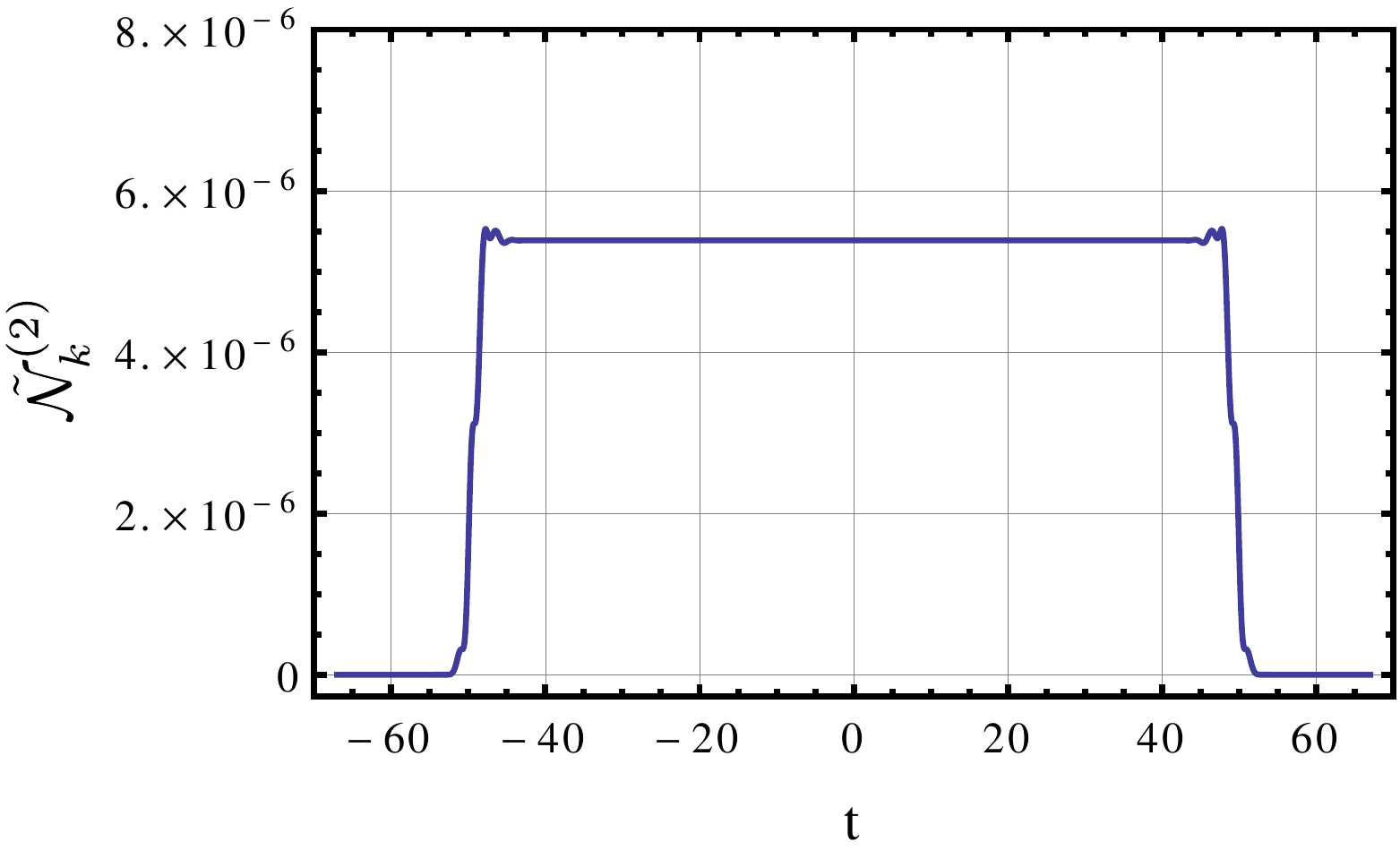} 
& 
\includegraphics[width=8cm]{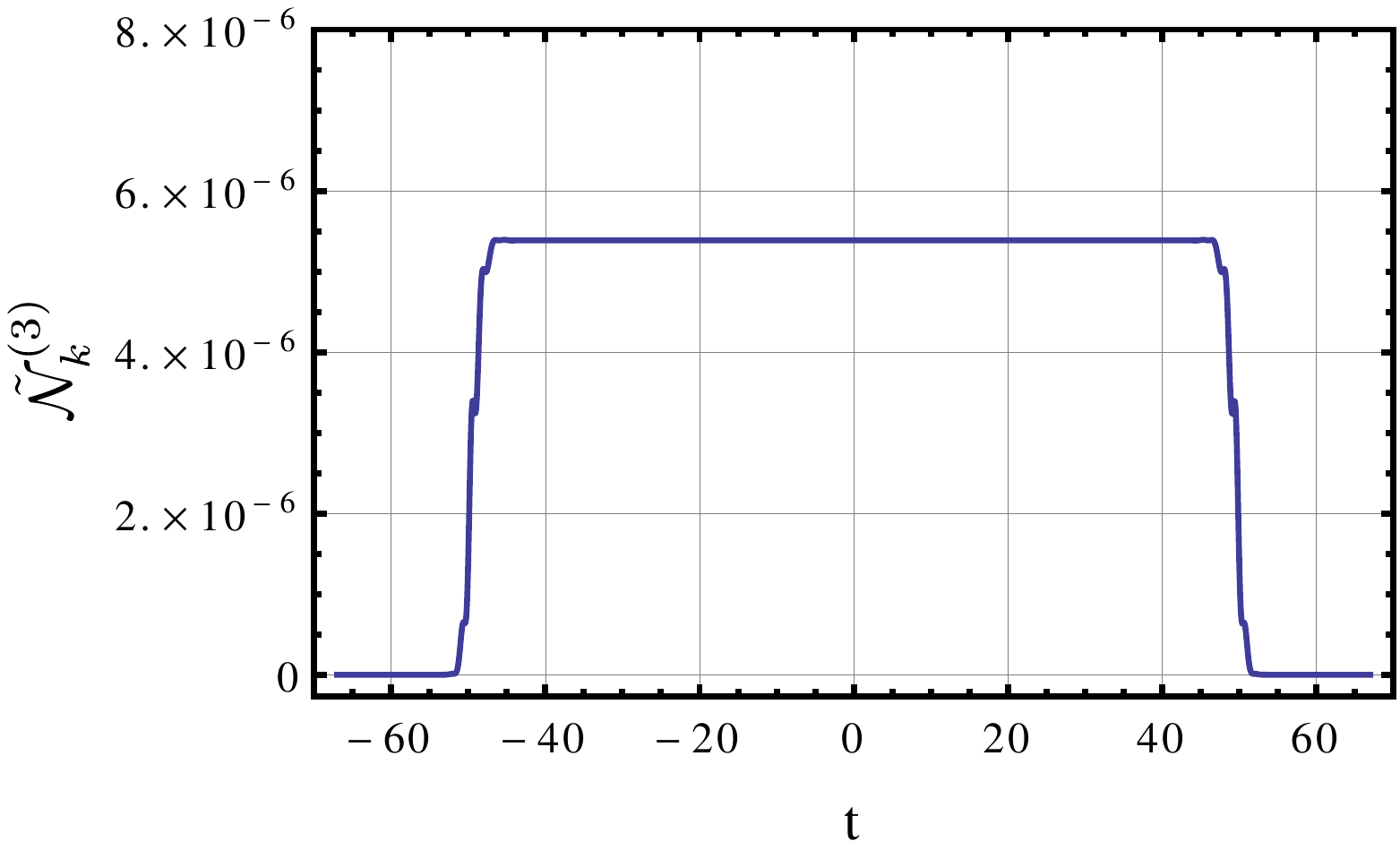} \\
\includegraphics[width=8cm]{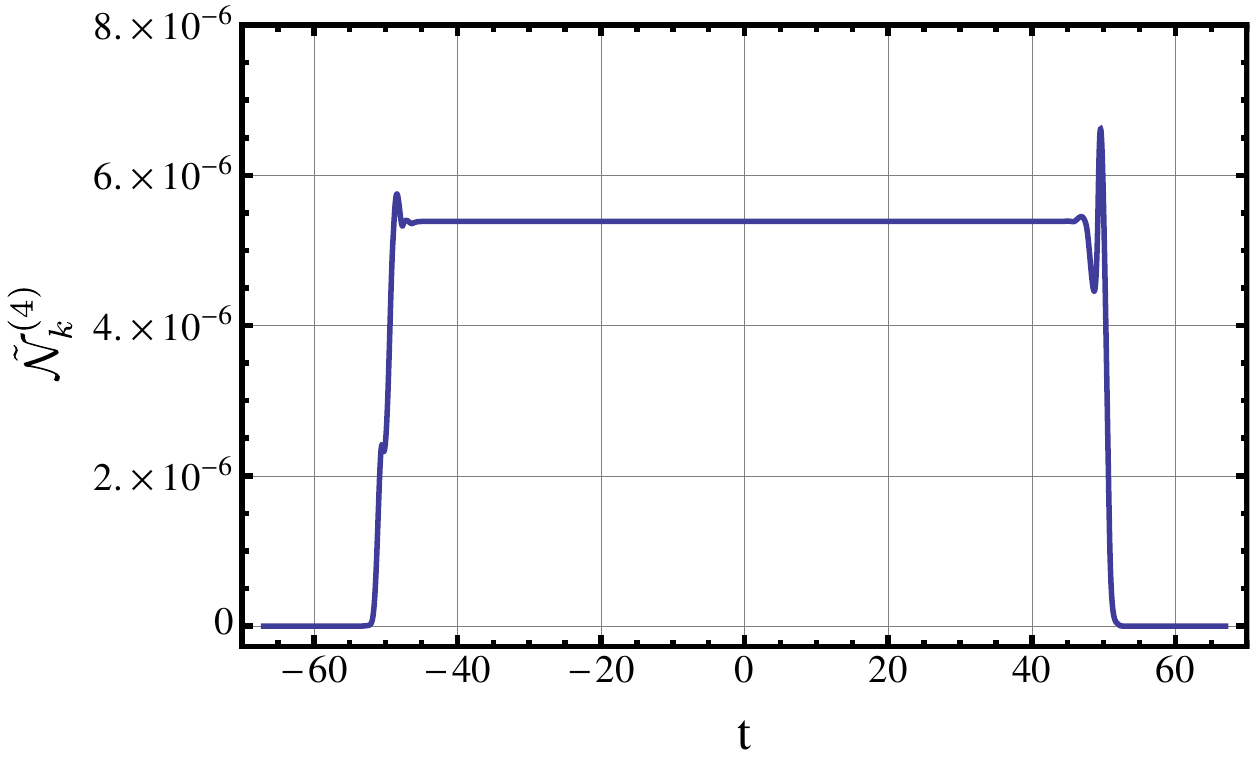} 
& 
\includegraphics[width=8cm]{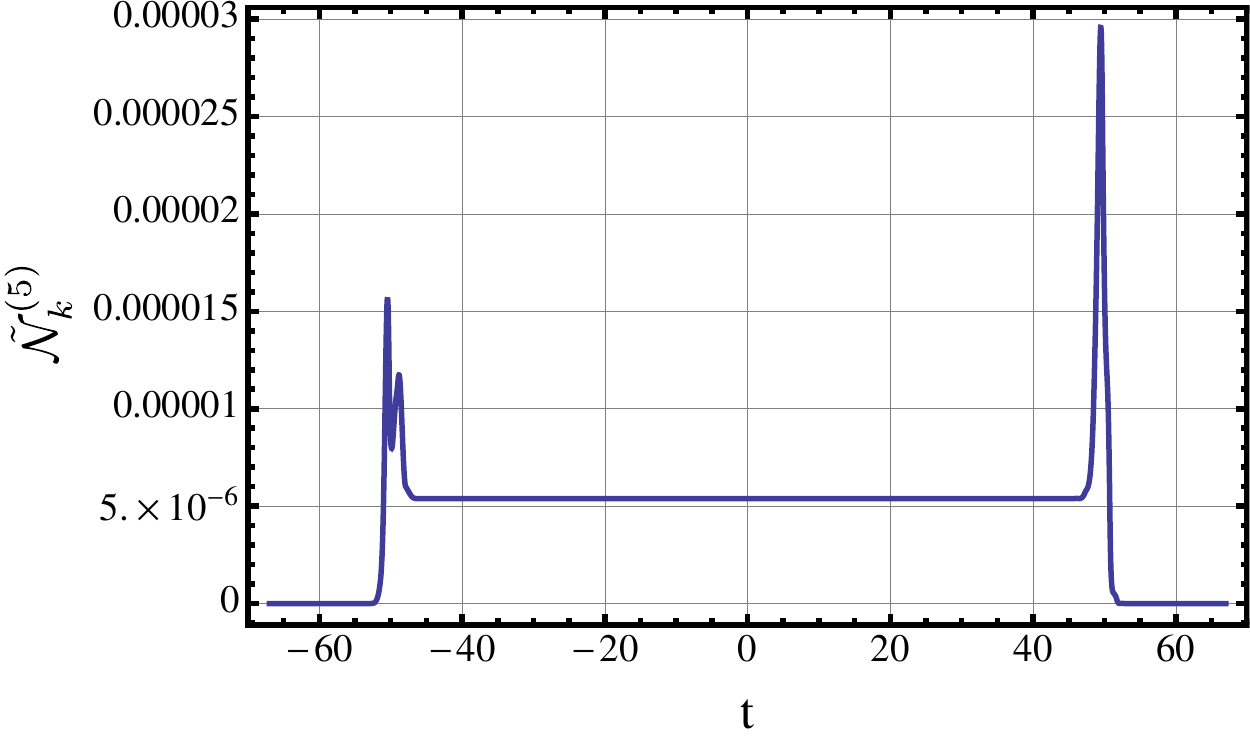} \\
\end{tabular}
\caption{
As in Figure \ref{2-pulse-constructive}, but  with parameters: $E=0.25$, $a=0.1$, $b=50$, transverse momentum $k_{\perp}=0$, and longitudinal momentum $k_\parallel=2.49887$, in units with $m=1$. The longitudinal momentum is chosen to correspond to destructive interference \cite{dd1,Akkermans:2011yn}. The final asymptotic value of the particle number, at future infinity, vanishes for each order of truncation. This vanishing of the final asymptotic value of the particle number is characteristic of coherent destructive quantum interference \cite{Akkermans:2011yn}.
}
\label{2-pulse-destructive} 
\end{figure}

We consider an electric field consisting of the two successive pulses, of alternating sign:
\begin{eqnarray}
E(t)=-E\, {\rm sech}^2\left[a(t+b)\right]+E\, {\rm sech}^2\left[a(t-b)\right]
\label{2-pulse-field}
\end{eqnarray}
for which we can choose a time-dependent vector potential:
\begin{eqnarray}
A_\parallel(t) = - \frac{E}{a} \Big( - \tanh \big[a (t + b)\big] + \tanh\big[ a (t - b) \big] \Big)
\end{eqnarray}
\begin{figure}[h!]
\begin{tabular}{cc}
\centering
\includegraphics[width=7cm]{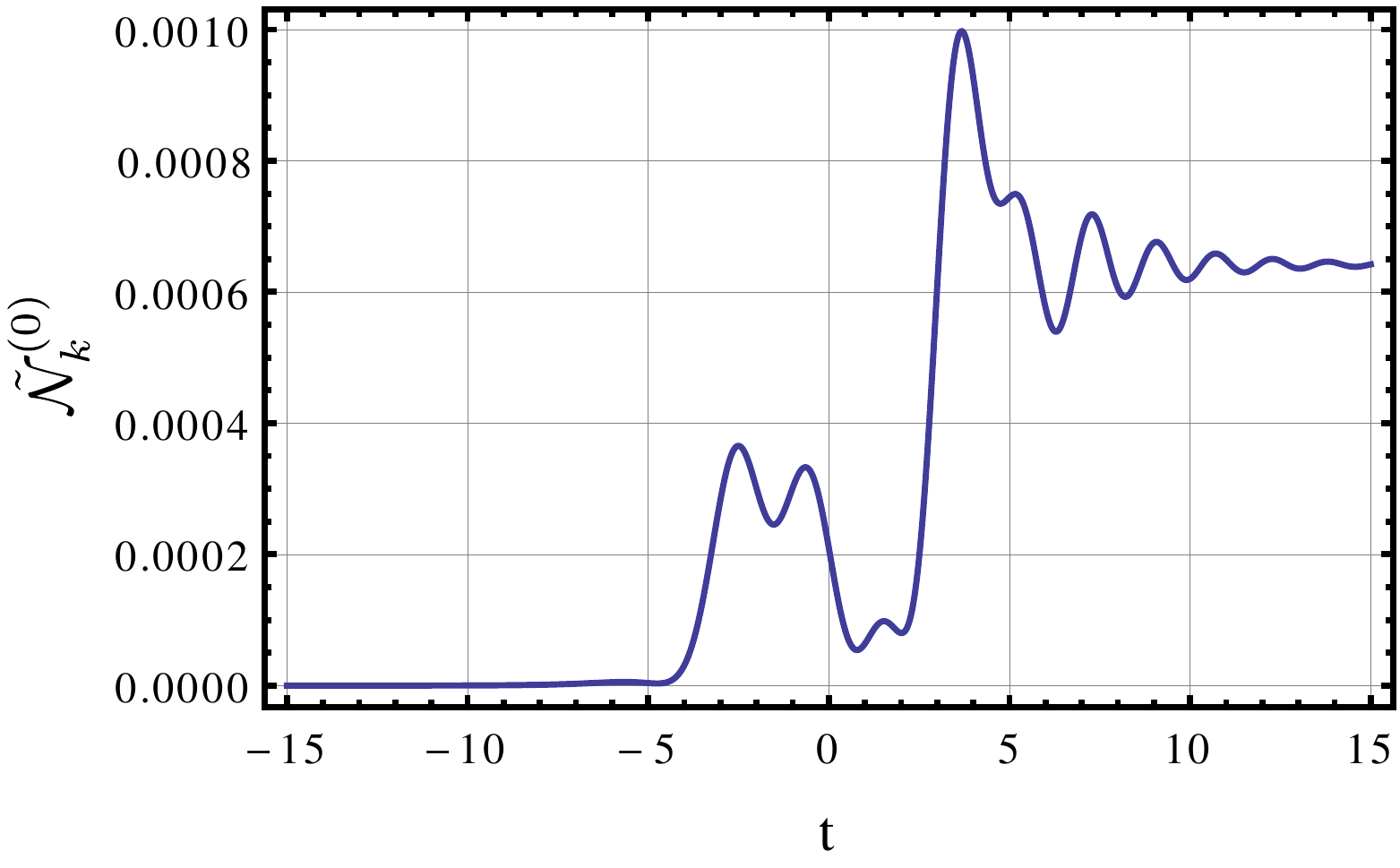} 
& 
\includegraphics[width=7cm]{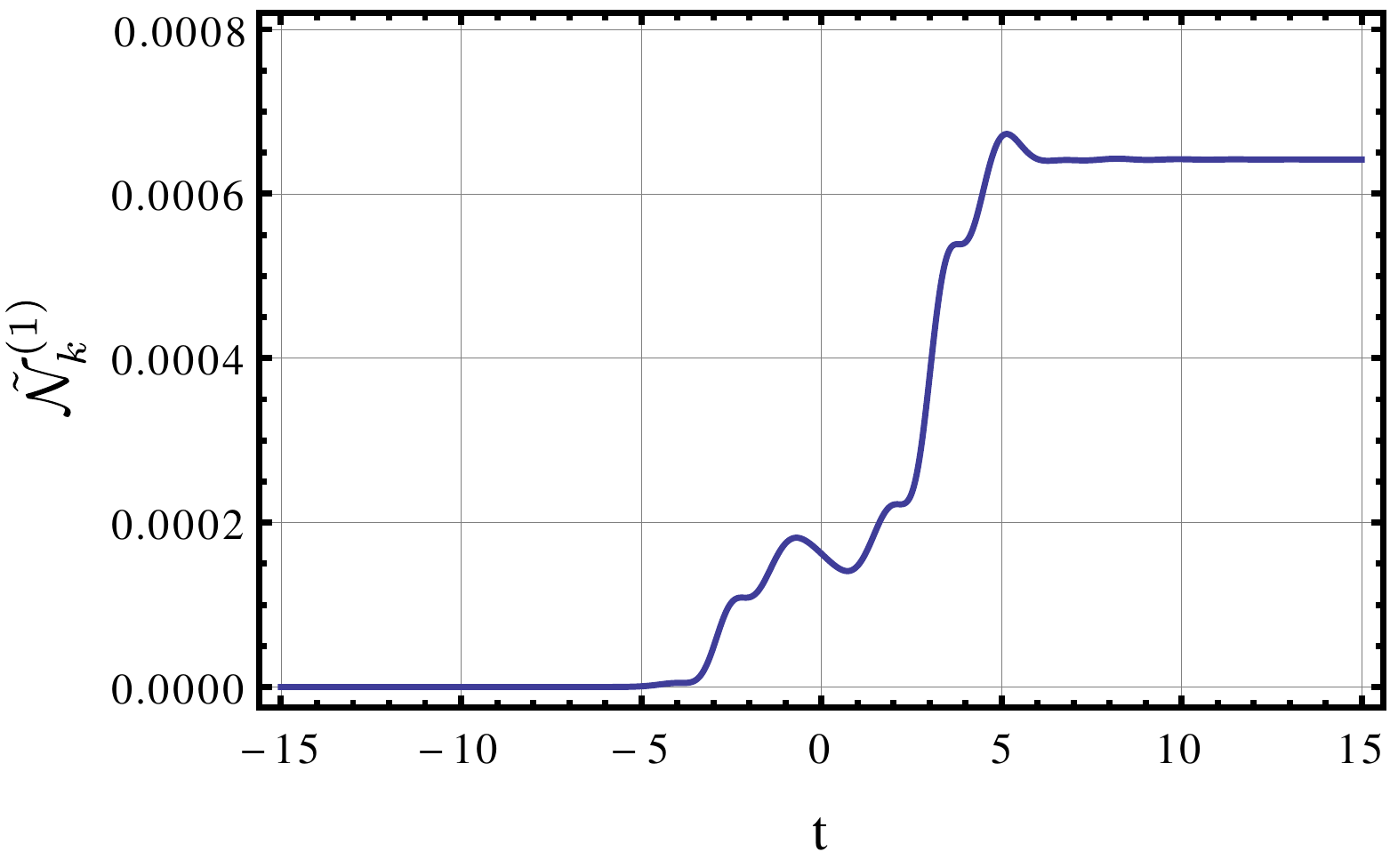} \\
\includegraphics[width=7cm]{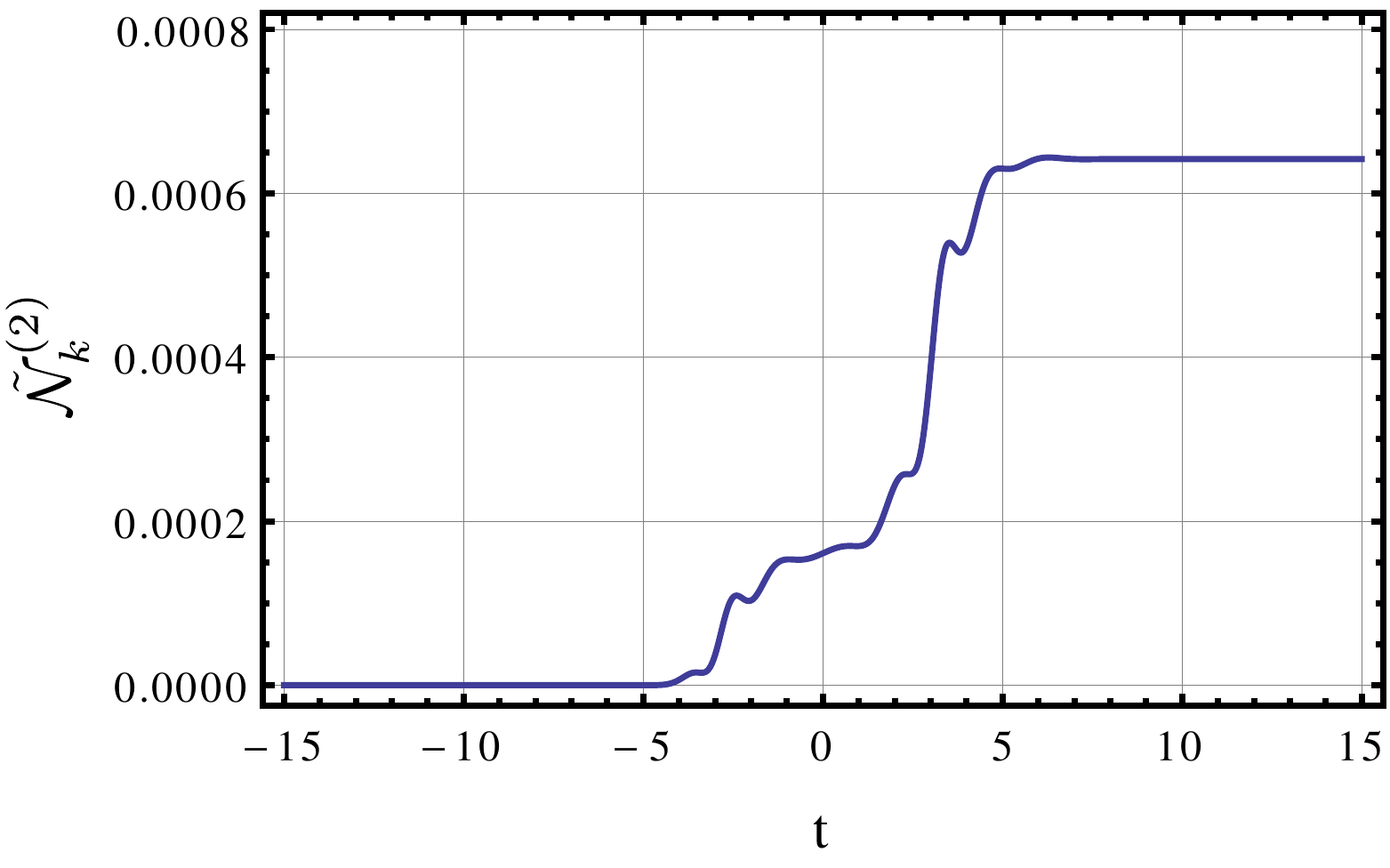} 
& 
\includegraphics[width=7cm]{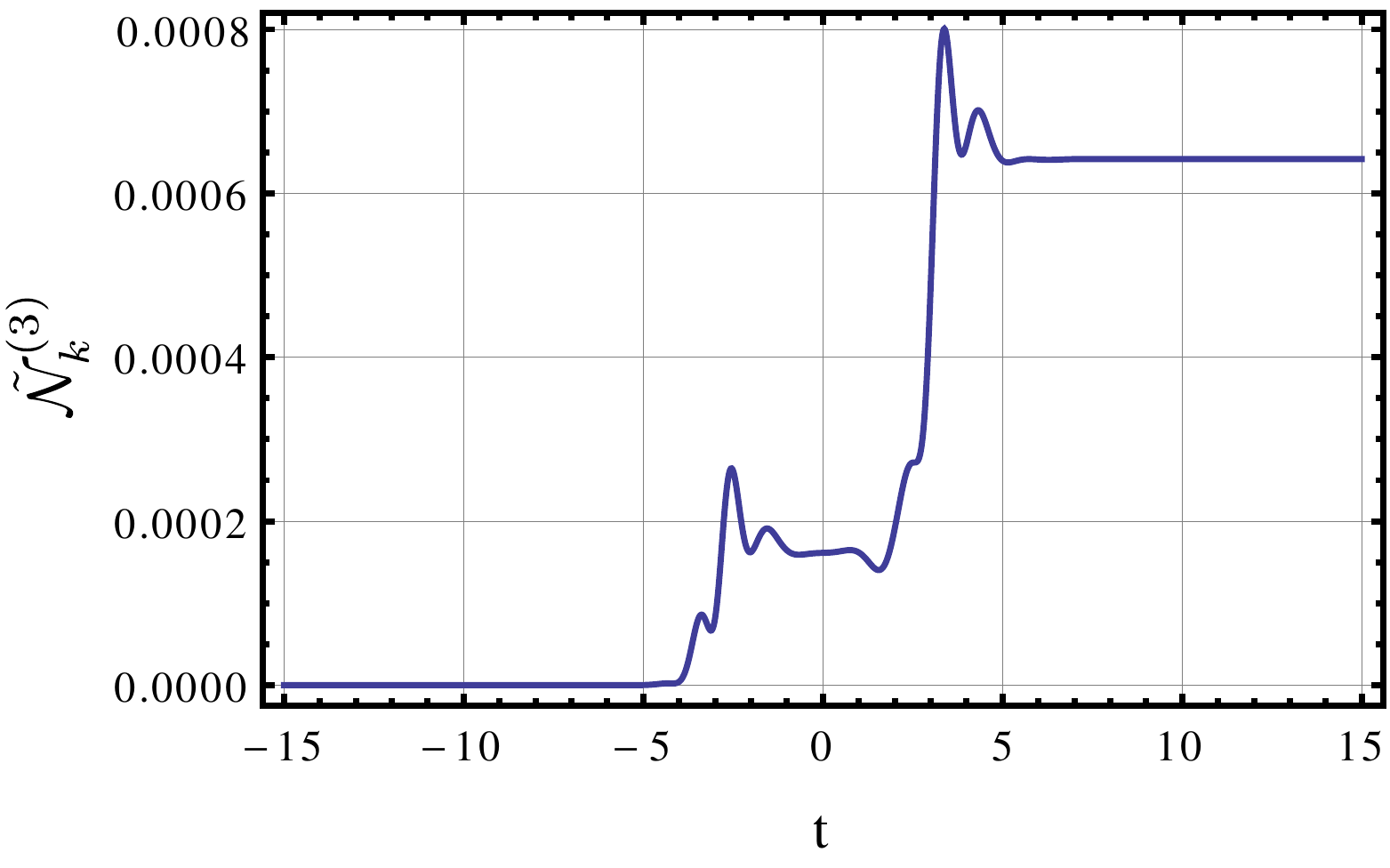} 
\end{tabular}
\caption{Time evolution of the adiabatic particle number for a  sequence of two alternating-sign pulses,  of the form  (\ref{2-pulse-field}), with parameters: $E=0.5$, $a=0.2$, $b=2.5$, transverse momentum $k_{\perp}=0$, and longitudinal momentum $k_\parallel=1.85$, in units with $m=1$. Here the pulses are much closer together than in Figure \ref{2-pulse-constructive}, and we observe in the first plot that at leading order of the adiabatic approximation it is more difficult to resolve the situation into two pulses with coherent constructive interference between the produced particles. For these parameters the optimal order is $j=2$, and we see clearly the two-step coherent constructive interference with plateaux in the ratio $1:4$.
}
\label{2-pulse-close-max}
\end{figure}
\begin{figure}[h!]
\begin{tabular}{cc}
\centering
\includegraphics[width=7cm]{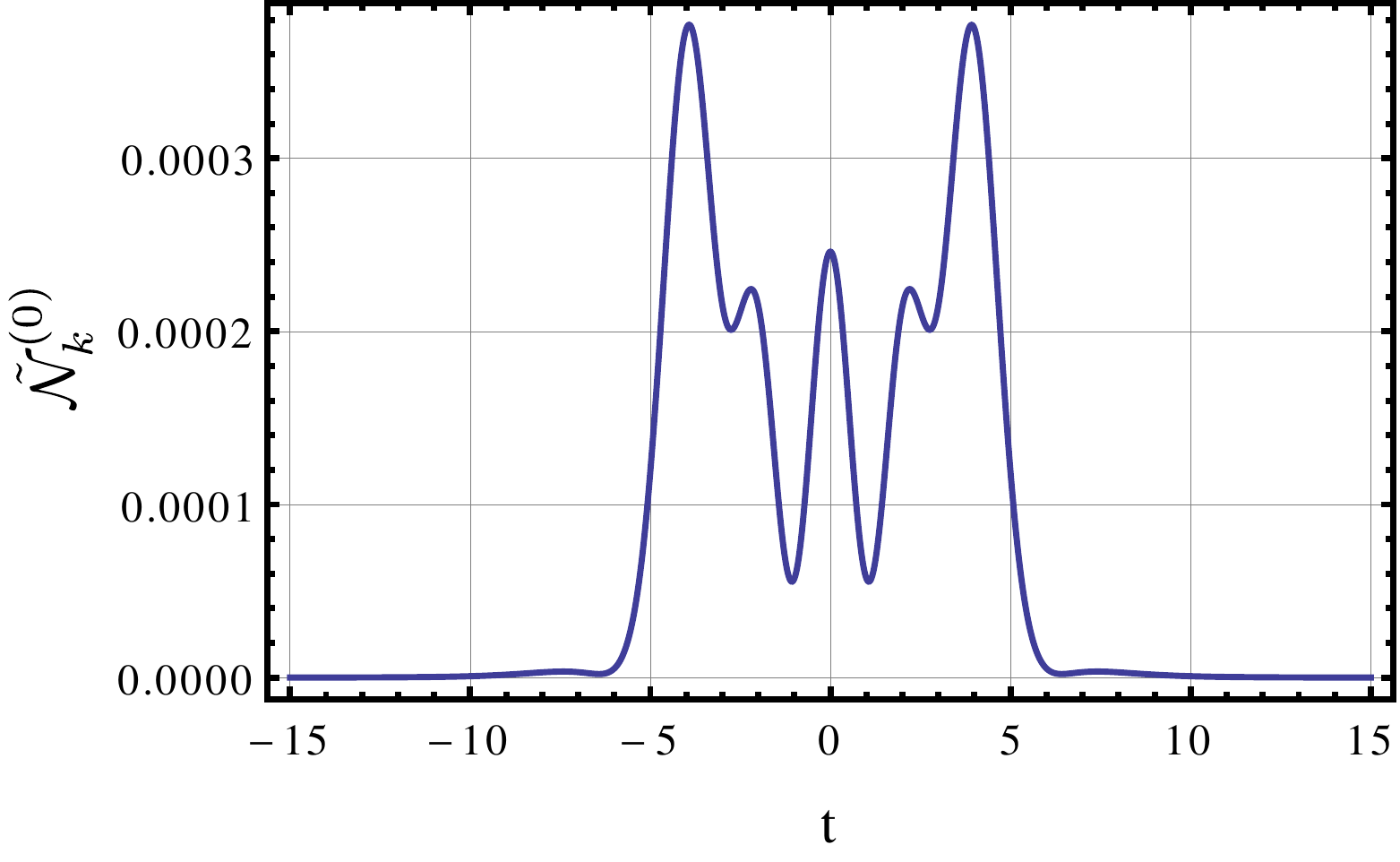} 
& 
\includegraphics[width=7cm]{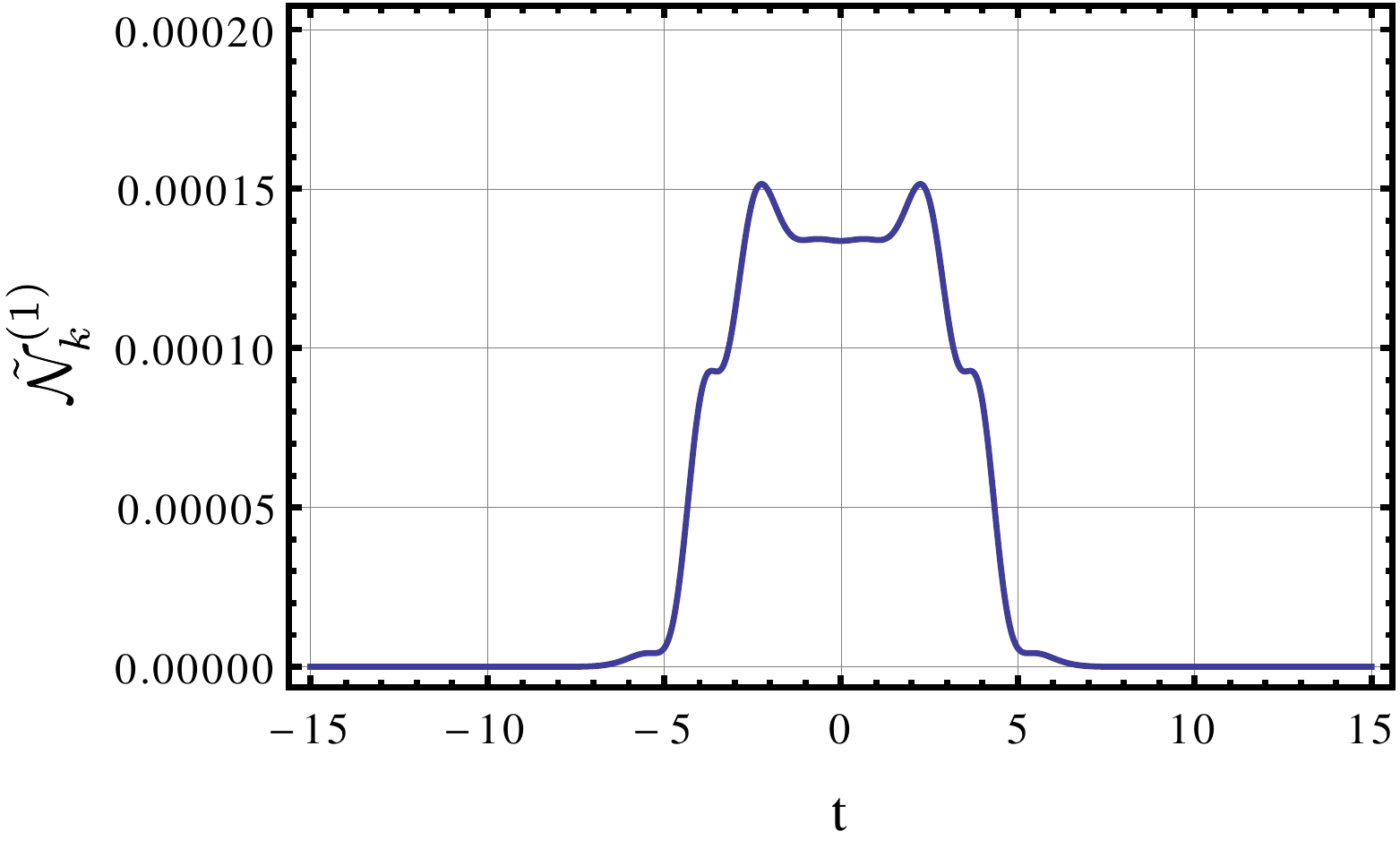} \\
\includegraphics[width=7cm]{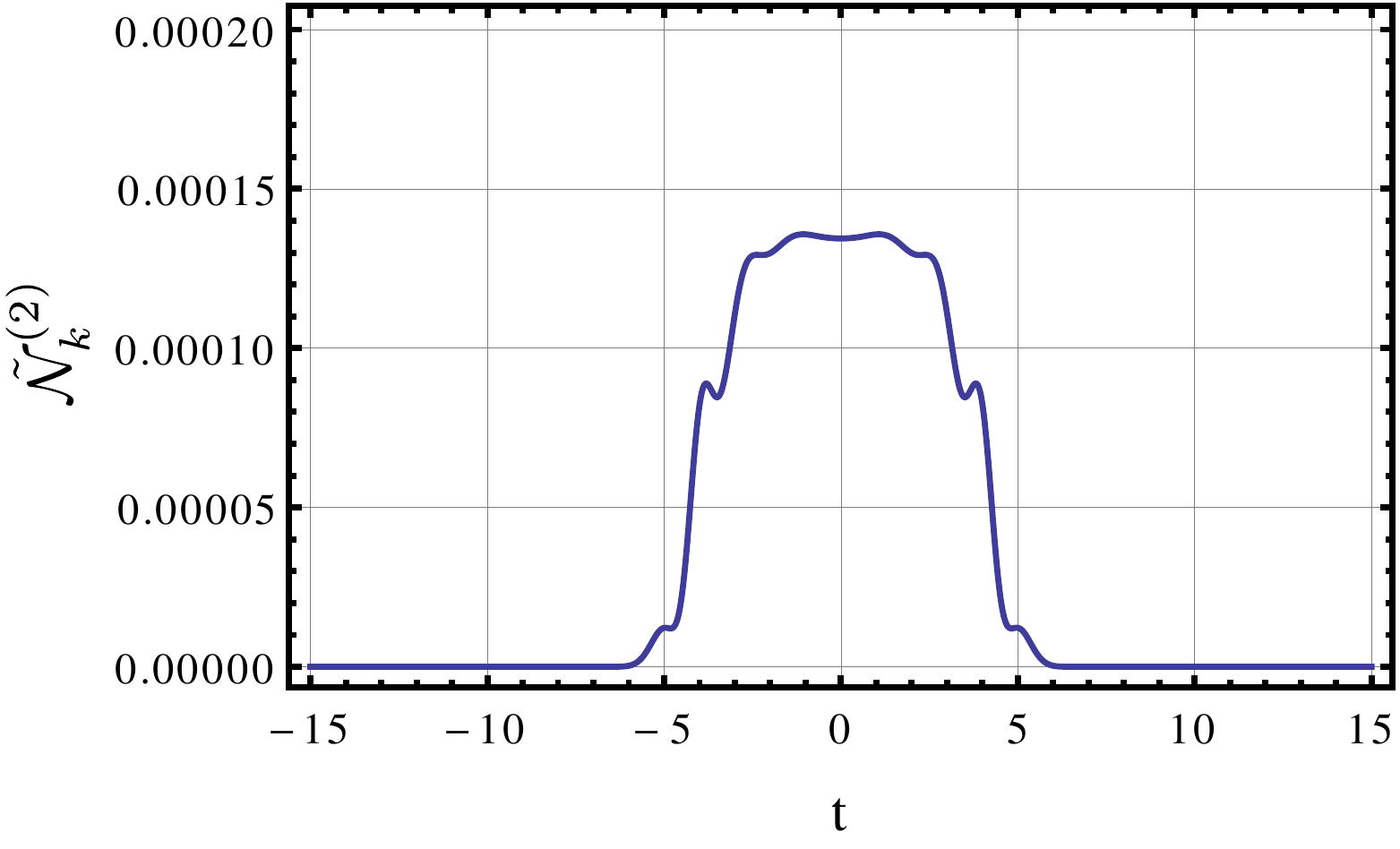} 
& 
\includegraphics[width=7cm]{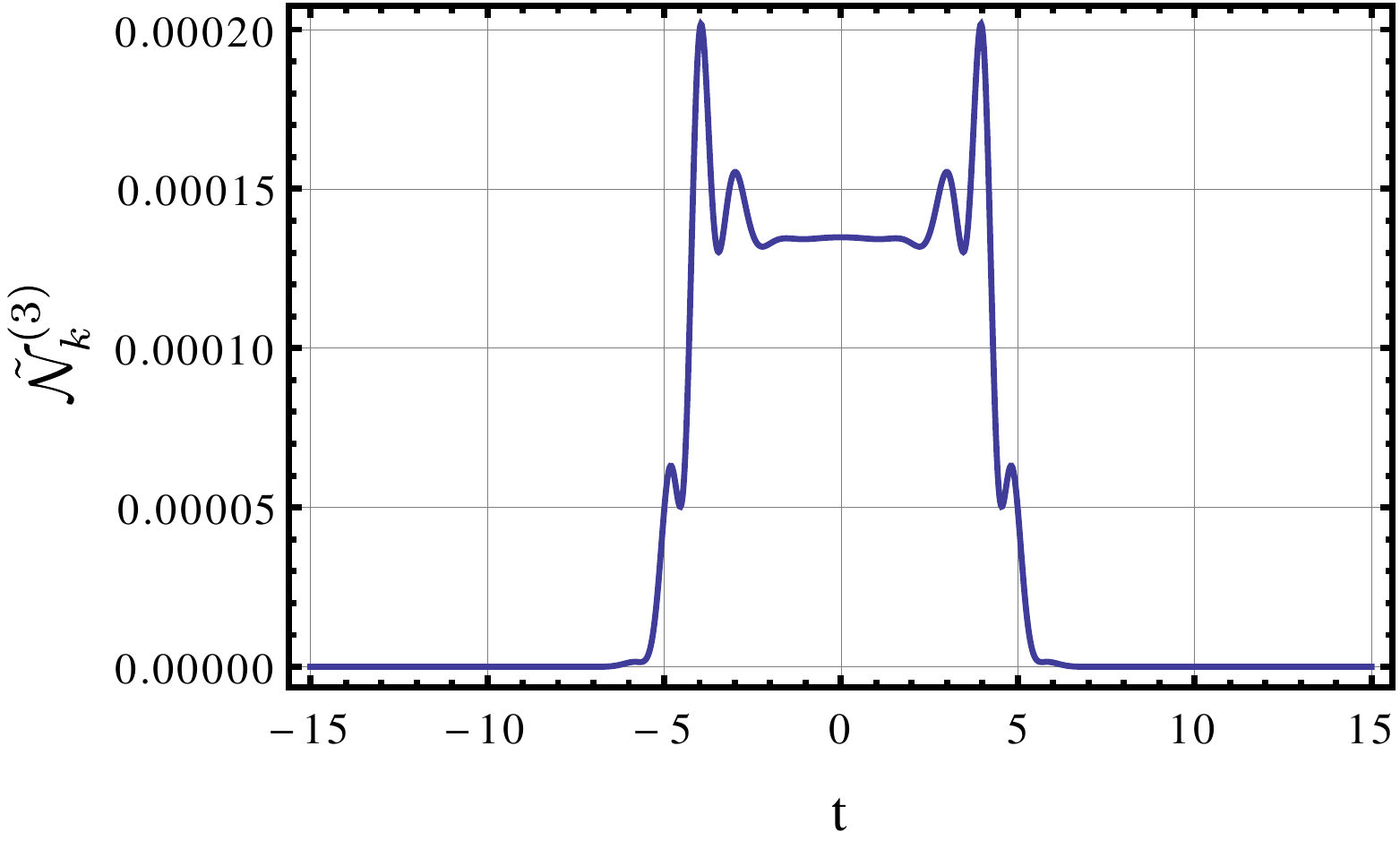}
\end{tabular}
\caption{
As in Figure \ref{2-pulse-close-max}, but with 
longitudinal momentum $k_\parallel=1.22175$
Here the pulses are much closer together than in Figure \ref{2-pulse-destructive}, and we observe in the first plot that at leading order of the adiabatic approximation it is more difficult to resolve the situation into two pulses with coherent destructive interference between the produced particles. The optimal order is $j=2$, and we see clearly the two-step coherent destructive interference with vanishing asymptotic particle number.}
\label{2-pulse-close-min}
\end{figure}

In addition to choosing the field parameters, we can choose to probe the momentum of the produced particles. Following \cite{dd1,Akkermans:2011yn}, we consider the cases of maximal constructive interference and maximal destructive interference by choosing two different values of the particle momenta, associated with the central maximum and the first minimum of the momentum spectrum. The results are shown in Figures \ref{2-pulse-constructive} and \ref{2-pulse-destructive}, respectively. In these plots the pulses occur at $t=\mp 50$. These plots show that in low orders of the adiabatic expansion there are large oscillations at non-asymptotic  times in the vicinity of the pulses, but at the optimal order of truncation (here $j=3$) the time evolution becomes smooth. Furthermore, we clearly see the coherent constructive interference in Figure \ref{2-pulse-constructive}, as the final plateau is four times the value of the plateau between the two pulses. On the other hand, Figure \ref{2-pulse-destructive} shows destructive interference, as the super-adiabatic particle number rises to a value corresponding to a single pulse, and then falls back to zero after the second pulse. In both Figure \ref{2-pulse-constructive} and  Figure \ref{2-pulse-destructive}, the physical parameters are such that $F_k^{(0)}\approx 6.050$, consistent with the estimate (\ref{joptimal}) for the optimal truncation order, and $\exp(-2F_k^{(0)})\approx 5.56\times 10^{-6}$, consistent with the intermediate plateau value of the particle number in both cases.

Examples with pulses  closer together in time, are shown in Figures \ref{2-pulse-close-max} and \ref{2-pulse-close-min}. Note that while  in Figures \ref{2-pulse-constructive} and \ref{2-pulse-destructive} it is easy to resolve the adiabatic particle number into two separate events, even at the lowest order of the adiabatic expansion, it is more difficult to make such a  distinction for the parameters of Figures \ref{2-pulse-close-max} and \ref{2-pulse-close-min}. But at the optimal order (here $j\approx 1- 2$) one can clearly resolve the situation of two separate creation events, with characteristic plateaux amplitudes in the ratio $1:4$ in the case of constructive interference (Figure \ref{2-pulse-close-max}), and destructive interference (Figure \ref{2-pulse-close-min}). 
In Figure \ref{2-pulse-close-max} and Figure \ref{2-pulse-close-min}, the physical parameters are such that $F_k^{(0)}\approx 4.303$ and $4.405$, respectively, and $\exp(-2F_k^{(0)})\approx 1.83\times 10^{-4}$ and $1.49\times 10^{-4}$, respectively, consistent with the estimates in (\ref{answer}) and (\ref{joptimal}).


\begin{figure}[h!]
\begin{tabular}{rcl}
\centering
\includegraphics[height=3.35cm,width=5.5cm]{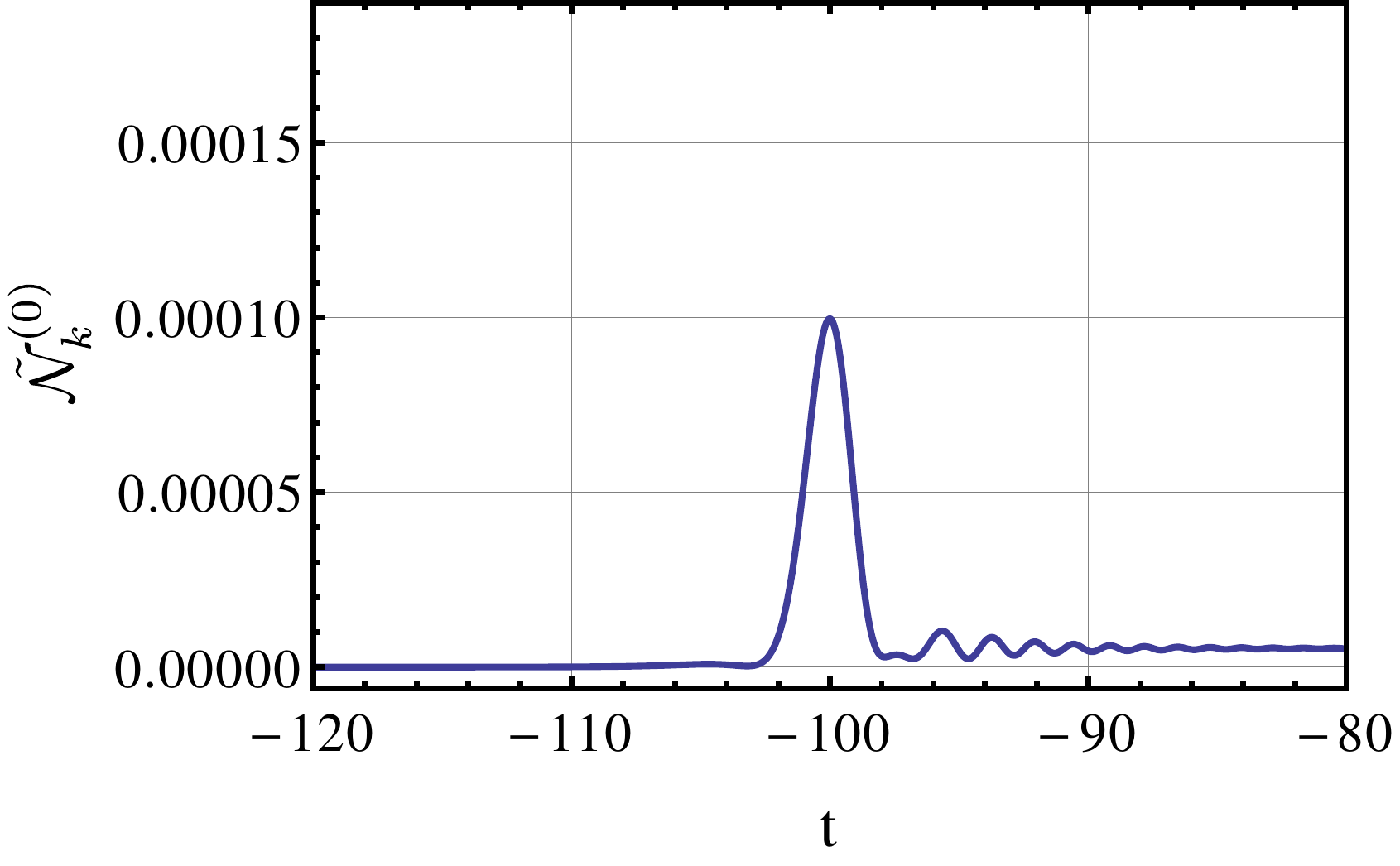} 
& \includegraphics[width=4.5cm]{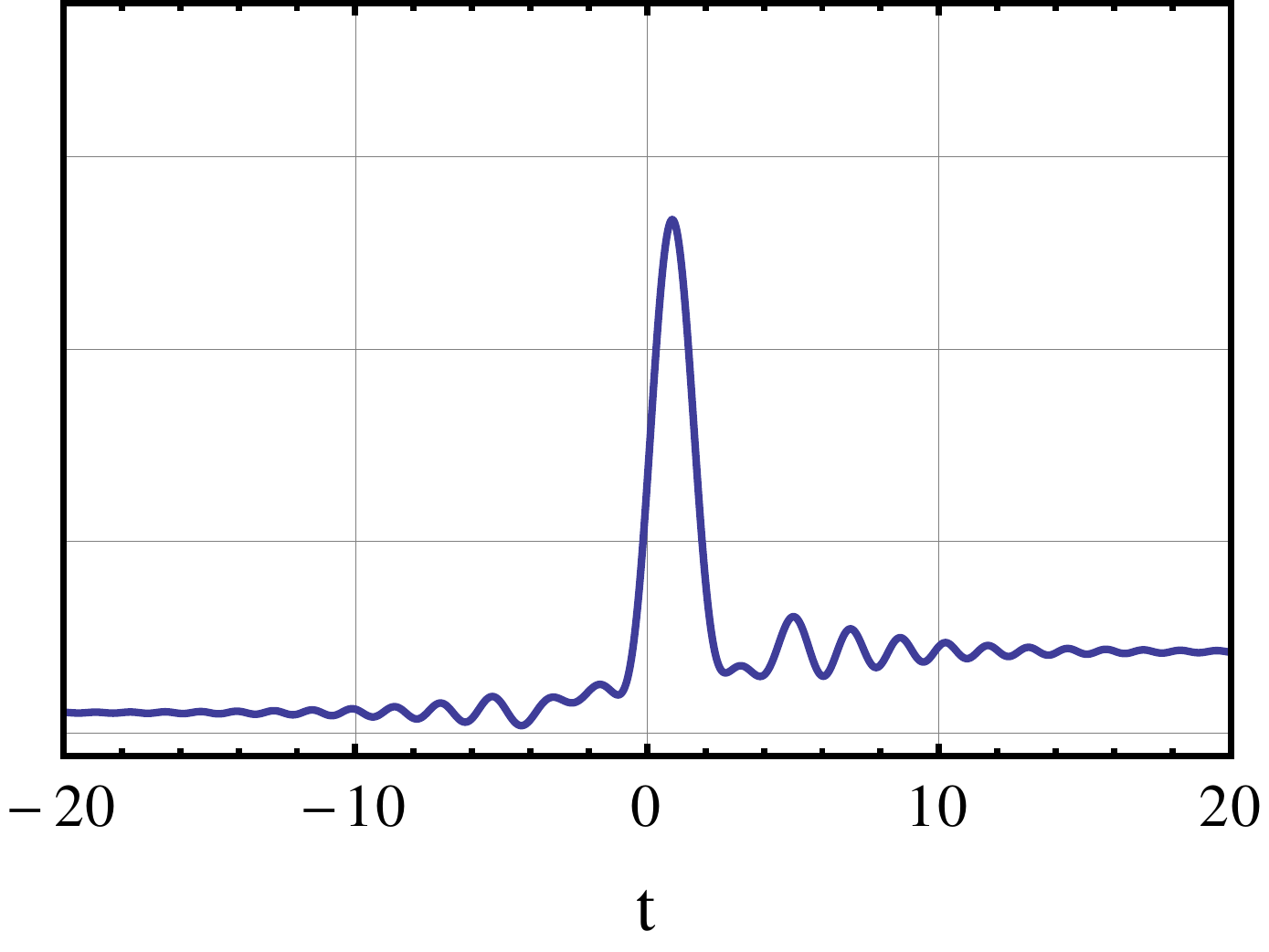}
& \includegraphics[width=4.5cm]{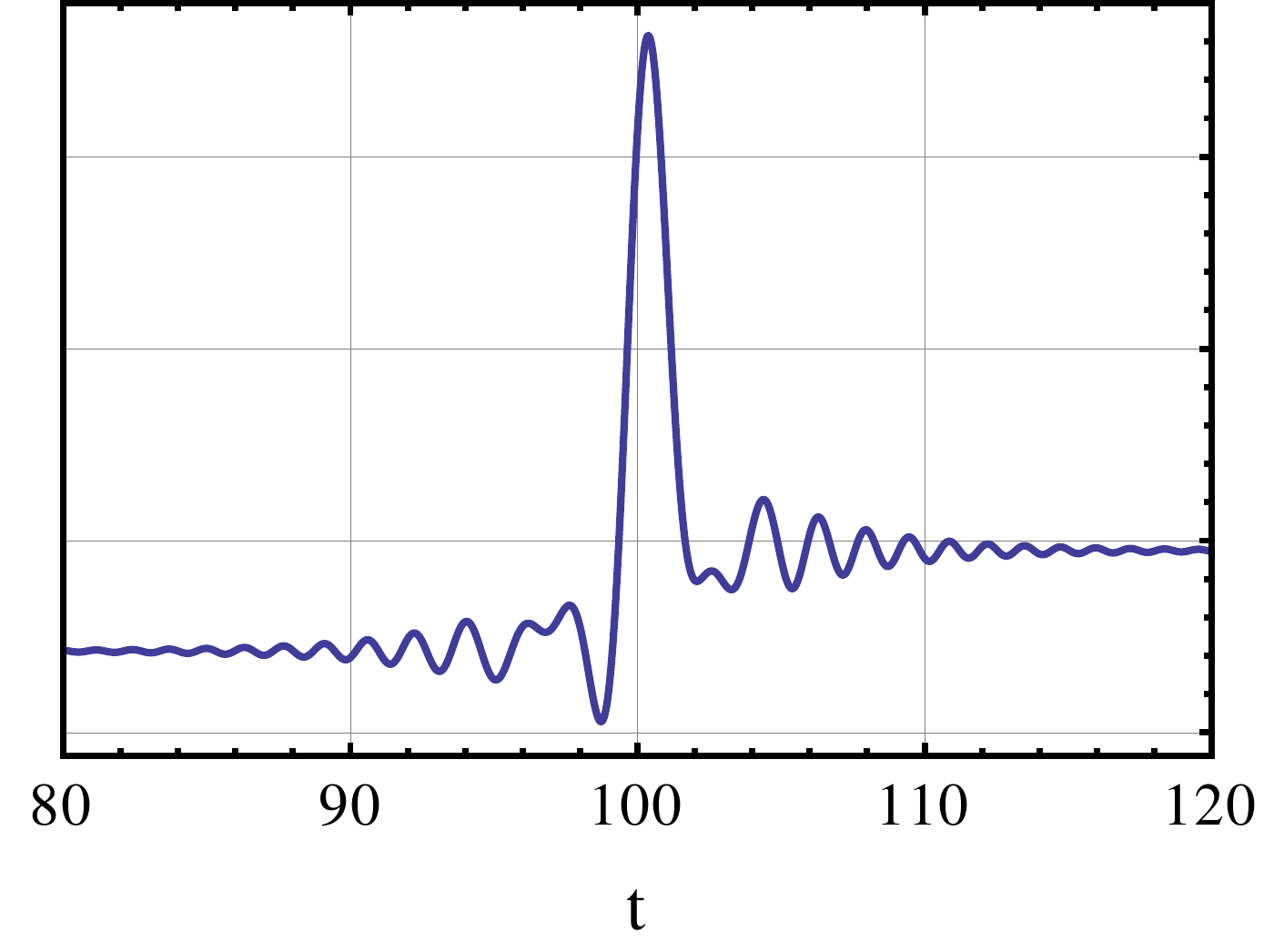} \\
\includegraphics[height=3.35cm,width=5.5cm]{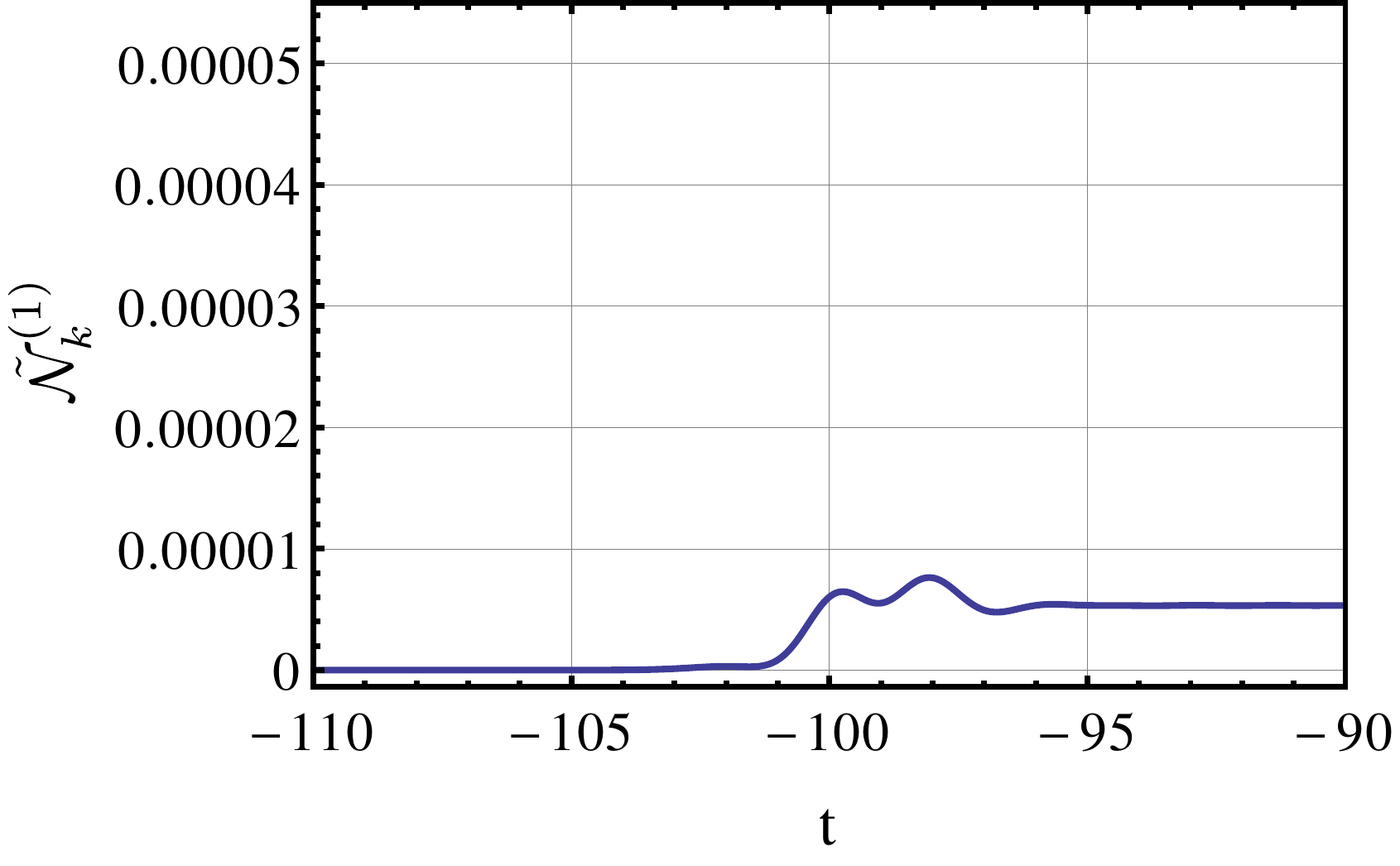} 
& \includegraphics[width=4.5cm]{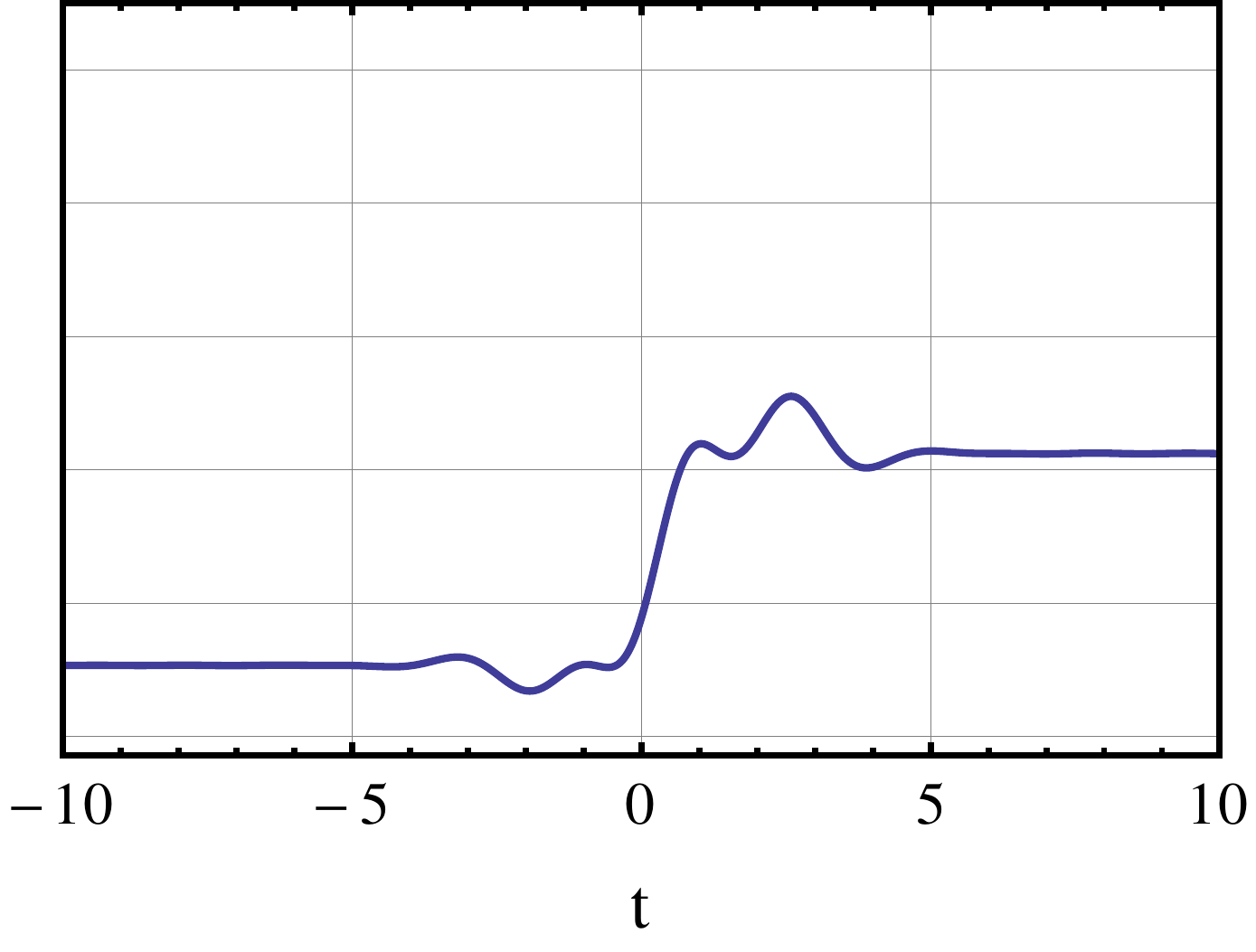} 
& \includegraphics[width=4.5cm]{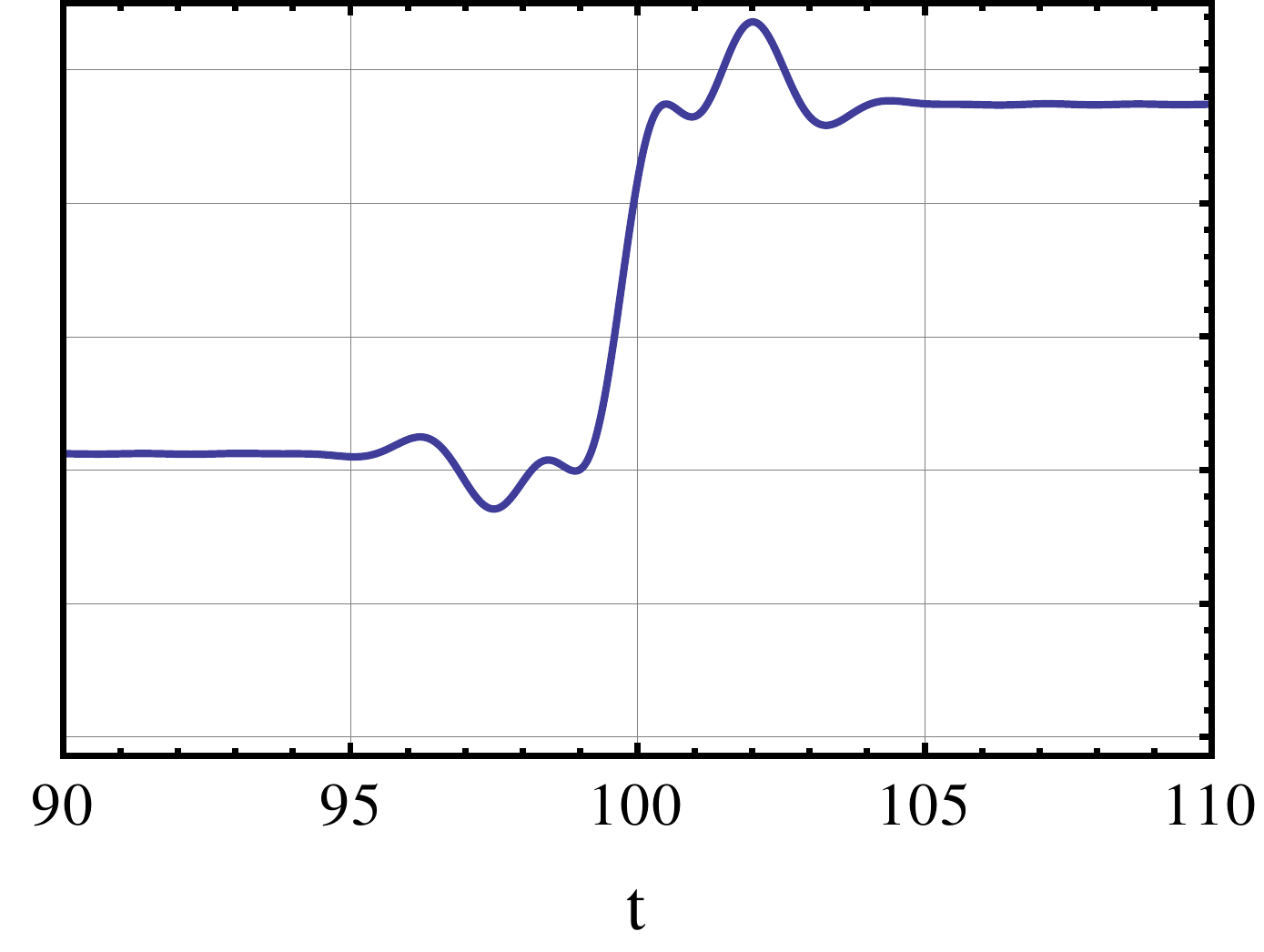} \\
\includegraphics[height=3.35cm,width=5.5cm]{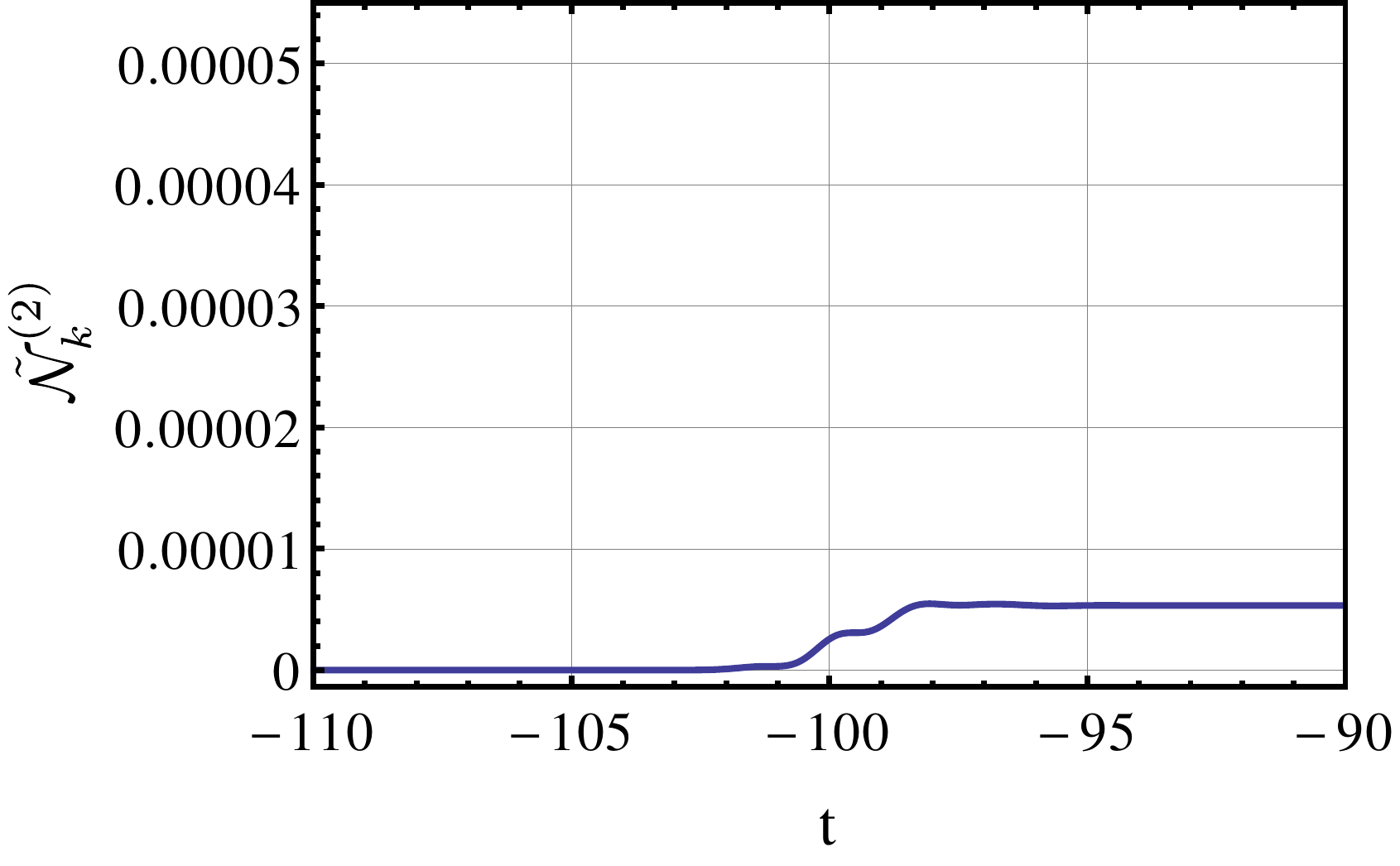}
& \includegraphics[width=4.5cm]{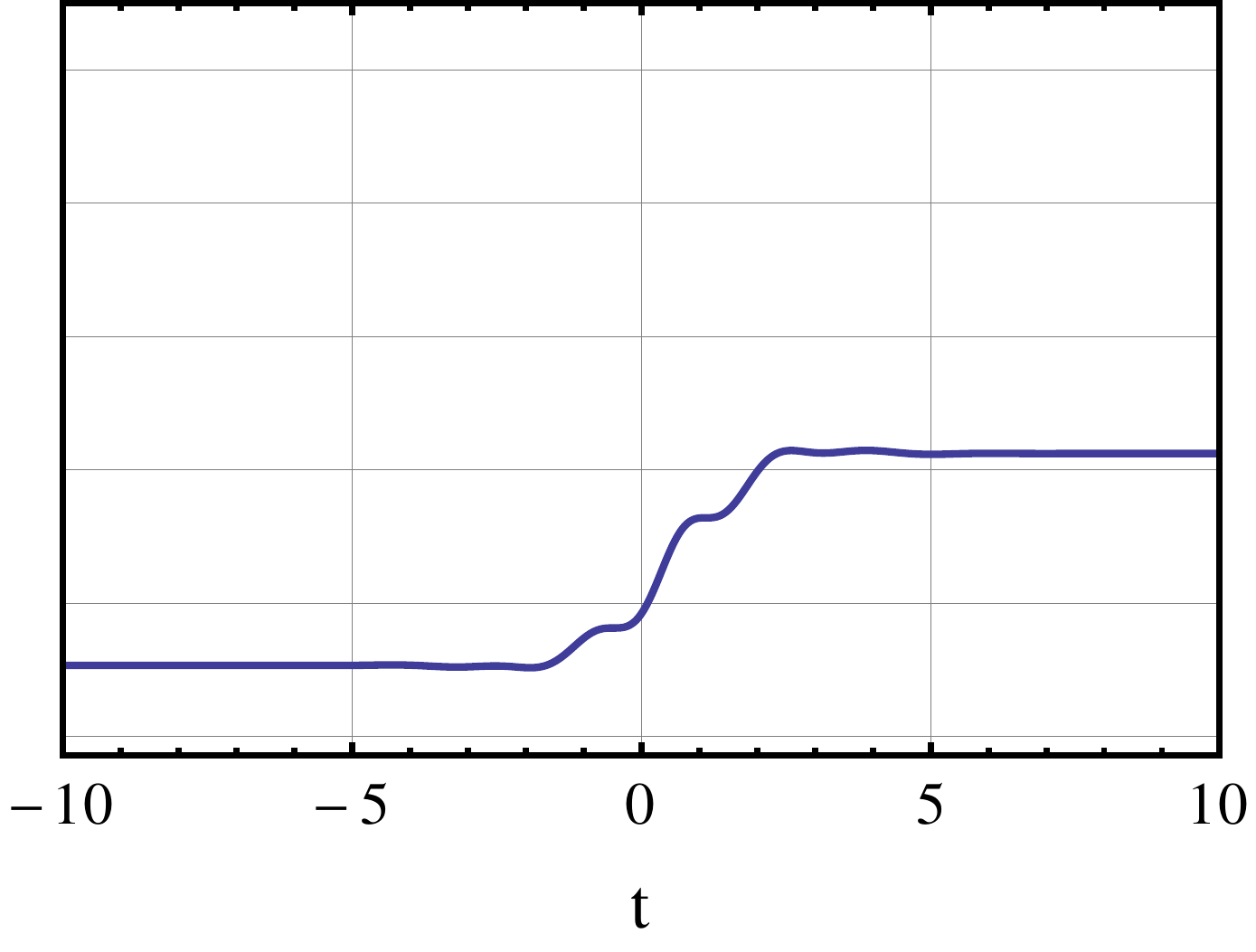}
& \includegraphics[width=4.5cm]{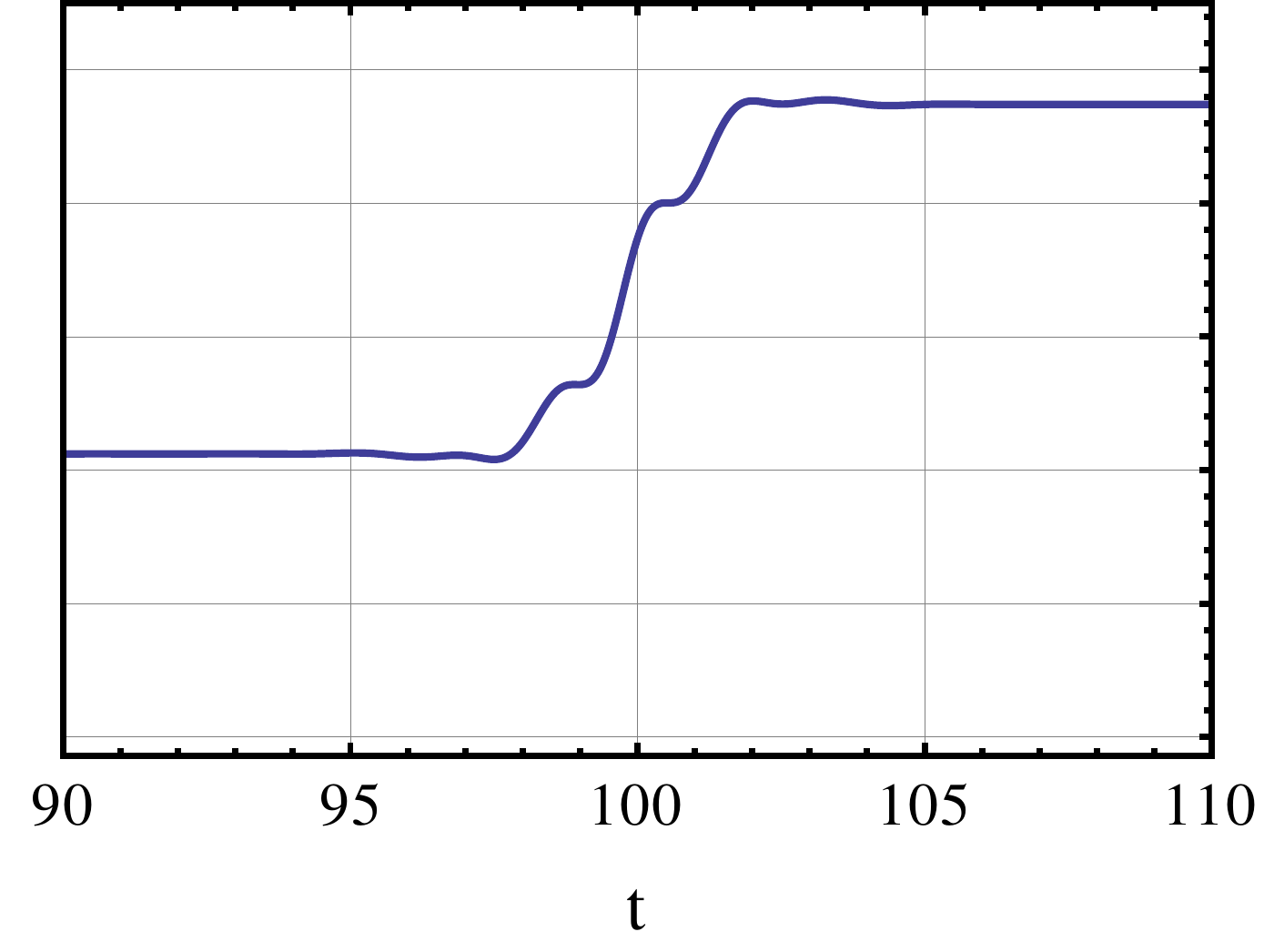} \\
\includegraphics[height=3.35cm,width=5.5cm]{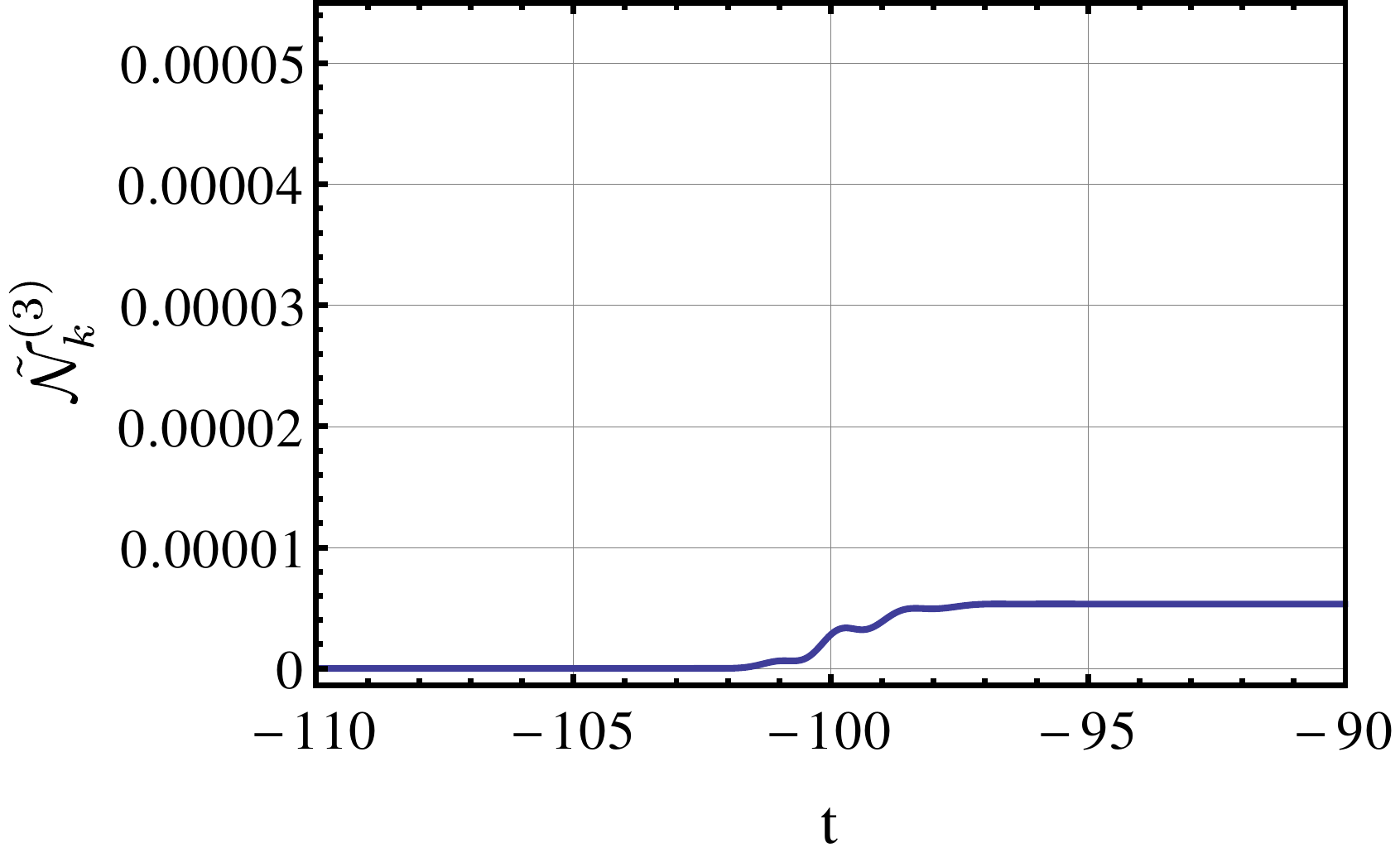}
& \includegraphics[width=4.5cm]{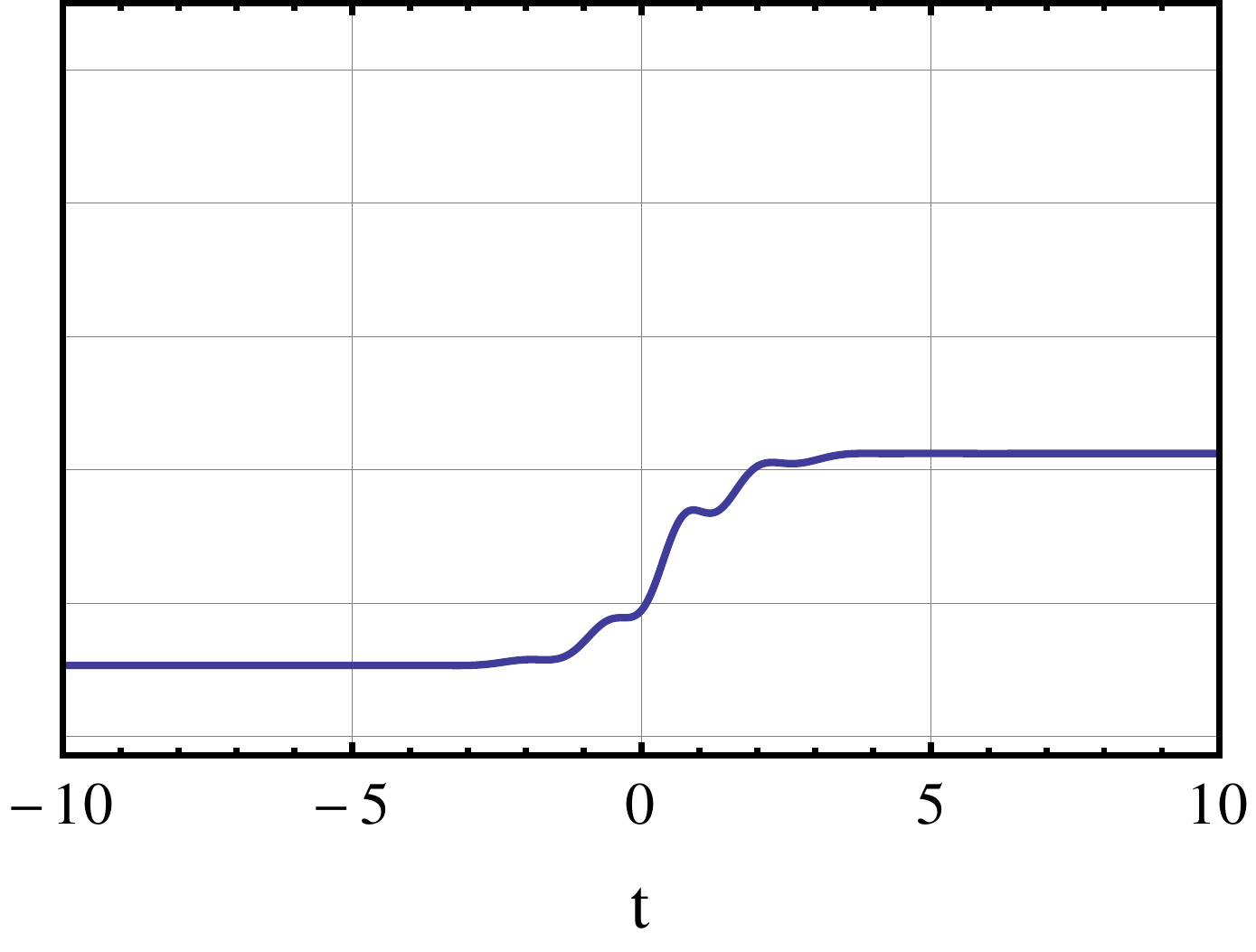}
& \includegraphics[width=4.5cm]{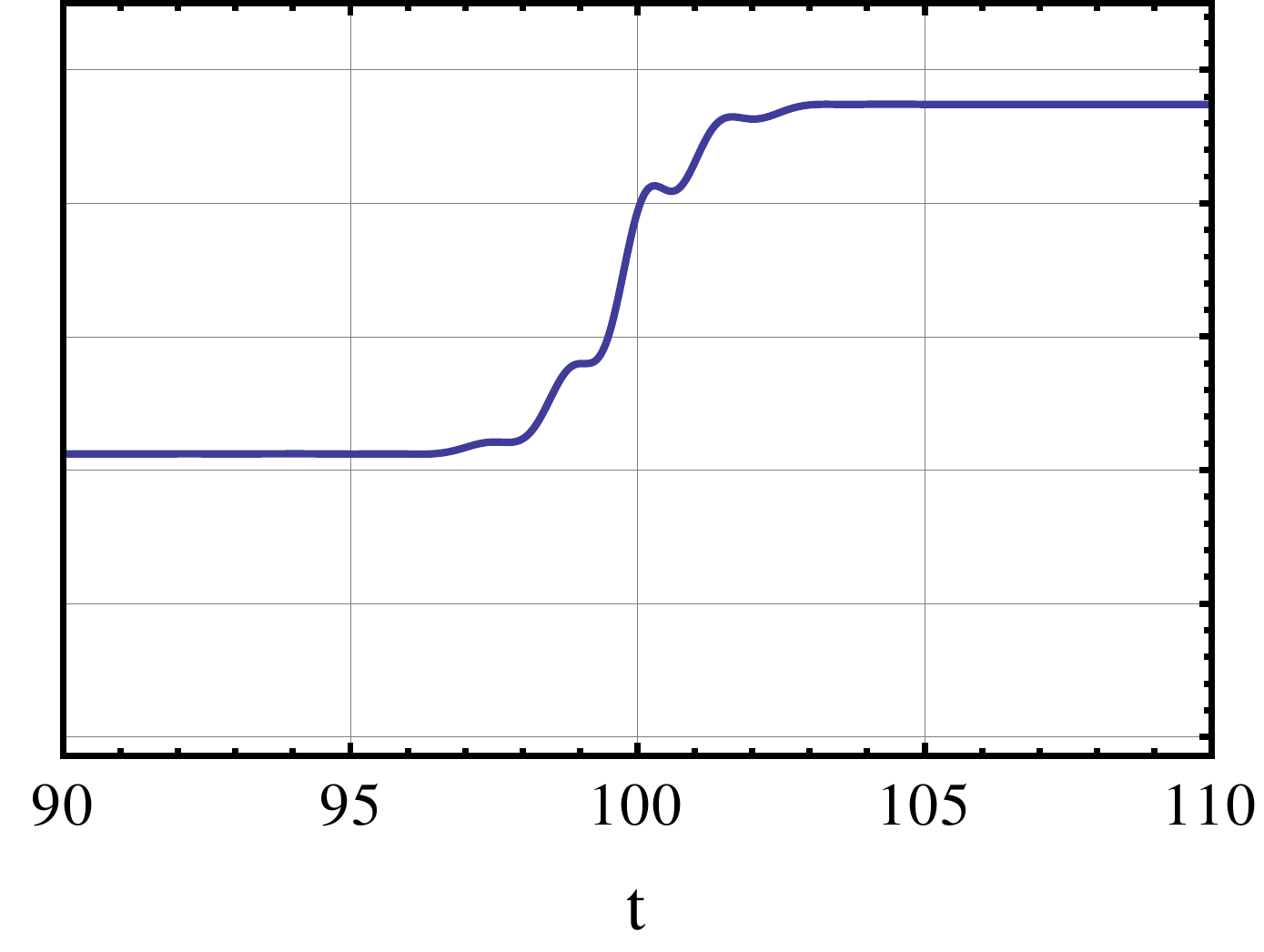} \\
\includegraphics[height=3.35cm,width=5.5cm]{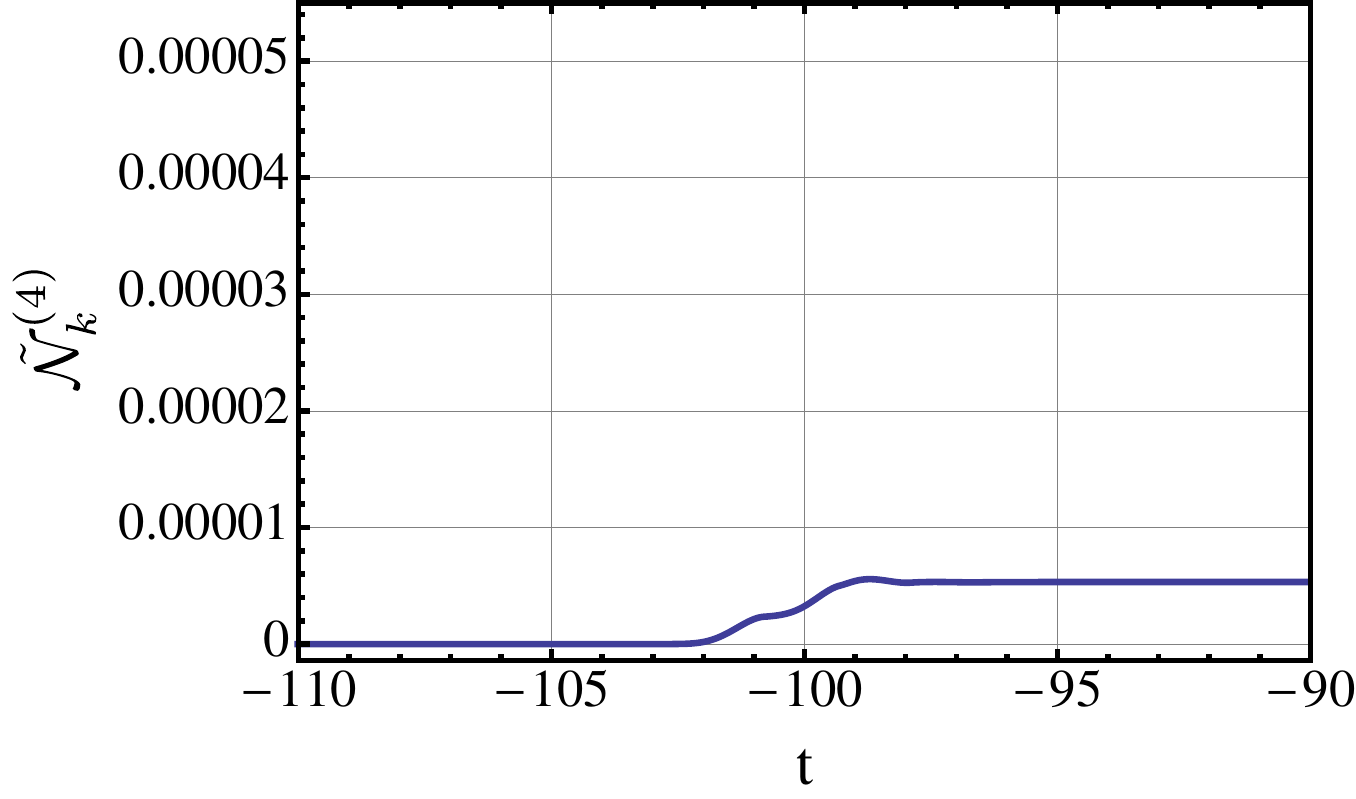}
& \includegraphics[width=4.5cm]{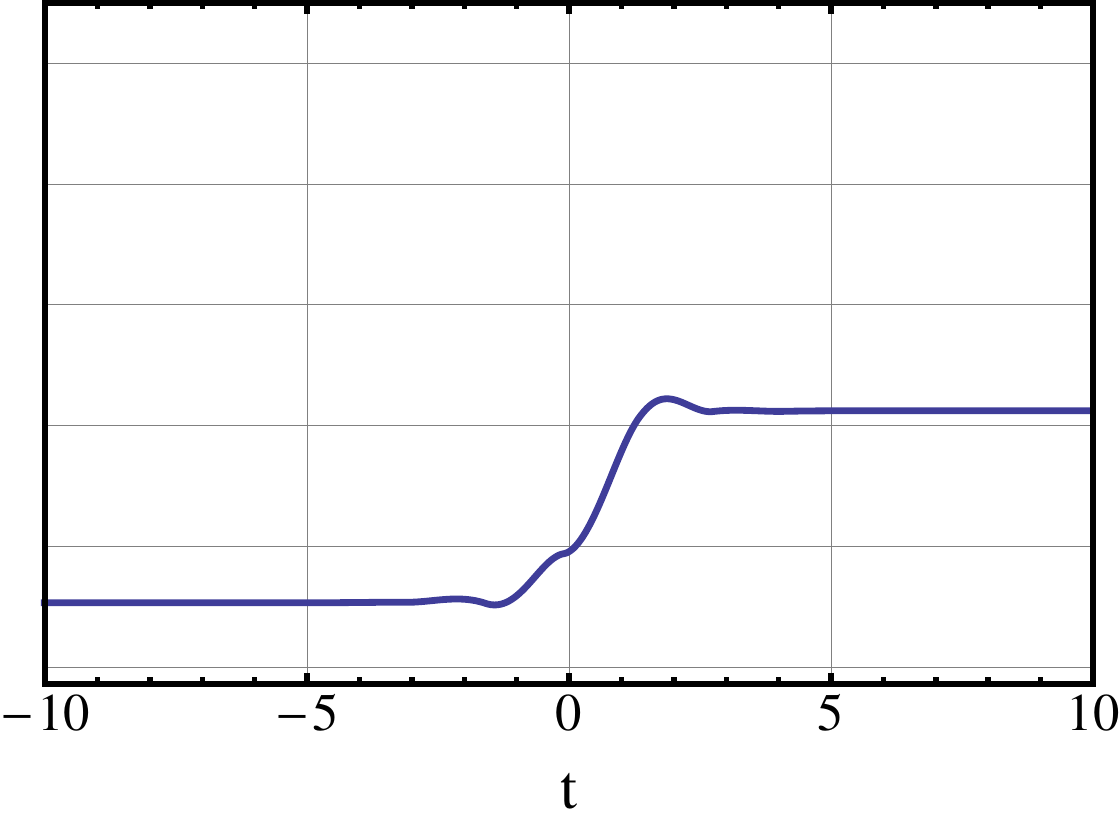}
& \includegraphics[width=4.5cm]{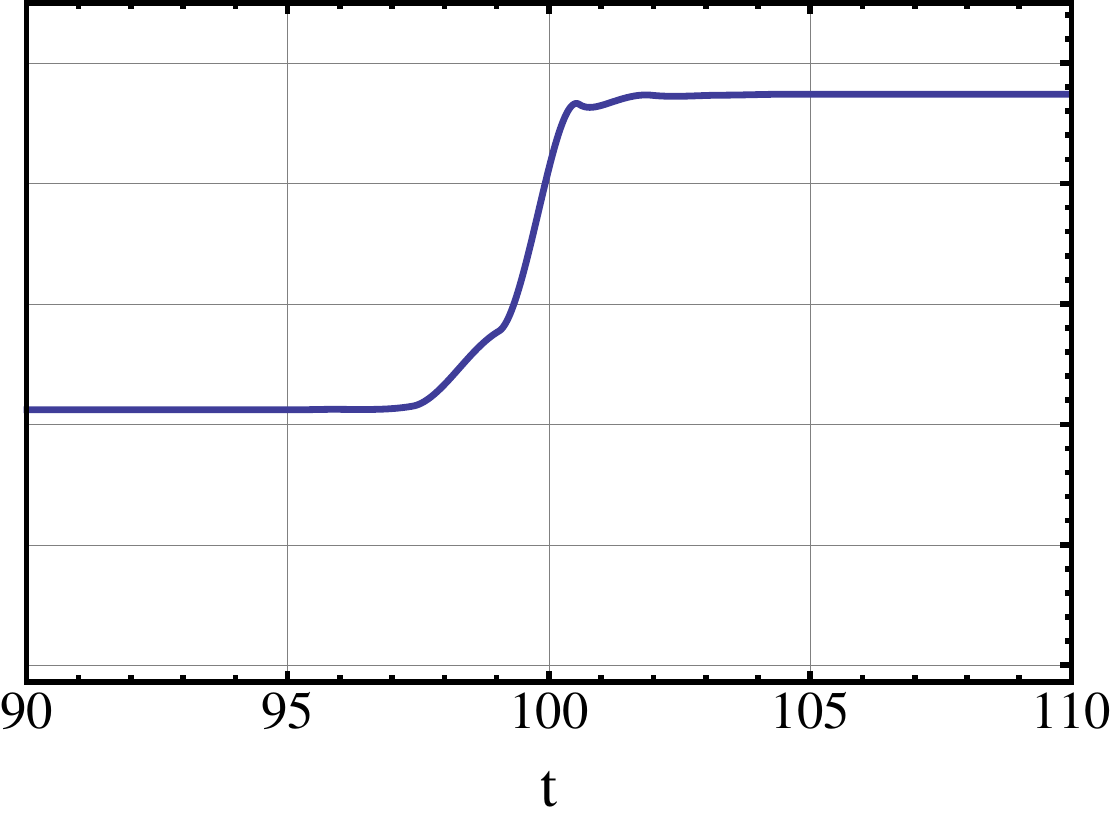} \\
\includegraphics[height=3.35cm,width=5.5cm]{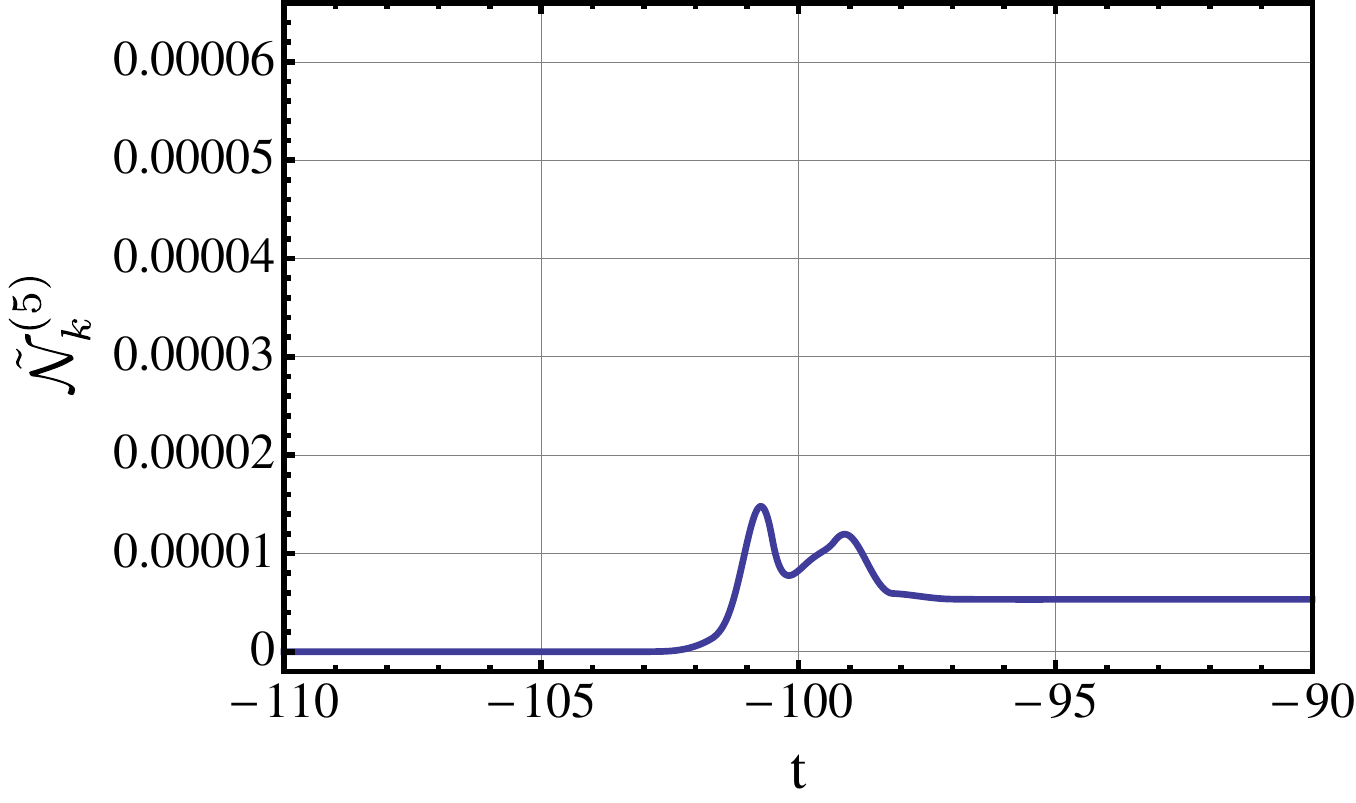}
& \includegraphics[width=4.5cm]{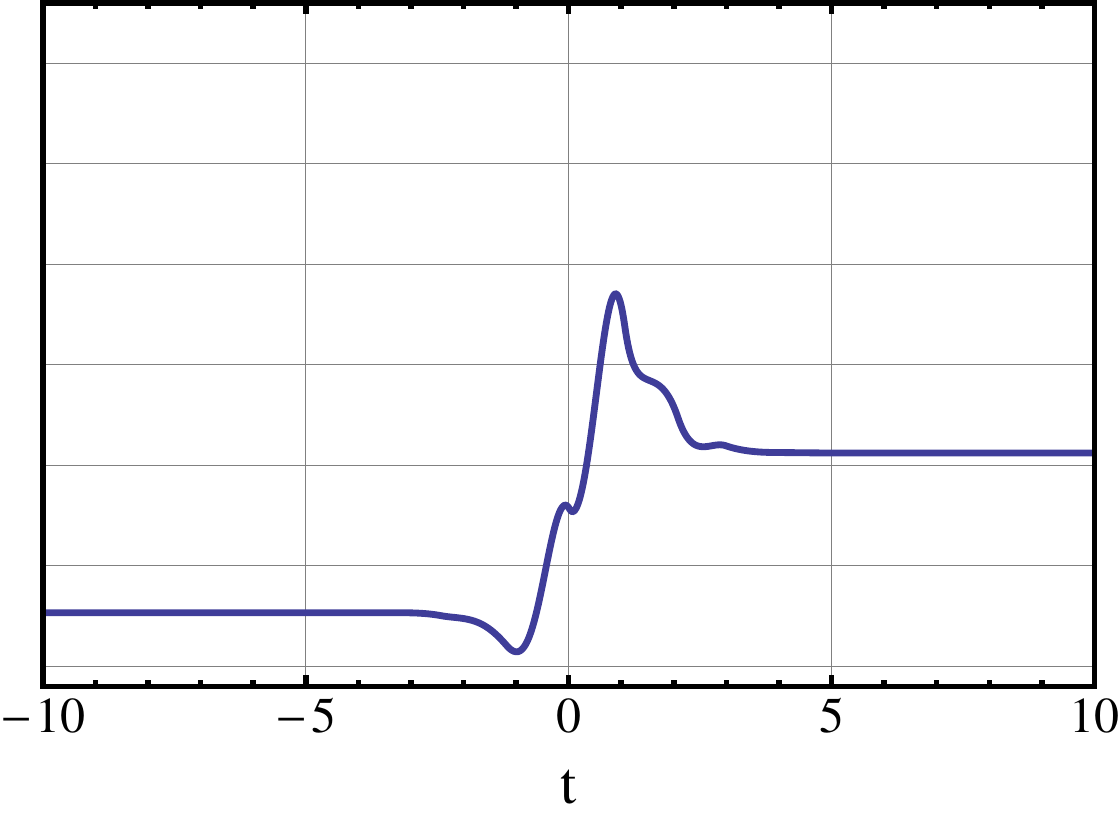}
& \includegraphics[width=4.5cm]{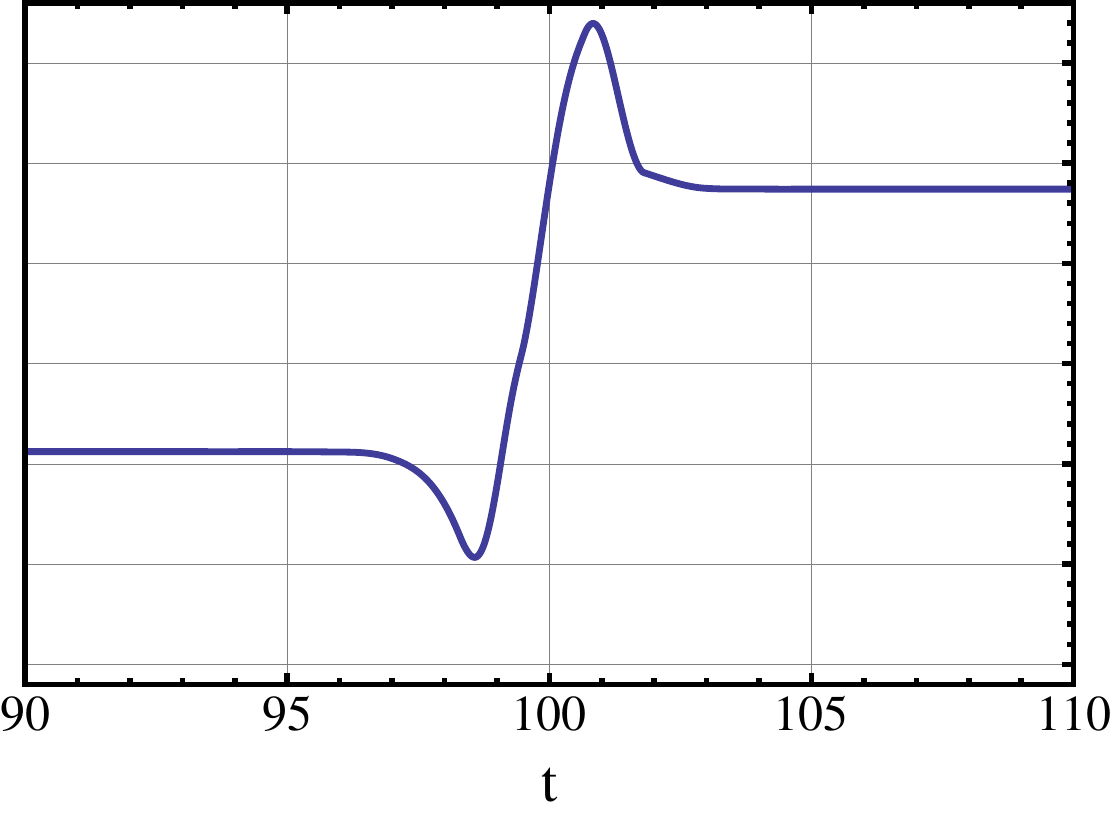}
\end{tabular}
\caption{Time evolution of the adiabatic particle number for a  sequence of two alternating-sign pulses,  of the form  (\ref{3-pulse-field}), with parameters: $E=0.25$, $a=0.1$, $b=50$, transverse momentum $p_\perp=0$, and longitudinal momentum $k_\parallel= 0.08336$, in units with $m=1$. The optimally truncated order is $j=3$, and the ratios of the plateaux are $1:3.99:8.9$, very close to the expected  $1:4:9$ for coherent constructive interference of three pulses.}
\label{3-pulse-figure}
\end{figure}
\subsection{Three Alternating-sign Electric Field Pulses}

 Consider an electric field consisting of the three successive pulses, of alternating sign:
\begin{eqnarray}
E(t)=E\, {\rm sech}^2\left[a(t+2b)\right]-E\, {\rm sech}^2\left[a t\right]+E\, {\rm sech}^2\left[a(t-2b)\right]
\label{3-pulse-field}
\end{eqnarray}
for which we can choose a time-dependent vector potential:
\begin{eqnarray}
A_\parallel(t) = - \frac{E}{a} \Big( \tanh[a (t + 2b)] - \tanh( a t ) + \tanh[ a ( t - 2b) ]
\Big)
\end{eqnarray}
The results are shown in Figure \ref{3-pulse-figure}, with longitudinal momentum associated with  coherent constructive interference. Notice the large oscillations that shrink and smooth out as the optimal order of truncation ($j=3$) is approached, and return after this order is passed. Also notice the coherence effect that the three plateaux are in the ratio $1: 4: 9$.

\begin{figure}[h!]
\begin{tabular}{cc}
\centering
\includegraphics[width=7.5cm]{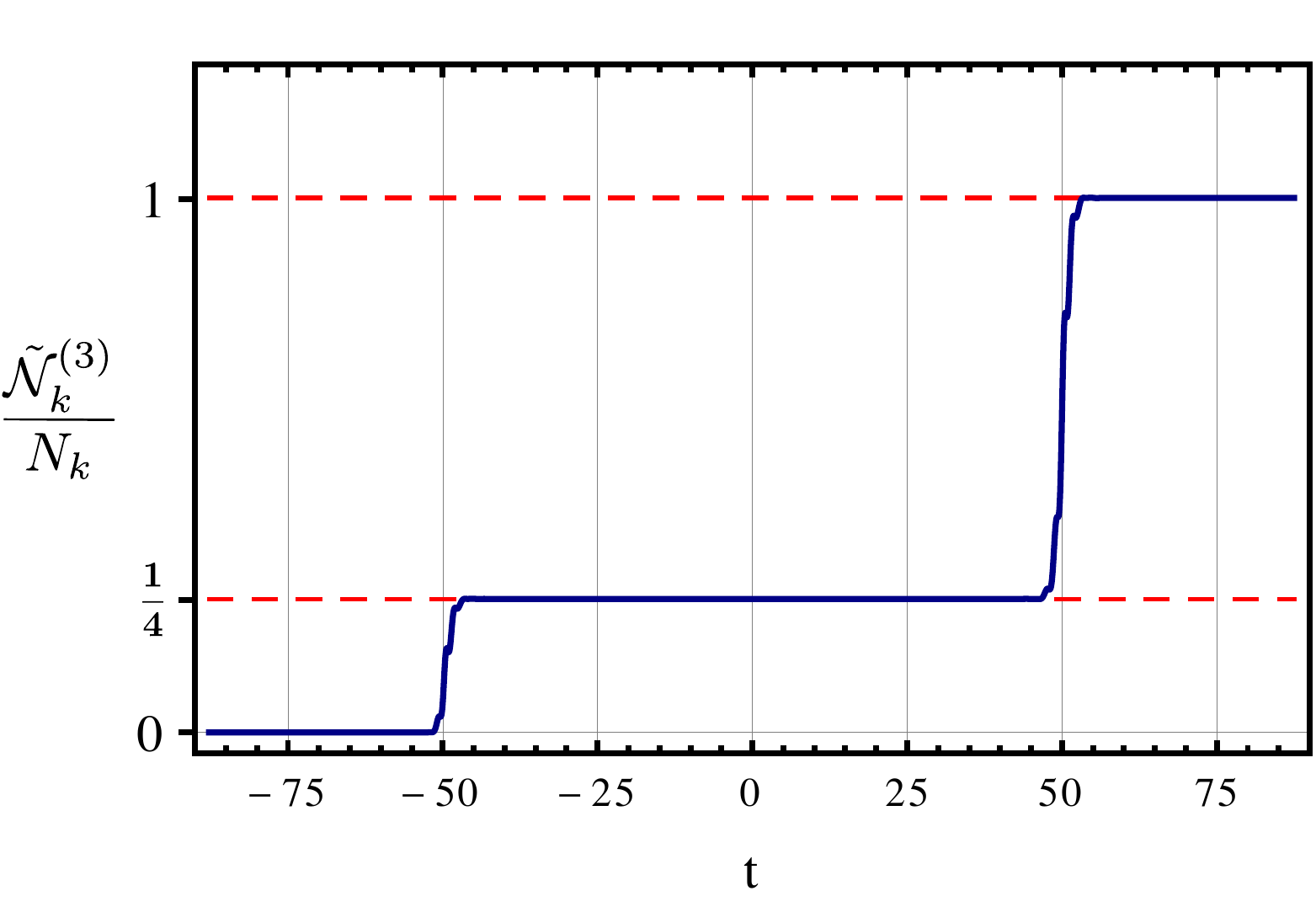} 
& 
\includegraphics[width=7.5cm]{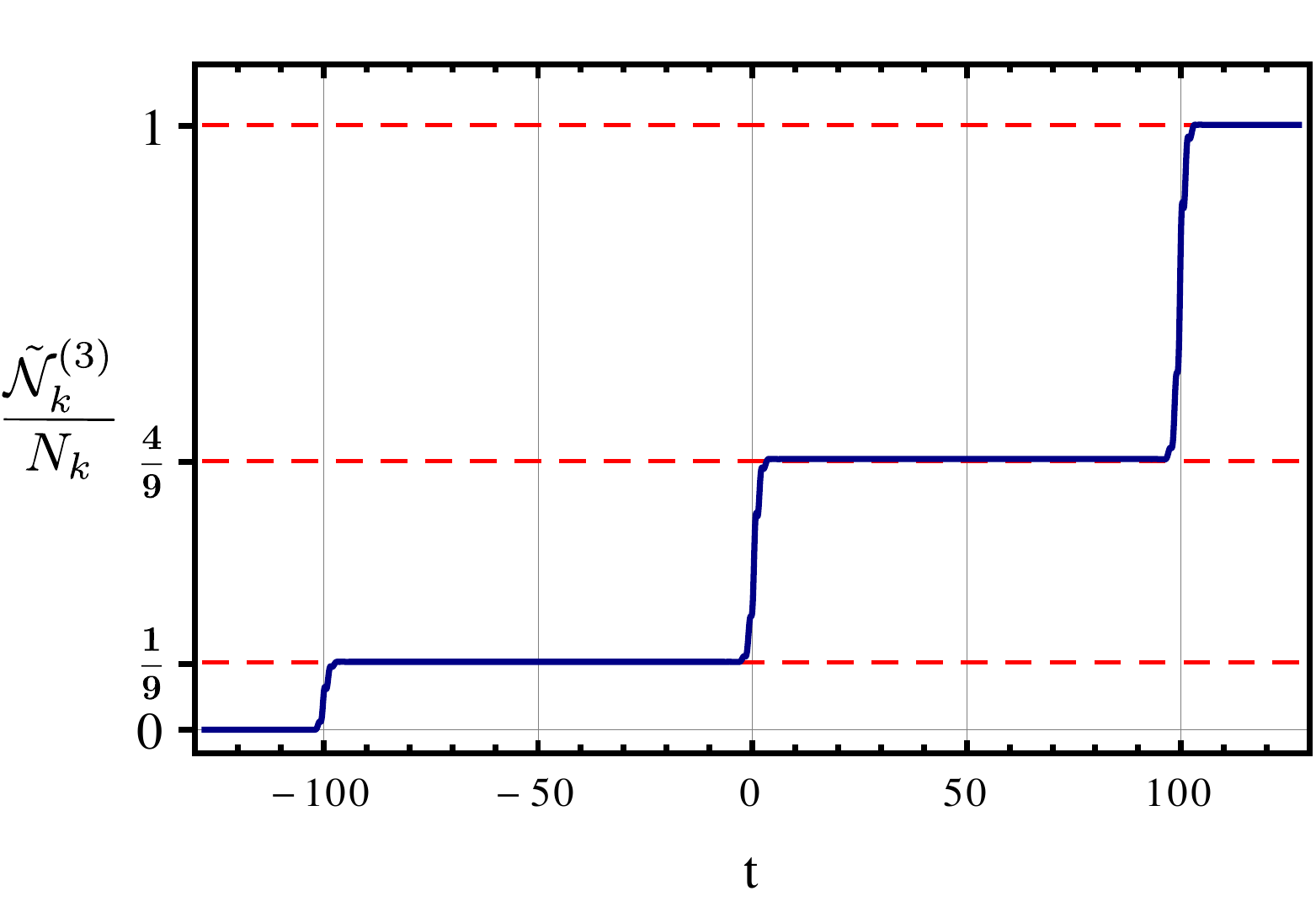}
\end{tabular}
\caption{Time evolution of the super-adiabatic particle number, normalized relative to the final value at future infinity, showing the coherent nature of the interference. The first plot is for a two-pulse field (\ref{2-pulse-field}), with parameters: $E=0.25$, $a=0.1$, $b=50$, transverse momentum $k_\perp=0$, and longitudinal momentum $k_\parallel= 2.515556$, in units with $m=1$; the second is for a three-pulse field  (\ref{3-pulse-field}), with parameters: $E=0.25$, $a=0.1$, $b=50$, $k_\perp=0$, and $k_\parallel= 0.08336$, in units with $m=1$. The optimally truncated order is $j=3$ in each case, and the ratios of the plateaux follow $1:4$ for two pulses, and $1:4:9$ for  three pulses.}
\label{coherence}
\end{figure}

Figure \ref{coherence} emphasizes this coherence effect, as the successive plateaux occur in the ratio $1:4$ for two pulses, and in the ratio $1:4:9$ for three pulses, when the momentum of the produced particles corresponds to a maximum in the momentum spectrum \cite{dd1,Akkermans:2011yn}.

\subsection{Approximate Super-Adiabatic Particle Number for Pulse Sequences}

These interference effects arise due to phase differences between different sets of complex-conjugate turning points, and have a significant effect on the final total particle number \cite{dd1,Akkermans:2011yn}, providing a simple semiclassical interpretation of the numerical results in \cite{Hebenstreit:2009km} which showed an intricate dependence of the momentum spectrum of the produced particles on the carrier phase of the sub-cycle structure of a time-dependent laser pulse. The results of \cite{dd1,Akkermans:2011yn} are that the final particle number can be expressed as a sum over contributions from each set of turning points. For scalar QED we have
\begin{eqnarray}
N_{k}\equiv \tilde{\mathcal N}_k(t=+\infty)&\approx& \left| \sum_{t_p} \exp\!\left(2i \theta_k^{(p)}\right) \exp\!\left(-F_{k,t_p}^{(0)}\right) \right|^2
\label{app-scalar}
\end{eqnarray}
where the exponent of the magnitude of the contribution of the turning point $t_p$ is
\begin{eqnarray}
F_{k,t_p}^{(0)}&\equiv &  i \int^{t_p^*}_{t_p}\omega_{k}(t) \, dt
\label{kt1}
\end{eqnarray}
while the accumulated phase (measured relative to the first real Stokes point $s_1$, the point where the Stokes line connecting $t_1$ and $t_1^*$ crosses the real axis) for turning point $t_{p}$ is
\begin{eqnarray}
\theta^{(p)}_{k}\equiv  \int_{s_1}^{s_p} \omega_{k}(t)\,dt \approx \int_{{\rm Re}[t_1]}^{{\rm Re}[t_{p}]} \omega_{k}(t)\,dt
\label{kt2}
\end{eqnarray}
\begin{figure}[h!]
\centering
\includegraphics[scale=0.4]{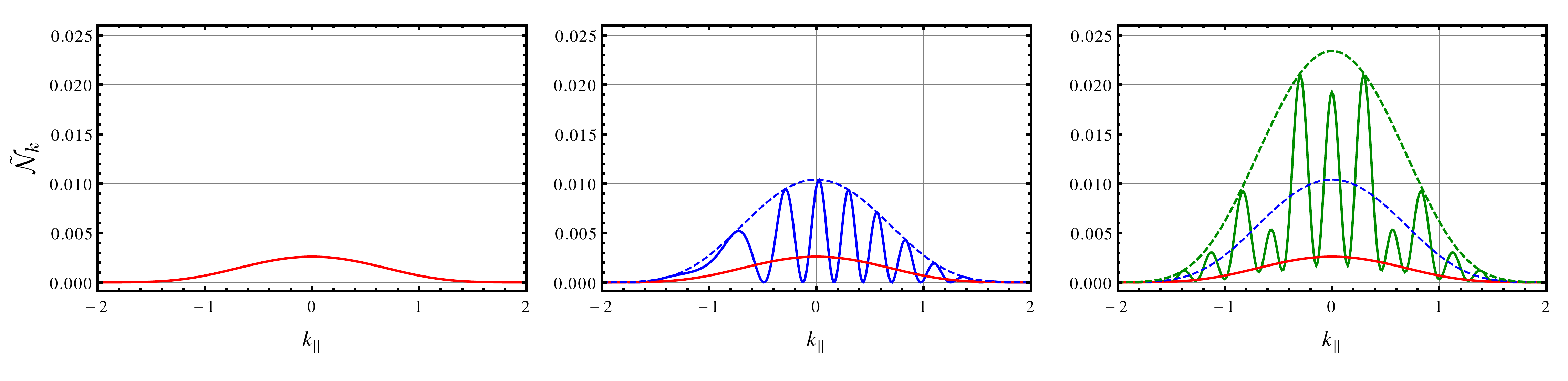}
\caption{The longitudinal momentum spectrum $N_k$ as a function of the produced particle's longitudinal momentum, for the cases of  one, two and three electric field pulses, of the forms (\ref{one-pulse}), (\ref{2-pulse-field}), and (\ref{3-pulse-field}), respectively. The distribution is of the $n$-slit interference form, with the envelope being $n^2$ times the envelope for one pulse. When integrated over $k_\parallel$, the total particle number is just $n$ times that for a single pulse, but the modes are redistributed into the $n$-slit form, with constructive enhancement in some modes, and destructive interference in other modes.}
\label{cross-sections}
\end{figure}
\begin{figure}[h!]
\includegraphics[scale=0.55]{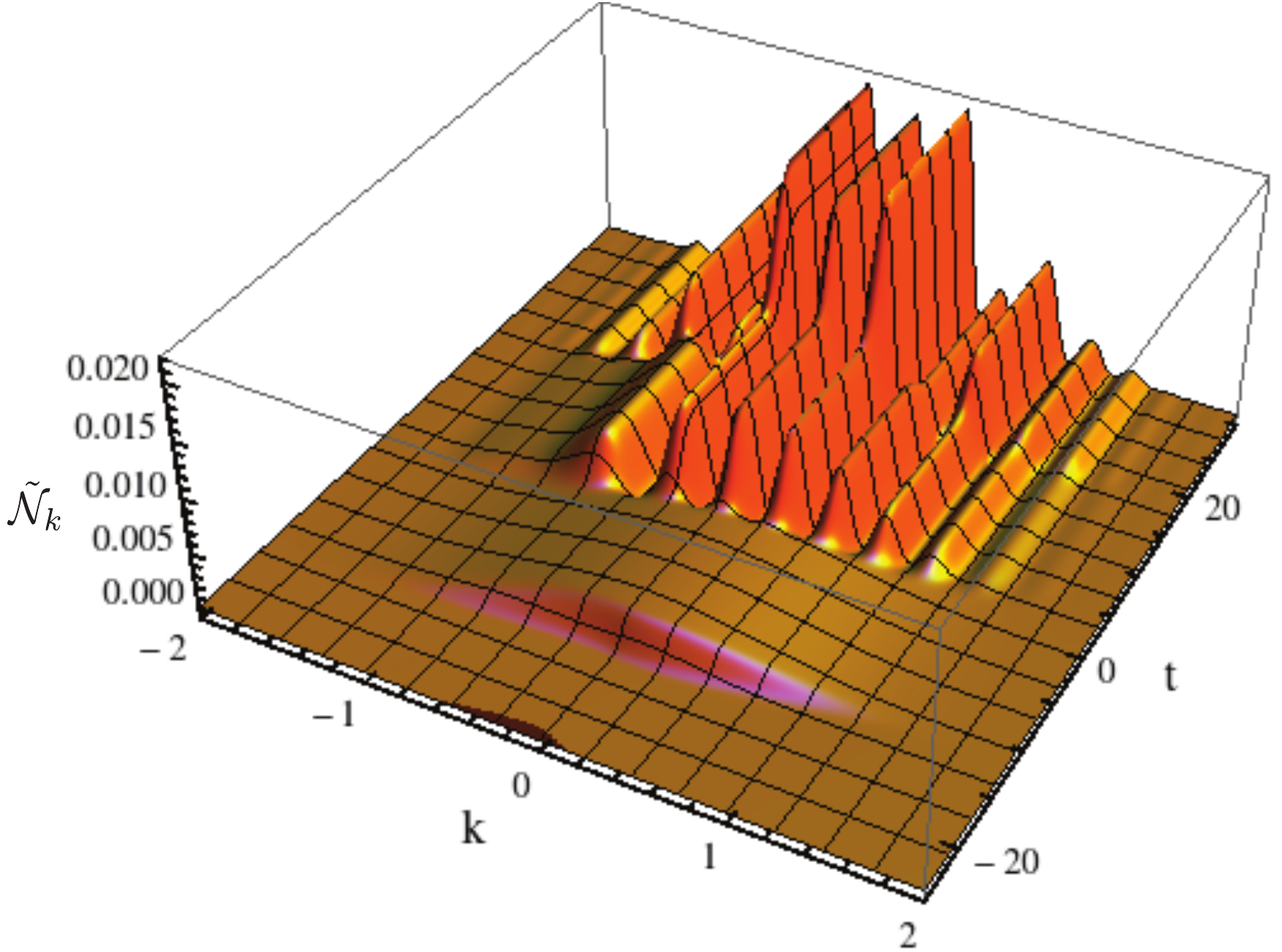}
\caption{
The approximate super-adiabatic particle number (\ref{answer2}) plotted as a function of time and the produced particle's longitudinal momentum $k_\parallel=k$, for Schwinger pair production due to three alternating-sign pulses of the form (\ref{3-pulse-field}) with parameters: $E=0.5$, $a=0.25$, $b=7.5$, and transverse momentum $k_\perp=0$, in units with $m=1$. 
Note that in the time region $-15\ll t\ll 0$, between the first and second pulses,
 the momentum distribution correponds to the asymptotic momentum distribution  for a single pulse; while  in the time region $0\ll t\ll 15$, between the second and third pulses,  the momentum distribution correponds to the asymptotic momentum distribution for a sequence of two alternating sign pulses; and finally for $t\gg 15$, after the third pulse, the momentum distribution correponds to the asymptotic momentum distribution for a sequence of three alternating sign pulses.
 This can also be seen in the accompanying Figure \ref{3dplotstimeslices}, which shows cross-sections of the momentum distributions at times in-between the pulses.
Notice that the time evolution of the super-adiabatic particle number critically depends on the longitudinal momentum: for certain $k$ there is enhancement due to constructive interference, while for other $k$ the interference is destructive. }
\label{3dplotsfull}
\end{figure}
If the pulse-sequence consists of alternating sign pulses of the same shape, then all the $\Big|e^{-F_{k,t_p}^{(0)}}\Big|^2$ factors are approximately equal, and we can have constructive or destructive interference depending on the relative phases. These phases depend both on the pulse parameters and on the produced particle's longitudinal momentum $k$. In particular, for equally-spaced alternating-sign pulses, we have approximately equal phase differences between successive turning point pairs, leading to coherent interference:
\begin{eqnarray}
N_{k}^{\text{n-pulse}}\approx 
\begin{cases}
N_{k}^{\text{1-pulse}}\, \sin^2\left[ n\, \theta_{k}\right]/\cos^2\left[ \theta_{k}\right]
 \quad, \quad n\,\,{\rm even}\cr
N_{k}^{\text{1-pulse}}\, \cos^2\left[ n\, \theta_{k}\right]/\cos^2\left[\theta_{k}\right]
 \quad, \quad n\,\,{\rm odd}
 \end{cases}
 \label{multipair}
\end{eqnarray}
Note that as a function of the longitudinal momentum $k$, these particle spectra represent the $n$-slit interference pattern, here probed in the time-domain from the quantum vacuum \cite{Akkermans:2011yn}. In Figure \ref{cross-sections} we show these momentum spectra (for the final asymptotic particle number $N_k=\tilde{\mathcal N}_k(t=+\infty)$), for the cases of one, two and three pulses, of the forms (\ref{one-pulse}), (\ref{2-pulse-field}), and (\ref{3-pulse-field}), respectively.
\begin{figure}[h!]
\centering
\includegraphics[scale=1]{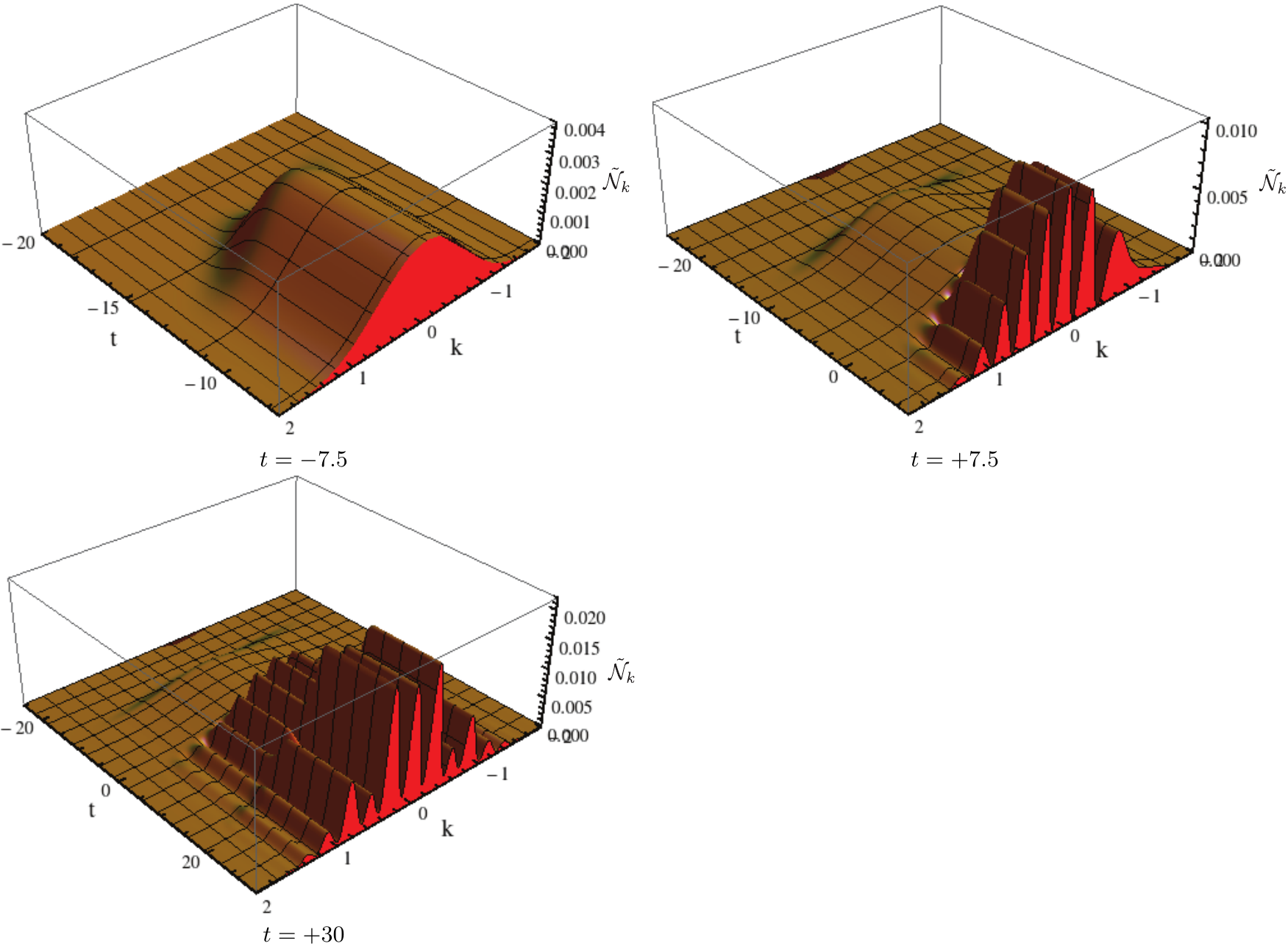}
\caption{
As in (\ref{3dplotsfull}), but plotted over specific time ranges, ending with cross-sections at $t=-7.5$, $t=+7.5$, and $t=+30$, respectively, to illustrate the time evolution of the longitudinal momentum spectrum (red-filled) of the produced particles during the time periods between the particle creation events due to  the pulses at $t=0,\pm15$. The highlighted longitudinal momentum spectrum of each subplot follows the $n$-slit interference pattern (\ref{multipair}), which can be compared to the asymptotic momentum spectra shown in Figure \ref{cross-sections}, for one, two and three pulses, respectively.
}
\label{3dplotstimeslices}
\end{figure}

In this paper we consider not just these final values for the asymptotic particle number, but also the {\it time evolution} of the adiabatic particle number. Since Berry's universal time-evolution corresponds to the behavior in the vicinity of a single turning point pair, we need to generalize the result (\ref{answer}) to the case with several (complex conjugate pairs of) turning points, as occurs for electric fields with nontrivial temporal substructure \cite{dd1}. Some steps in this direction were taken for non-relativistic two-level systems \cite{lim-berry}.

The natural generalization of (\ref{answer}) is the following:
\begin{eqnarray}
\tilde{\mathcal N}_k(t)\approx \frac{1}{4} \left | \sum_{t_p} \exp\!\left(2i \theta_k^{(p)}\right) \exp\!\left(-F_{k,t_p}^{(0)}\right) {\rm Erfc}\!\left(-\sigma^{(p)}_k(t)\right)\right|^2
\label{answer2}
\end{eqnarray}
We have verified numerically that this form fits well with the time evolution of the super-adiabatic particle number for sequences of time-dependent pulses for different field parameters and different momenta. In Figures \ref{3dplotsfull} and \ref{3dplotstimeslices}  we show plots of the super-adiabatic particle number (\ref{answer2}) as a function of both time and the longitudinal momentum of the produced particles. These plots show the smooth evolution of the quantum interference effects due to three separate particle creation events. After the first pulse, and before the second pulse, the momentum distribution has the form that would be obtained asymptotically at future infinity from just a single pulse. But after the second pulse, and before the third pulse, the momentum distribution has the form that would be obtained asymptotically at future infinity from a sequence of two alternating-sign pulses. Finally, after the third pulse, we observe a momentum distribution of the form that would be obtained asymptotically at future infinity from the full sequence of three alternating-sign pulses.
In Figure \ref{3dplotstimeslices} we emphasize the interference effects of the time dependent super-adiabatic particle number with cross-sections of the momentum distributions at times in-between the pulses. These time-slices coincide with the final future infinity particle number momentum distributions shown in Figure \ref{cross-sections}  for one, two and three pulses, respectively.

\section{Super-Adiabatic Particle Production in de Sitter Space}

In this section we give numerical examples to illustrate how the super-adiabatic particle number evolves in time for (eternal) de Sitter space in even and odd dimensions. It is known \cite{emil,Anderson:2013ila,Anderson:2013zia} that in even space-time dimensions there is particle production in de Sitter space, but that there is no particle production in  odd space-time de Sitter space \cite{Bousso:2001mw,Lagogiannis:2011st}. It has been argued in \cite{Kim:2010xm} that this difference between particle production in even and odd dimensional de Sitter space can also be understood in terms of quantum interference between two sets of complex turning points, and the associated Stokes phenomenon.\footnote{There is also a distinct difference in the large order behavior of perturbation theory in even and odd dimensional de Sitter space-time \cite{Das:2006wg}.} Here we analyze this question for the full time evolution of the super-adiabatic particle number. We find that in even dimensions there is coherent constructive interference, while for odd dimensions destructive interference leads to the vanishing of the net particle number. The resulting behavior is similar to that of the two-pulse electric field example, with either constructive (even dimensions) or destructive (odd dimensions) interference.
\begin{figure}[h!]
\begin{tabular}{cc}
\centering
\includegraphics[width=7.5cm]{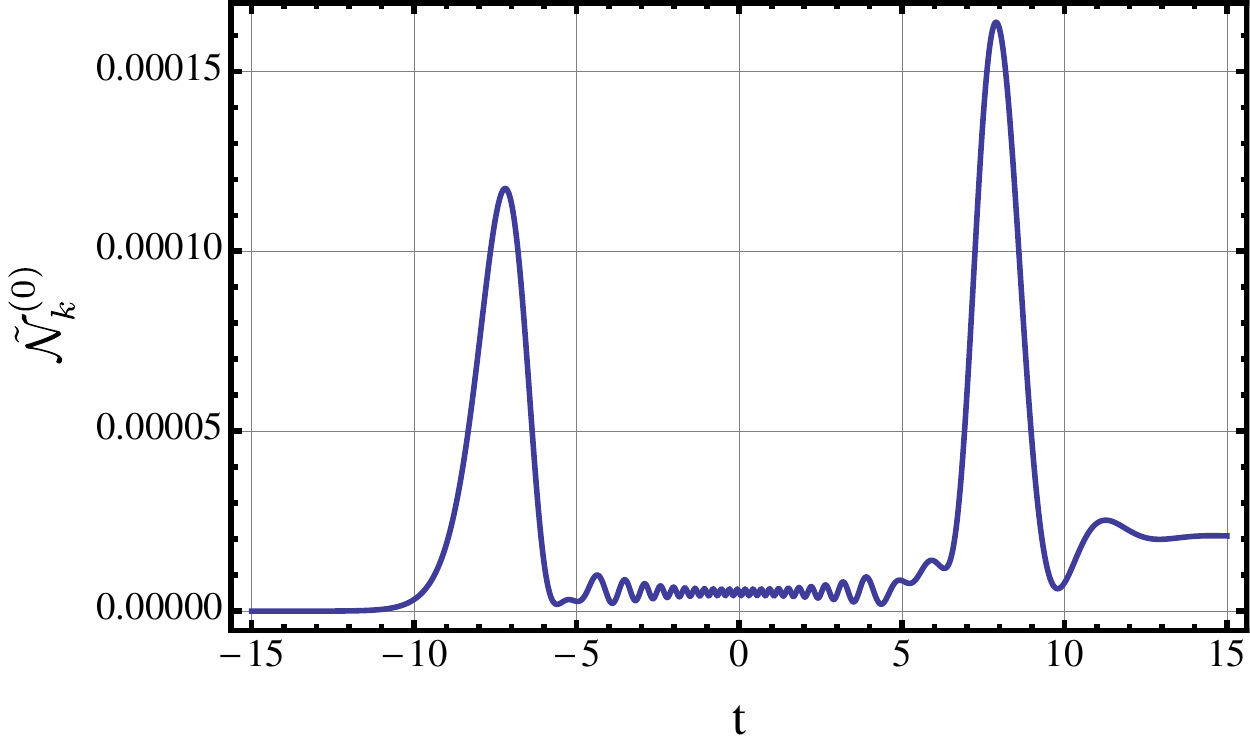} 
& 
\includegraphics[width=7.5cm]{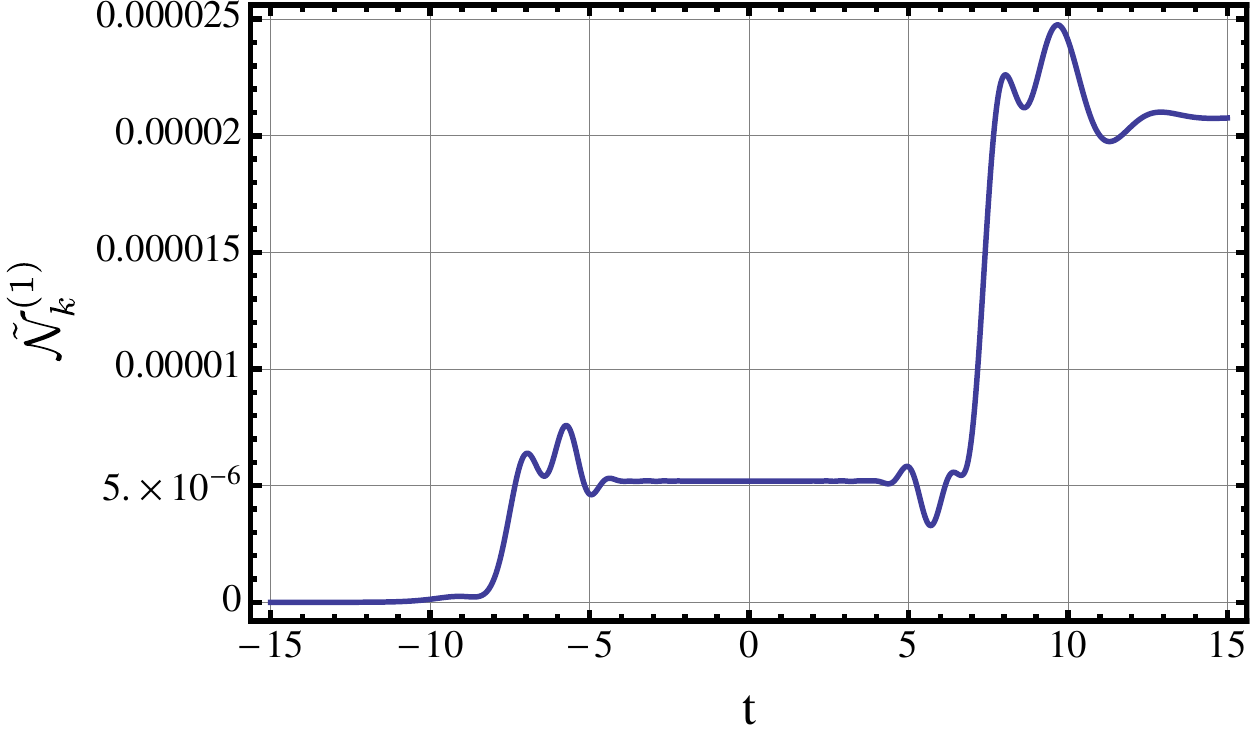} \\
\includegraphics[width=7.5cm]{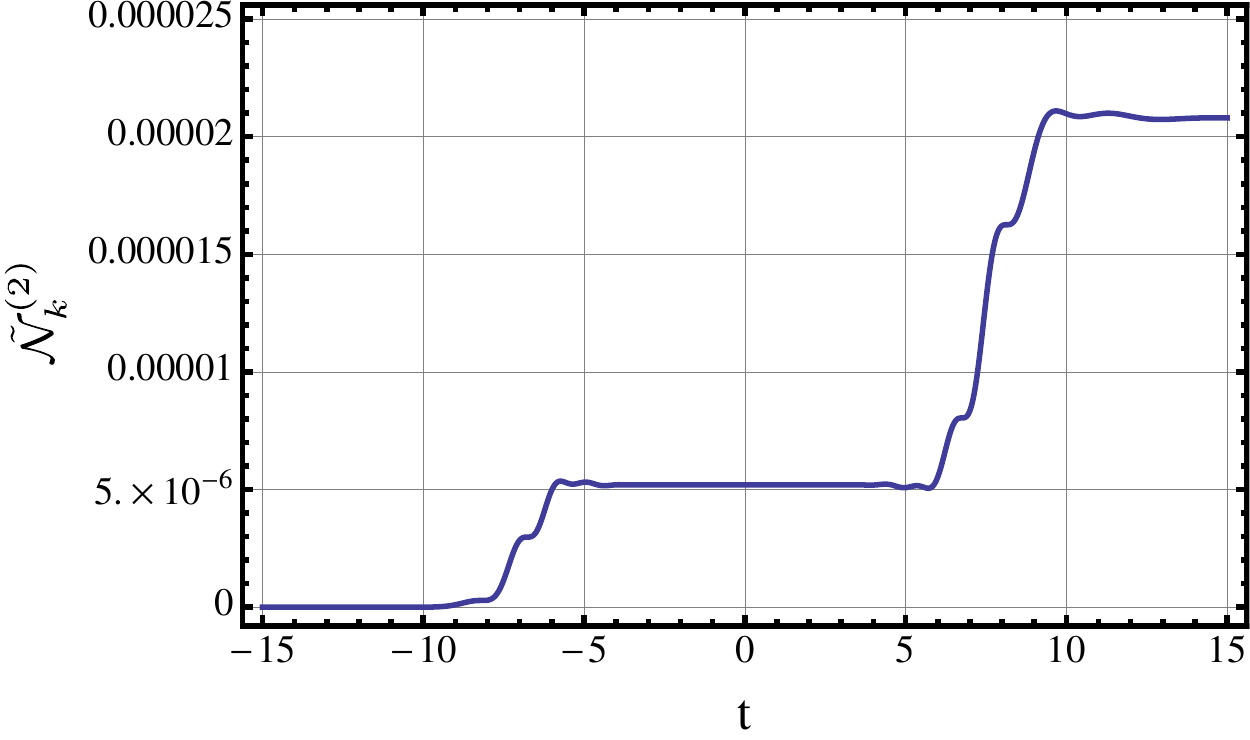} 
& 
\includegraphics[width=7.5cm]{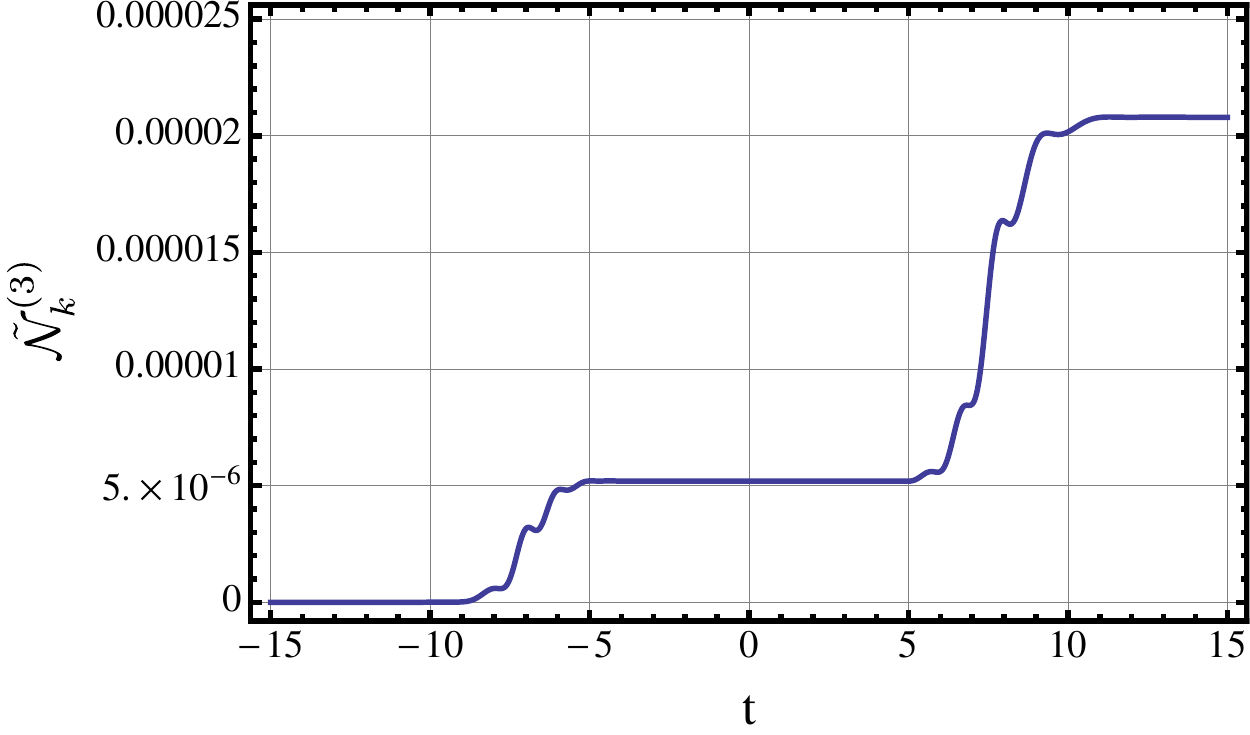} \\
\includegraphics[width=7.5cm]{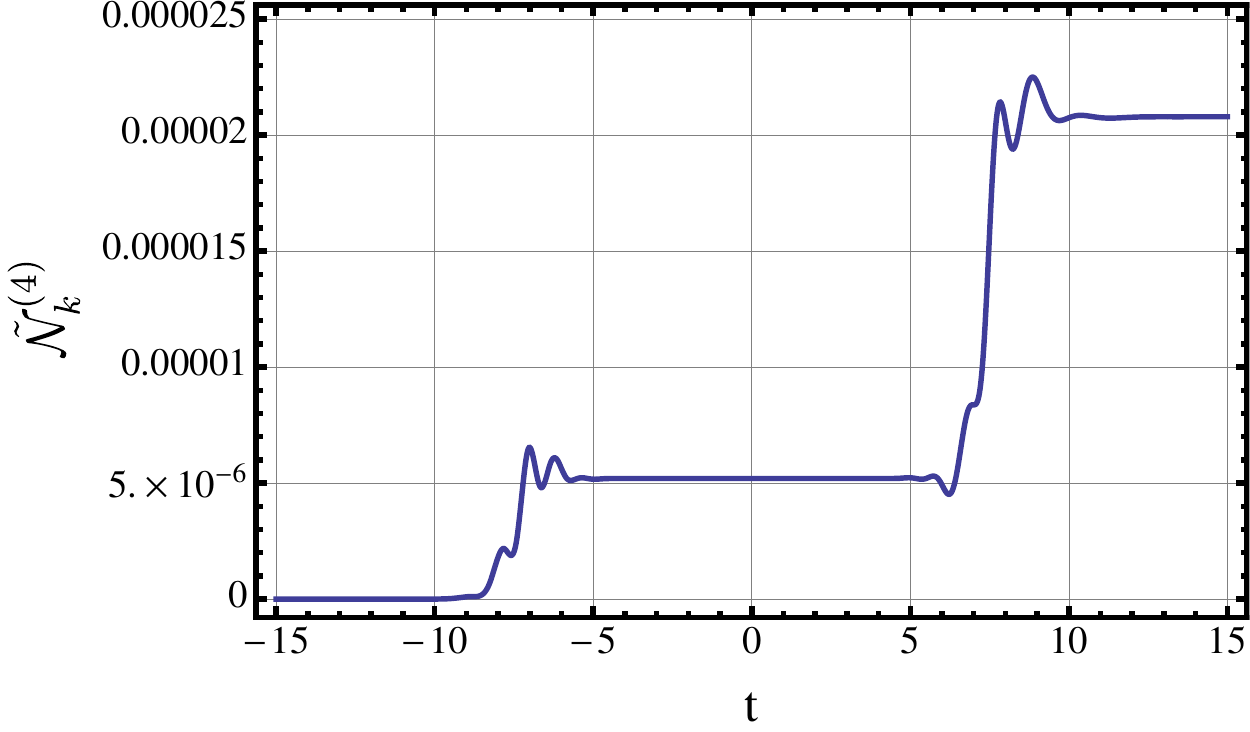} 
& 
\includegraphics[width=7.5cm]{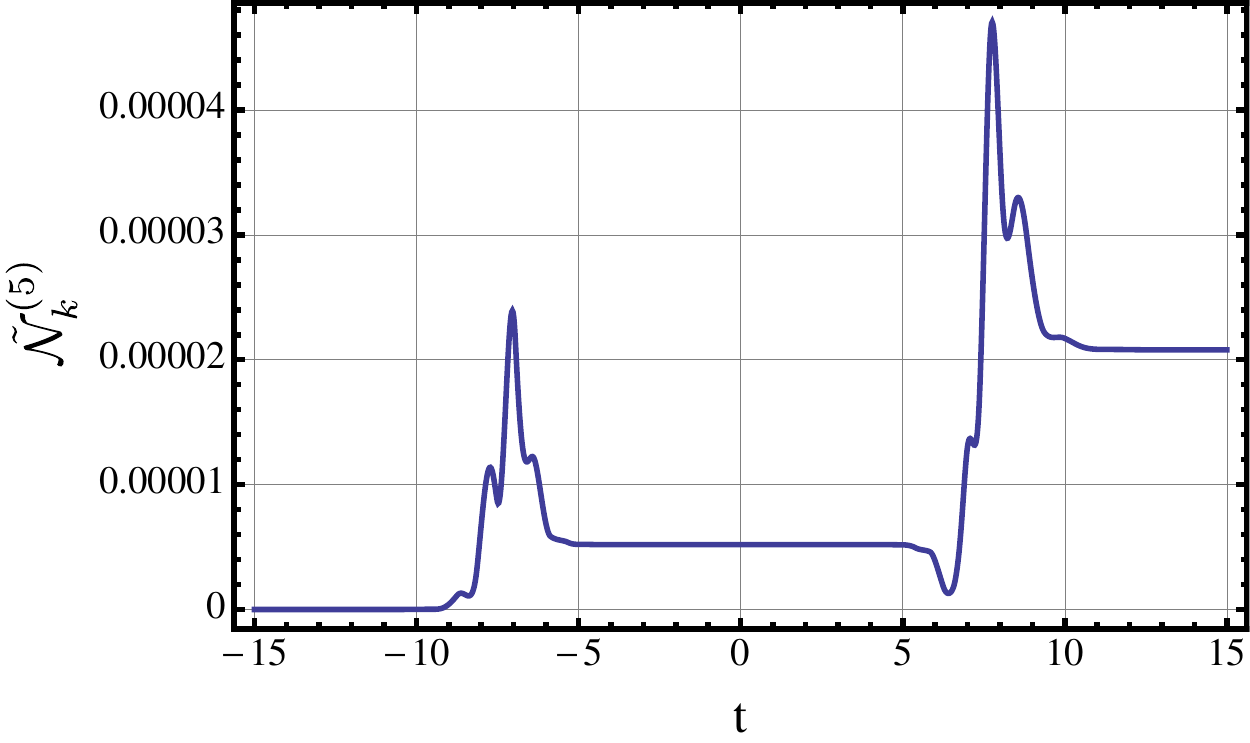}
\end{tabular}
\caption{
Time evolution of the adiabatic particle number for the first 6 orders of the adiabatic expansion, for  particle production in four-dimensional de Sitter space  (\ref{parker-osc1}, \ref{parker-osc2}), with parameters: $k = 25$ and $H = 0.5$, in units with $m=1$. We use conformal coupling so that $\xi=\frac{1}{6}$ in $d=4$. Note the similarity with the case of a two-alternating-sign-pulse electric field in the Schwinger effect, with constructive interference, as in Figure \ref{2-pulse-constructive}. The final asymptotic value of the particle number, at future infinity, is the same for all orders of truncation. Note that the final asymptotic value of the particle number is 4 $(=2^2)$ times that of the intermediate plateau, which is the $n^2$ enhancement factor for coherent constructive quantum interference, and is consistent with the analysis of \cite{Anderson:2013ila,Anderson:2013zia}. At intermediate times there are large oscillations in the particle number, which become much smaller as the optimal order ($j=3$) is reached, and then grow again rapidly beyond this optimal order of truncation. Such behavior is characteristic of {\it asymptotic} expansions, where the order of truncation depends  on the size of the expansion parameter, and going beyond this optimal order typically yields increasingly worse results.  
}
\label{fig:ds4}
\end{figure}

\subsection{Super-Adiabatic Particle Number in 4d de Sitter space: coherent constructive interference}

We compute the adiabatic particle number numerically using (\ref{abWdot2}, \ref{abWdot2factor}), for various orders $j$ of the adiabatic expansion, starting with the time-dependent frequency for four dimensional de Sitter space, from (\ref{parker-osc1}, \ref{parker-osc2}). We consider conformal coupling, so that $\xi=\frac{1}{6}$ in 4 dimensions.

The results are shown in Figure \ref{fig:ds4}. Note the strong similarity to the two-alternating-sign-pulse Schwinger effect, with particle momentum such that the interference is constructive, as shown in Figure \ref{2-pulse-constructive}. In the leading order of the adiabatic expansion, there are two large oscillatory peaks, several orders larger than the final asymptotic particle number. As the order of the adiabatic expansion increases these oscillations become smaller, and at the optimal order ($j=3$) the super-adiabatic particle number evolves much more smoothly. Moreover, we clearly see the two-step structure, with the final plateau being 4 times the height of the intermediate plateau. This is indicative of coherent constructive interference. Note also that this is completely consistent with the analysis of 
\cite{Anderson:2013ila,Anderson:2013zia}. See, for example, Figure 6 of  \cite{Anderson:2013ila}, where two creation events can be clearly seen, again associated with the two towers of complex-conjugate turning points. Moreover, the first plateau is given in \cite{Anderson:2013ila} by
\begin{eqnarray}
\Delta N_1=\frac{1}{e^{2\pi\,\gamma}-1}\approx e^{-2\pi \gamma}
\label{dn1}
\end{eqnarray}
while  the second plateau is given by
\begin{eqnarray}
\Delta N_2=\frac{1}{\sinh^2(\pi\,\gamma)}\approx 4\, e^{-2\pi \gamma}
\label{dn2}
\end{eqnarray}
which is 4 times as large. Furthermore, in Figure 8 of \cite{Anderson:2013ila} one also sees an indication that higher orders of the adiabatic expansion lead to a smoother time evolution of the adiabatic particle number. As we see from our Figure \ref{fig:ds4}, if the order of the adiabatic expansion increases beyond the optimal order, the large oscillations return. At the optimal order, the time evolution of the super-adiabatic particle number is given by Berry's universal error function form, generalized to two sets of turning points, as in (\ref{answer2}).

\begin{figure}[h!]
\begin{tabular}{cc}
\centering
\includegraphics[width=7.5cm]{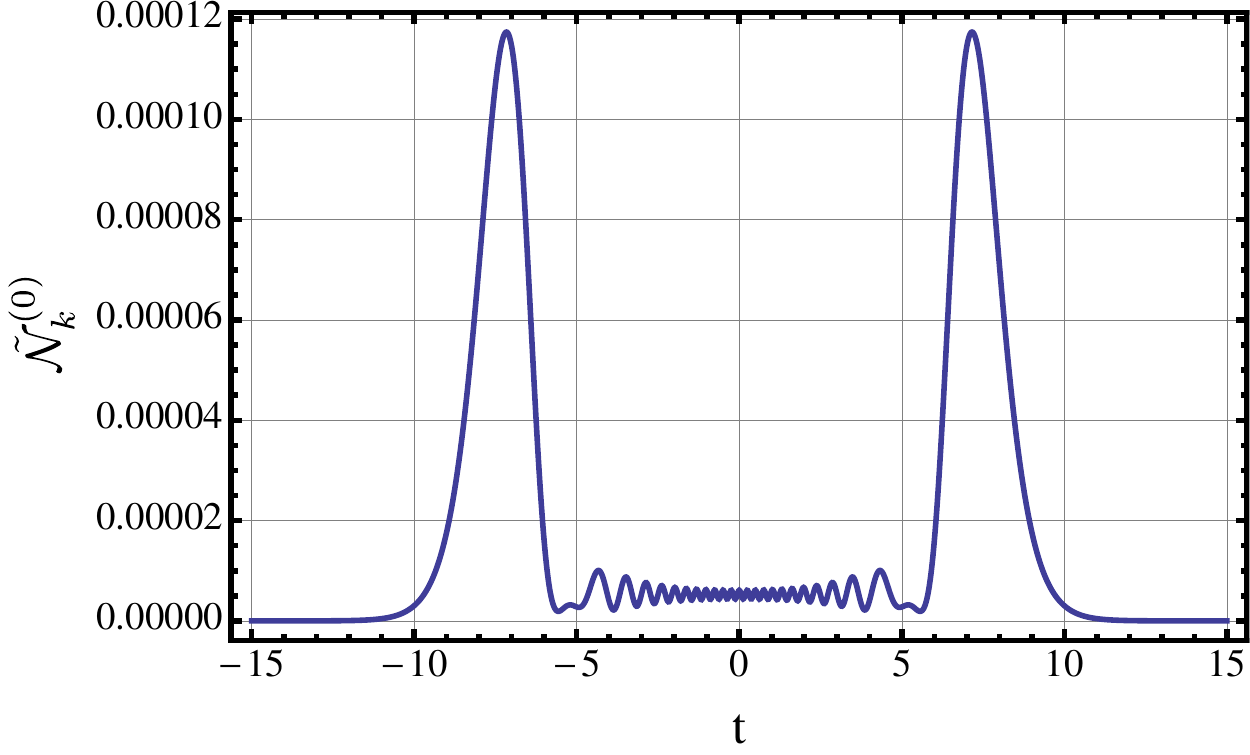} 
& 
\includegraphics[width=7.5cm]{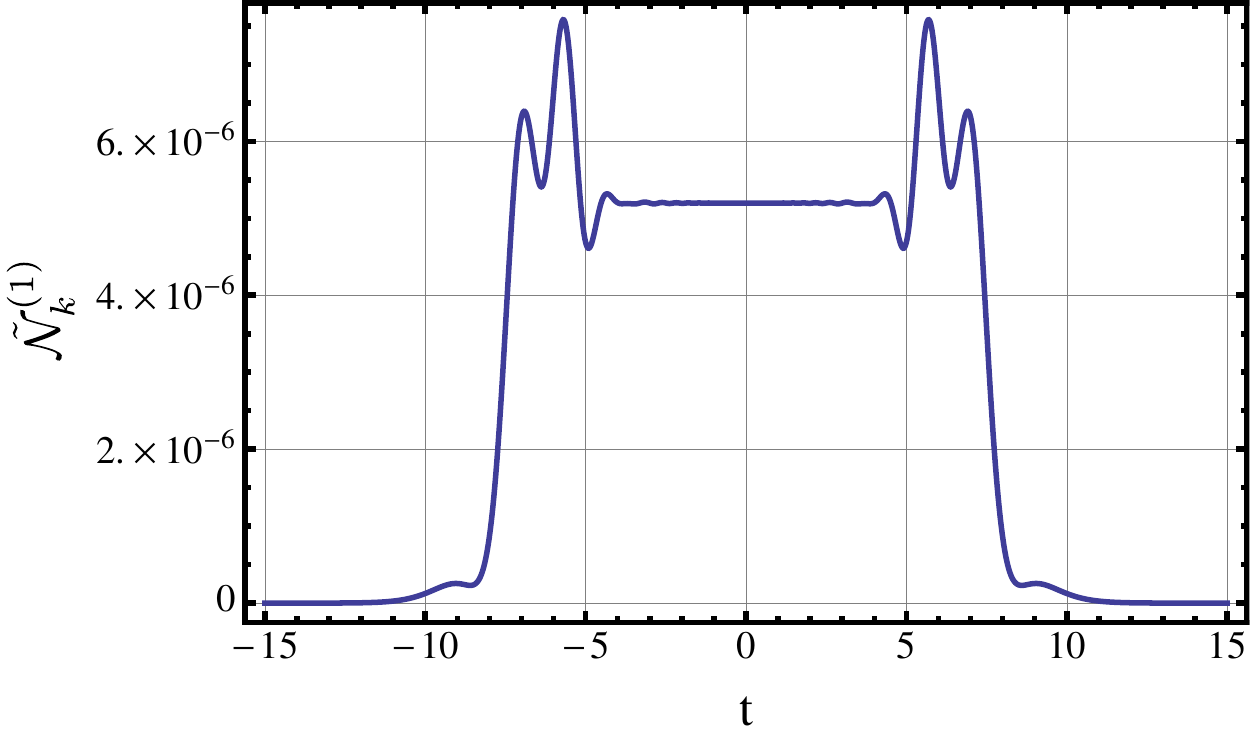} \\
\includegraphics[width=7.5cm]{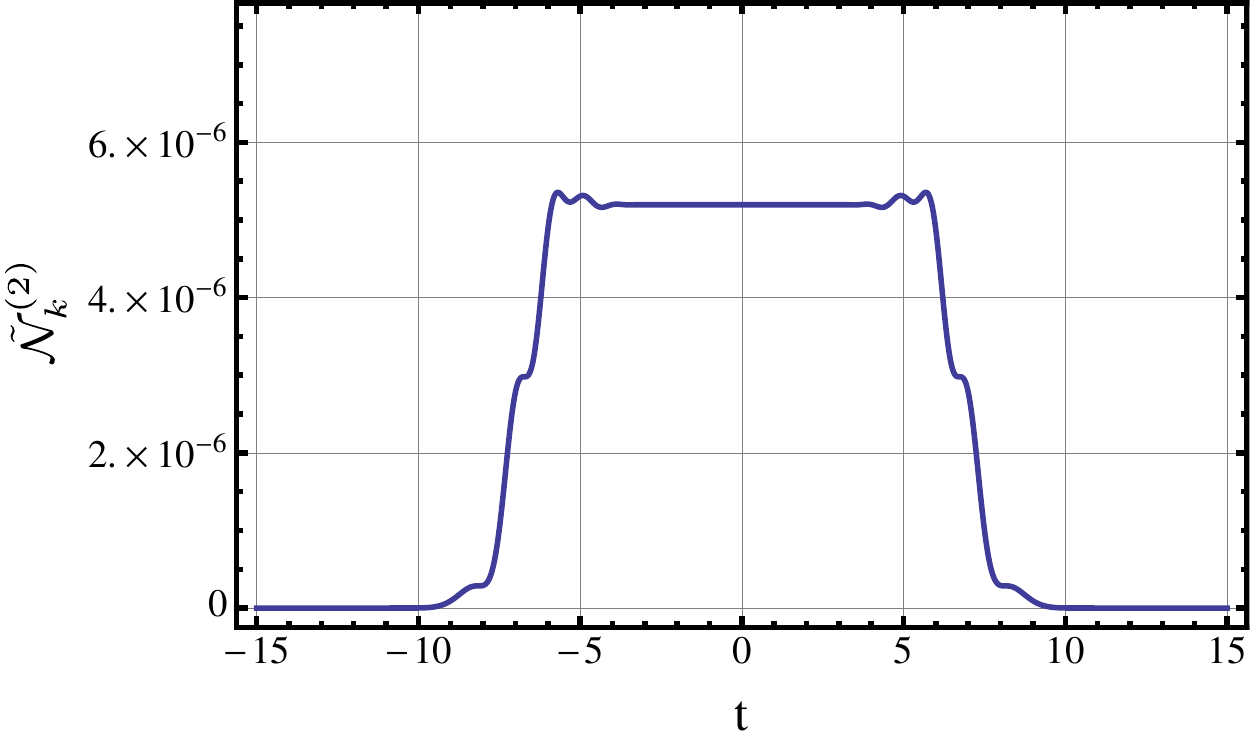} 
& 
\includegraphics[width=7.5cm]{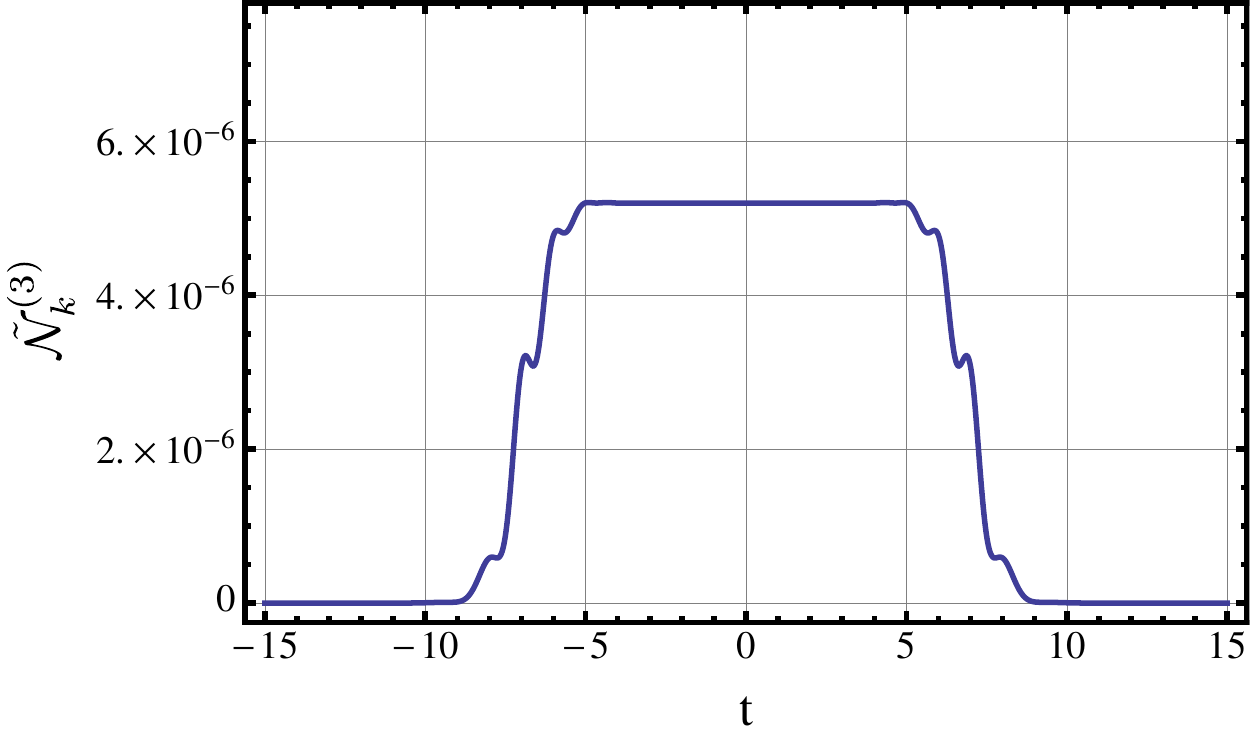} \\
\includegraphics[width=7.5cm]{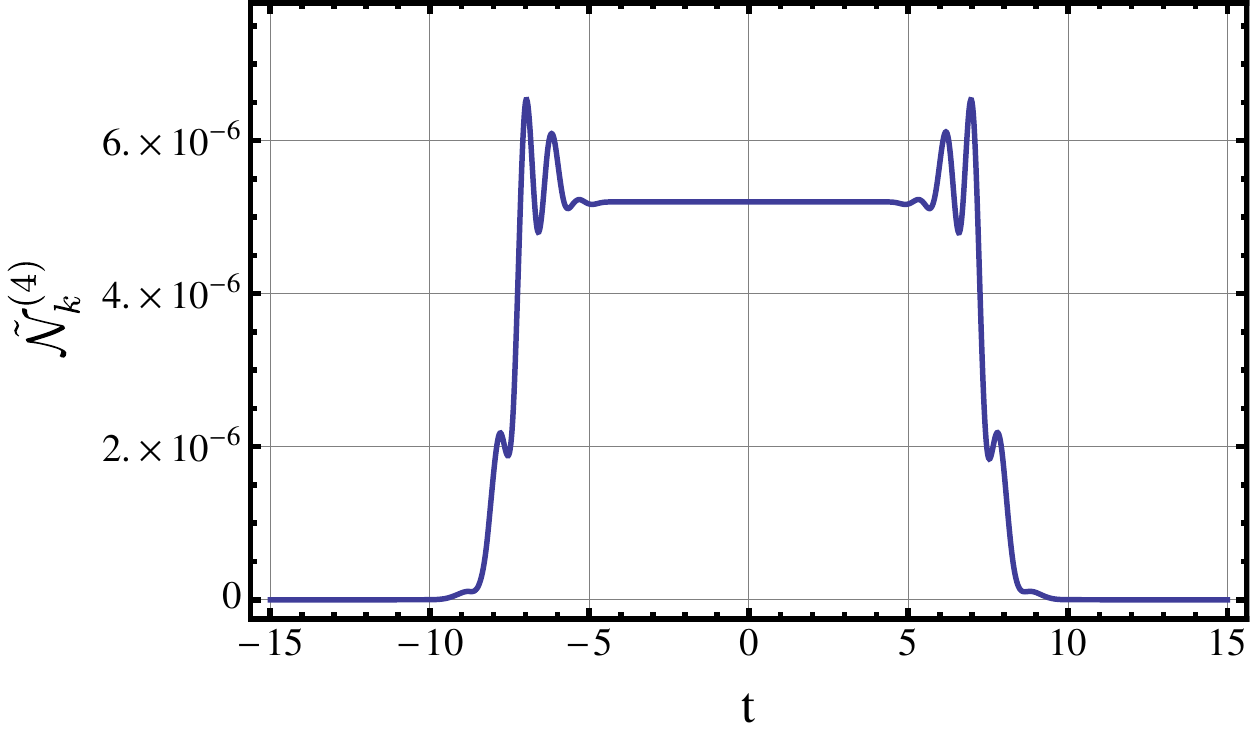} 
& 
\includegraphics[width=7.5cm]{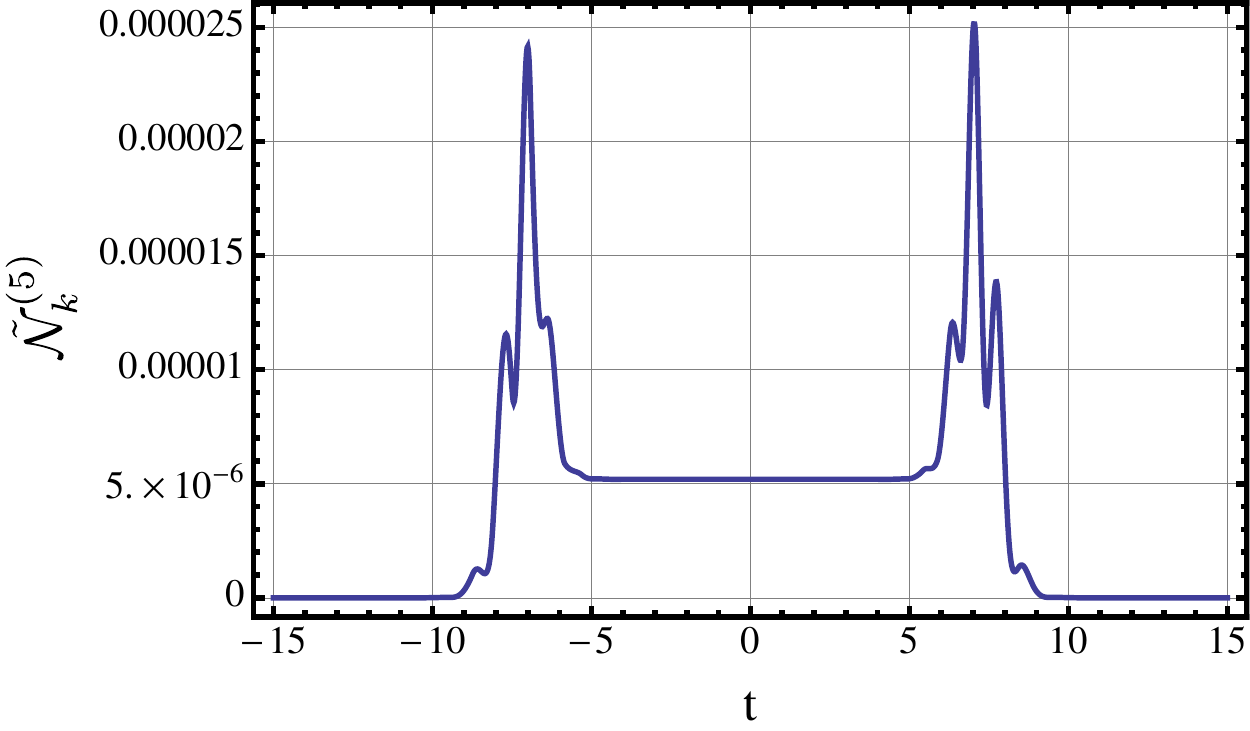}
\end{tabular}
\caption{
Time evolution of the adiabatic particle number for the first 6 orders of the adiabatic expansion, for  particle production in three-dimensional de Sitter space  (\ref{parker-osc1}, \ref{parker-osc2}), with parameters: $k = 25$ and $H = 0.5$, in units with $m=1$. We use conformal coupling so that $\xi=\frac{1}{8}$ in $d=3$. Note the similarity with the case of a two-alternating-sign-pulse electric field in the Schwinger effect, with destructive interference, as in Figure \ref{2-pulse-destructive}. The final asymptotic value of the particle number, at future infinity, is the same for all orders of truncation. The final asymptotic value of the particle number vanishes, due to coherent destructive quantum interference,  consistent with the analysis of \cite{Kim:2010xm}. At intermediate times there are large oscillations in the particle number, which become much smaller as the optimal order ($j=3$) is reached, and then grow again rapidly beyond this optimal order of truncation. Such behavior is characteristic of {\it asymptotic} expansions, where the order of truncation depends  on the size of the expansion parameter, and going beyond this optimal order typically yields increasingly worse results.  
}
\label{fig:ds3}
\end{figure}

\subsection{Super-Adiabatic Particle Number in 3d de Sitter space: coherent destructive interference}

We compute the adiabatic particle number numerically using (\ref{abWdot2}, \ref{abWdot2factor}), for various orders $j$ of the adiabatic expansion, starting with the time-dependent frequency for three dimensional de Sitter space, from (\ref{parker-osc1}, \ref{parker-osc2}). We consider conformal coupling, so that $\xi=\frac{1}{8}$ in 3 dimensions.
The results are shown in Figure \ref{fig:ds3}. Note the similarity to the two-alternating-sign-pulse Schwinger effect, with particle momentum such that the interference is destructive, as shown in Figure \ref{2-pulse-destructive}. In the leading order of the adiabatic expansion, there are two large oscillatory peaks, several orders larger than the final asymptotic particle number. As the order of the adiabatic expansion increases these oscillations become smaller, and at the optimal order ($j=3$) the super-adiabatic particle number evolves much more smoothly. Moreover, we clearly see the two-step structure, with destructive interference leading to the final asymptotic result of zero net particle production. Also note that, as in the case of the Schwinger effect shown in Figure  \ref{2-pulse-destructive}, the smooth evolution and interference is most evident at the optimal order, namely for the super-adiabatic particle number, with the large oscillations returning as one goes beyond the optimal order.

\section{Conclusions}

In this paper we have studied the time evolution of the adiabatic particle number for particle production in time-dependent electric fields (the Schwinger effect) and in de Sitter space. We examined various orders of the adiabatic expansion, noting  the well-known fact that the  time evolution at intermediate times is highly sensitive to the truncation order, even though the final particle number at future infinity is independent of this order. Nevertheless, defining the {\it super-adiabatic} particle number as that corresponding to the optimal truncation of the adiabatic expansion, we found very good agreement with Berry's universal error-function form of the time evolution (\ref{answer}) for a single pulse creation event, and with our generalization (\ref{answer2}) for backgrounds with structure corresponding to multiple particle creation events. The phase differences between different sets of turning points incorporate the physics of quantum interference, which has a significant impact on the final particle number. The universality of these super-adiabatic results means that in fact one does not need to make the explicit adiabatic expansion to the optimal order, which  is complicated at high orders, and moreover for which the optimal order depends on the physical parameters. Instead, equations (\ref{answer}) and (\ref{answer2}) express the universal time-evolution form at the optimal order in a simple and compact way. As  illustrations, we have verified the accuracy of this result in a variety of circumstances.

This super-adiabatic particle number typically evolves in time much more smoothly than at low orders of the adiabatic expansion. Furthermore, the resulting time evolution reveals the quantum interference processes at work in the Stokes phenomenon, which is ultimately responsible for particle production.
Depending on the phase accumulated by the Bogoliubov coefficients between successive turning points, the net particle number at future infinity can be understood in terms of quantum interference between different particle creation events. For the Schwinger effect we illustrated this for sequences of two and three alternating-sign electric field pulses, which can exhibit both constructive or destructive interference, depending on the momentum of the produced particles. For particle production in global de Sitter space, the distinction between constructive and destructive interference lies in the space-time dimensionality, with even dimensional de Sitter space producing a net particle number by coherent constructive interference, and odd  dimensional de Sitter space producing zero net particle number, due to coherent destructive interference.

Some important open questions include: (i) the behavior of the induced current and the energy momentum tensor in this super-adiabatic basis; (ii) the relation to the observer and possible measurement processes; (iii) back reaction effects; (iv) the relation to other formulations  of particle production, other than this Bogoliubov transformation formalism. These will be addressed in future work.

\bigskip

We acknowledge support from the U.S. DOE grant DE-FG02-13ER41989,  the University of Connecticut Research Foundation, the German DFG through the Mercator Guest Professor Program (G.D.), and the German Fulbright Commission (G.D.).
Both authors thank the Theoretisch-Physikalisches Institut in Jena, especially Holger Gies and Andreas Wipf, for their hospitality and support, and G.D. thanks Eric Akkermans and the Technion Physics Department for hospitality and support in Haifa where part of this work was done. 

%

\end{document}